\documentclass[traditabstract,longauth]{aa}  
\usepackage{graphicx}

\usepackage{txfonts}

\usepackage{multirow}

\usepackage{hyperref}
\hypersetup{colorlinks,citecolor=blue}
\newcommand{\Hii}{H\thinspace\textsc{ii}\xspace}
\newcommand{\Ha}{\ifmmode \mathrm{H}\alpha \else H$\alpha$\fi\xspace}
\newcommand{\Hb}{\ifmmode \mathrm{H}\beta \else H$\beta$\fi\xspace}
\newcommand{\nii}{\ifmmode [\mathrm{N}\,\textsc{ii}] \else [N~{\scshape ii}]\fi\xspace}
\newcommand{\oiii}{\ifmmode [\rm{O}\,\textsc{iii}] \else [O\,{\sc iii}]\fi\xspace}
\newcommand{\sii}{\ifmmode [\mathrm{S}\,\textsc{ii}] \else [S~{\scshape ii}]\fi\xspace}
\newcommand\logMt{\log M_\star}
\newcommand\ageMt{\langle \log\,\mathrm{t}\rangle_\mathrm{L}}

\newcommand{\photoz}{photo-$z$}

\begin{document} 

\titlerunning{miniJPAS emission line properties in the AEGIS field}
\title{The miniJPAS survey: Identification and characterization of the emission line galaxies down to $z < 0.35$ in the AEGIS field}

\author{G. Mart\'inez-Solaeche \inst{\ref{1}}
 \and R. M. Gonz\'alez Delgado\inst{\ref{1}} \and
R. Garc\'ia-Benito \inst{\ref{1}} \and L.~A.~D\'iaz-Garc\'ia   \inst{\ref{1}} \and J. E. Rodr\'iguez-Mart\'in  \inst{\ref{1}} \and E. P\'erez \inst{\ref{1}} \and A. de Amorim \inst{\ref{2}} \and S. Duarte Puertas \inst{\ref{3},} \inst{\ref{1}} \and Laerte Sodr\'e Jr. \inst{\ref{4}} \and David Sobral \inst{\ref{5}} \and Jon\'as Chaves-Montero \inst{\ref{6}} \and J. M.~V\'ilchez \inst{\ref{1}} \and A. Hern\'an-Caballero \inst{\ref{7}} \and C.~L\'opez-Sanjuan \inst{\ref{8}} \and A.~Cortesi 
\inst{\ref{9},} \inst{\ref{20}} \and S.~Bonoli \inst{\ref{6},} \inst{\ref{10}} \and A.~J.~Cenarro\inst{\ref{8}} \and R.~A.~Dupke\inst{\ref{11},}\inst{\ref{12},}\inst{\ref{13}} \and A.~Mar\'in-Franch\inst{\ref{8}} \and J.~Varela\inst{\ref{8}} \and
H.~V\'azquez~Rami\'o\inst{\ref{8}} \and L.~R.~Abramo\inst{\ref{14}} \and
D.~Crist\'obal-Hornillos\inst{\ref{7}} \and
M.~Moles\inst{\ref{7},}\inst{\ref{1}} \and J.~Alcaniz\inst{\ref{11}}
\and N.~Benitez\inst{\ref{1}} \and
A.~Ederoclite\inst{\ref{7},} \inst{\ref{15}}\and
V.~Marra\inst{\ref{16},} \inst{\ref{17},} \inst{\ref{18}}\and
C.~Mendes~de~Oliveira \inst{\ref{4}} \and K.~Taylor\inst{\ref{19}} \and
J.A.~Fern\'andez-Ontiveros  \inst{\ref{7}} }

\institute{Instituto de Astrof\'isica de Andaluci\'a (CSIC), PO Box 3004, 18080 Granada, Spain (\email{gimarso@iaa.es}) \label{1} \and 
Departamento de F\'isica, Universidade Federal de Santa Catarina, P.O. Box 476, 88040-900, Florian\'opolis, SC, Brazil \label{2} \and
D\'epartement de Physique, de G\'enie Physique et d'Optique, Universit\'e Laval, and Centre de Recherche en Astrophysique du Qu\'ebec (CRAQ), Qu\'ebec, QC, G1V 0A6, Canada \label{3} \and
Universidade de SÃ£o Paulo, Instituto de Astronomia, Geof\'isica e Ci\^encias Atmosf\'ericas, R. do Matão 1226, 05508-090, São
Paulo,Brazil \label{4} \and
Department of Physics, Lancaster University, Lancaster, LA1 4YB, UK \label{5} \and
Donostia International Physics Center, Paseo Manuel de Lardizabal 4, E-20018 Donostia-San Sebastian, Spain \label{6} \and
Centro de Estudios de F\'isica del Cosmos de Arag\'on (CEFCA), Plaza San Juan, 1, 44001 Teruel, Spain \label{7} \and
Centro de Estudios de F\'isica del Cosmos de Arag\'on (CEFCA), Unidad Asociada al CSIC, Plaza San Juan, 1, E-44001 Teruel, Spain\label{8} \and
 Observat\'orio do Valongo, Universidade Federal do Rio de Janeiro, 20080-090, Rio de Janeiro, RJ, Brazil \label{9} \and
 Ikerbasque, Basque Foundation for Science, E-48013 Bilbao, Spain \label{10} \and
 Observat\'orio Nacional, Rua General Jos\'e Cristino, 77, São Crist\'ovão, 20921-400, Rio de Janeiro, Brazi  \label{11} \and
Department of Astronomy, University of Michigan, 311 West Hall, 1085 South University Ave., Ann Arbor, USA \label{12} \and
Department of Physics and Astronomy, University of Alabama, Box 870324, Tuscaloosa, AL, USA \label{13} \and
Instituto de F\'isica, Universidade de São Paulo, Rua do Matão 1371, CEP 05508-090, São Paulo, Brazil \label{14} \and
Departamento de Astronomia, Instituto de Astronomia, Geof\'isica e Ciências Atmosféricas da USP, Cidade Universit\'aria, 05508-
900, São Paulo, SP, Brazil  \label{15} \and
PPGCosmo \& Departamento de F\'sica, Universidade Federal do Esp\'irito Santo, 29075-910, Vit\'oria, ES, Brazil
\label{16} \and
INAF, Osservatorio Astronomico di Trieste, via Tiepolo 11, 34131 Trieste, Italy\label{17} \and
IFPU, Institute for Fundamental Physics of the Universe, via Beirut 2, 34151, Trieste, Italy
\label{18} \and
Instruments4, 4121 Pembury Place, La Cañada-Flintridge, Ca 91011, USA  \label{19} \and
Centro Brasileiro de Pesquisas F\'isicas, Rua Dr. Xavier Sigaud 150, CEP 22290-180, Rio de Janeiro, RJ, Brazil \label{20}} 
\date{}

   \abstract{The Javalambre-Physics of the Accelerating Universe Astrophysical Survey (J-PAS) is expected to map thousands of square degrees of the northern sky with 56 narrowband filters (spectral resolution of $R \sim 60$) in the upcoming years. This resolution allows us to study emission line galaxies (ELGs) with a minimum equivalent width of 10 \AA $ $ in the \Ha emission line for a median signal-to-noise ratio (S/N) of 5. This will make J-PAS a very competitive and unbiased emission line survey compared to spectroscopic or narrowband surveys with fewer filters. The miniJPAS survey covered 1 deg$^2$ , and it used the same photometric system as J-PAS, but the observations were carried out with the pathfinder J-PAS camera. In this work, we identify and characterize the sample of ELGs from miniJPAS with a redshift lower than $0.35$, which is the limit to which the \Ha line can be observed with the J-PAS filter system. Using a method based on artificial neural networks, we detect the ELG population and measure the equivalent width and flux of the \Ha, \Hb, \oiii, and \nii emission lines. We explore the ionization mechanism using the diagrams [OIII]/H$\beta$ versus [NII]/H$\alpha$ (BPT) and EW(H$\alpha$) versus [NII]/H$\alpha$ (WHAN). We identify 1787 ELGs ($83$\%) from the parent sample (2154 galaxies) in the AEGIS field. For the galaxies with reliable EW values that can be placed in the WHAN diagram (2000 galaxies in total), we obtained that $72.8 \pm 0.4$~\%, $17.7 \pm 0.4$~\% , and $9.4 \pm 0.2$~\% are star-forming (SF), active galactic nucleus (Seyfert), and quiescent galaxies, respectively. The distribution of EW(H$\alpha$) is well correlated with the bimodal color distribution of galaxies. Based on the rest-frame $(u-r)$--stellar mass diagram, 94\% of the blue galaxies are SF galaxies, and 97\% of the red galaxies are LINERs or passive galaxies.
   The nebular extinction and star formation rate (SFR) were computed from the \Ha and \Hb fluxes. We find that the star formation main sequence is described as $\log$ SFR $[M_\mathrm{\odot} \mathrm{yr}^{-1}] = 0.90^{+ 0.02}_{-0.02} \log M_{\star} [M_\mathrm{\odot}] -8.85^{+ 0.19}_{-0.20}$ and has an intrinsic scatter of $0.20^{+ 0.01}_{-0.01}$. The cosmic evolution of the SFR density ($\rho_{\text{SFR}}$) is derived at three redshift bins: $0 < z \leq 0.15$, $0.15 < z \leq 0.25$, and $0.25 < z \leq 0.35$, which agrees with previous results that were based on measurements of the \Ha emission line. However, we find an offset with respect to other estimates that were based on the star formation history obtained from fitting the spectral energy distribution of the stellar continuum. We discuss the origin of this discrepancy, which is probably a combination of several factors: the escape of ionizing photons, the SFR tracers, and dust attenuation, among others.
   }
 
   \maketitle
\section{Introduction}
The \Ha emission line is an excellent tracer for estimating the current star formation rate (SFR) in galaxies because it is less affected by dust extinction than UV light \citep{1998ARA&A..36..189K,2010MNRAS.402.2017G,2015MNRAS.452.2018O,2015A&A...584A..87C}. The  \Ha line can be observed in the optical range up to $z \sim 0.4$. {Thus, it is very useful for the identification of emission line galaxies (ELGs) in spectroscopic and photometric surveys.}
The detection of other emission lines, such as \oiii$\lambda \lambda 4959, 5007$  \AA $ $ and the \nii$\lambda \lambda 6548, 6584 $  \AA $ $ doublets\footnote{In the remaining paper, \oiii$\lambda 5007$  and  \nii$\lambda6584$ are denoted \oiii and \nii, respectively.} {, is crucial to determine the main ionization mechanism of ELGs \citep[see, e.g.,][]{2011MNRAS.413.1687C,2016MNRAS.461.3111B,2018RMxAA..54..217S,2020MNRAS.492.3073L,2021A&A...648A..64K}. Diagrams such as the WHAN (EW(H$\alpha$) vs. [NII]/H$\alpha$) \citep{2011MNRAS.413.1687C} or the BPT \citep{1981PASP...93....5B} (e.g., [OIII]/H$\beta$ vs. [NII]/H$\alpha$)  can differentiate galaxies in which the gas is ionized by young stars or by an active galactic nucleus (AGN), from low ionization nuclear emission regions \citep[LINERs,][]{1980A&A....87..152H}, or extended low-ionization emission lines \citep[see, e.g.,][]{2018MNRAS.474.3727L}, in which the ionization might be attributed to old and hot stars.  }
Furthermore, the characterization of the galaxy populations through the SFR and its correlation with other galaxy properties, such as stellar mass, colors, ages, metallicity, and neutral gas content \citep{2019ARA&A..57..511K,2020ARA&A..58..661F}, is essential to obtain insight into the formation and evolution of galaxies.
\par Galaxies grow in mass mainly through star formation, which is fed by gas accretion from the cosmic web. While massive galaxies undergo a larger fraction of their star formation at early times, less massive galaxies are still forming stars at a high rate today. The star formation main sequence (SFMS), a tight quasi-linear relation between stellar mass, ($M_\star$), and the SFR in log scale \citep{2012ApJ...757...54Z,2015ApJ...801L..29R,2016ApJ...821L..26C,2017A&A...599A..71D,2018MNRAS.477.3014B,2018A&A...619A..27B,2019MNRAS.482.1557S,2019MNRAS.488.3929C,2021MNRAS.501.2231S,2021arXiv210104062V}, can reveal indications how this process takes place. Galaxies that are undergoing a starburst, for instance, lie above the SFMS, while galaxies that have already quenched their star formation lie below this relation.
\par The SFMS and its evolution with redshift are expected outcomes of hydrodynamical models. The currently best cosmological hydrodynamical simulations of galaxy formation such as Illustris \citep{2015MNRAS.447.3548S} or EAGLE \citep{2015MNRAS.450.4486F} predict a slope near unity. Semi-analytical models favor a sublinear slope that is generally higher than 0.8. For instance, \cite{2010MNRAS.405.1690D} predicted a slope of 0.96 for galaxies with stellar masses between $10^9$ and $10^{11} M_\mathrm{\odot} $. However, \cite{2014MNRAS.444.2637M} used GALFORM and retrieved a slope of 0.87 at $z = 0.1$. 
\par The slope of the SFMS in observations ranges from 0.6 to 1, depending on the data, the SFR tracer, and method used \citep[see, e.g., the study of][and references therein]{2014ApJS..214...15S}. The discrepancies found by different studies are expected. On the one hand, spectroscopic surveys such as the Sloan Digital Sky Survey \citep[SDSS,][]{2000AJ....120.1579Y} have aperture effects that can cause an underestimation of the total SFR within the galaxy \citep{2017A&A...599A..71D}. On the other hand, the SFR derived from photometric surveys throughout \Ha measurements needs to be corrected for the \nii and dust extinction, which become the main sources of uncertainty. 
\par The definition of the SFMS itself might also lead to significant differences between different works, even though they all trace the SFR through the \Ha line. Some authors (e.g., \citet[][$z \leq 0.017$]{2021arXiv210104062V} or \citet[][$ z \sim 0.07-0.5$]{2021MNRAS.501.2231S}) relied on color-color diagrams. Others selected star-forming (SF) galaxies based on the BPT diagrams with a cut in the equivalent width (EW) of \Ha or \Hb. For example, \citet[][$0.005 \leq z \leq 0.03$]{2016ApJ...821L..26C} imposed a minimum EW in \Ha of 6 \AA $ $ while \citet[][$0.005 \leq z \leq 0.22$]{2017A&A...599A..71D} used instead 3 \AA $ $ and \citet[][$z = 0.07, 0.8$ and $2.26$]{2012ApJ...757...54Z} adopted a EW of 4 \AA $ $ in \Hb. In addition, the SFMS has also been defined as the ridge line in the M$\star$-N-SFR- plane where N account for the number of galaxies in every M$\star$-SFR bin \citep[$0.02 \leq z \leq 0.085$,][]{2015ApJ...801L..29R}.
\par In essence, there is no unique and homogeneous definition of the galaxies that belong to the SFMS. Furthermore, any dividing line between star-forming and quiescent galaxies affects the analysis of the SFMS because it includes or excludes some of the galaxies in the the so-called `green valley' (GV), that is, galaxies that are in transition and are interpreted as a crossroads in galaxy evolution \citep[see, e.g.,][]{2011ApJ...736..110M,2012ApJ...759...67G,2014MNRAS.440..889S,2019A&A...631A.156D}.  \citet[$0.03 \leq z \leq 0.2$,][]{2019MNRAS.482.1557S} attributed the constancy of the SFMS slope across galaxy mass to the selection criterion  (based on sSFR cut). There is no drop in the SFR at high masses. In the same vein, \citet[$0.03 \leq z \leq 0.15$,][]{2018MNRAS.477.3014B}, who also used the \Ha line as an SFR tracer, found that the flattening in the slope of the SFMS only occurs if galaxies with quiescent central regions (cLIERs) are included in the fit. 
\par In addition, the detection limit and particularities of each study might lead to a specific bias in the selection criteria. For instance, a photometric survey that selects ELGs based on a minimum contrast would be limited to the minimum EW that can be measured and would therefore be biased toward highly actively SF galaxies. As a consequence, it produces an increase in normalization constant and a shallower slope \citep{2021MNRAS.503.5115K}. Finally, the minimization method employed in the fitting takes the uncertainties into account in different ways. It might therefore also have an impact on the shape of the SFMS.
\par Another important aspect that helps to understand how galaxies assemble their mass throughout cosmic time is estimating the intrinsic scatter of galaxies in the SFMS. It is expected that low-mass galaxies are more sensitive to stochastic events such as starbursts or feedback from supernovae. Theoretical simulations \citep{2014MNRAS.445..581H,2015MNRAS.451..839D,2019MNRAS.484..915M} and observations \citep{2007ApJS..173..267S,2019ApJ...881...71E,2018A&A...619A..27B,2020MNRAS.493..141S} have both found an increase in scatter for low-mass galaxies ($< 10^9 M_\mathrm{\odot}$~).
\par Other studies \citep{2015MNRAS.449..820W,2019MNRAS.483.1881D} found that the dispersion along the SFMS follows a U-shaped distribution, meaning that galaxies with high and low stellar masses scatter more from the SFMS. Interestingly, the U-shape depends on the way the SFMS is defined. While selecting SF galaxies based on $u - r$ colors or morphology causes the SFMS to have higher scatter for galaxies at high mass, a selection based on a minimum sSFR, which is equivalent to a minimum EW in \Ha, produces a decrease in scatter as the mass of the galaxy increases \citep[see, e.g., ][]{2019MNRAS.483.1881D}.
\par It has been proven by the analysis of stellar populations within galaxies through stellar continuum spectral energy distribution (SED) fitting that the SFMS holds true at high redshift with an increase in the global SFRs of galaxies \citep{2004ApJ...617..746D,2010MNRAS.405.2279O,2011ApJ...730...61K,2015A&A...579A...2I,2015A&A...575A..74S,2015A&A...581A..54T,2011ApJ...739L..40R}. In terms of the SFR density ($\rho_{\text{SFR}}$), the Universe reached a peak at $\sim 3$ Gyr after the Big Bang, and it has been decreasing ever since \citep{2014ARA&A..52..415M,2018MNRAS.475.2891D,2018A&A...615A..27L,2019MNRAS.482.1557S,2019ApJ...876....3L,2020MNRAS.498.5581B}. Through \Ha measurements, astronomers are also able to measure  $\rho_{\text{SFR}}$ both in the nearby Universe and at intermediate redshift, which has confirmed this trend \citep{1995ApJ...455L...1G,2007ApJ...657..738L,2008ApJS..175..128S,2010ApJ...712L.189D,2010ApJ...708..534W,2013MNRAS.433..796D,2013MNRAS.428.1128S,2013MNRAS.433.2764G,2015MNRAS.451.2303S,2015MNRAS.453..242S,2016ApJ...824...25V,2020MNRAS.493.3966K,2021arXiv210104062V}.
\par The incredible progress achieved in the past decades would not have been possible without the construction of large galaxy surveys. Multi-object spectroscopy (MOS) surveys such as the SDSS and the the Galaxy And Mass Assembly \citep[GAMA;][]{2011MNRAS.413..971D} or integral field unit (IFU) surveys such as the Calar Alto Legacy Integral Field Area \citep[CALIFA; ][]{sanchez2012, garcia-benito2015, sanchez2016} and the survey Mapping Nearby Galaxies at the Apache Point Observatory \citep[MaNGA; ][]{bundy2015, law2015} provide a very detailed description of the optical SED of galaxies. However, they are partially biased through their preselection of samples, which is driven by some properties such as redshift, fluxes, or a galaxy size that is constrained to a particular range. 
\par In contrast, narrowband photometric surveys such as HiZELS \citep{2013ASSP...37..235B,2013MNRAS.428.1128S,2017MNRAS.471..629M}, ALHAMBRA \citep{2008AJ....136.1325M,2014MNRAS.441.2891M},
DAWN \citep{2018ApJ...858...96C}, J-PLUS \citep{2019A&A...622A.176C}, S-PLUS \citep{2019MNRAS.489..241M}, the Deep and UDeep layers driven by the Subaru Strategic Program with the Hyper Suprime-Cam (HSC-SSP) \citep{2018PASJ...70S..17H,2020PASJ...72...86H}, LAGER \citep{2020MNRAS.493.3966K}, or SHARDS \citep{2013ApJ...762...46P,2019A&A...621A..52L}, experience these effects to a lesser degree. In particular, narrowband photometric surveys are able to detect fainter objects than their spectroscopic counterpart at a fixed exposure time. Furthermore, they can fully observe galaxies whose light cannot be captured entirely by IFU-like surveys \citep[see, e.g., Fig. 19 in ][]{2021A&A...653A..31B}. However, their SED in the optical, infrared, or UV is limited by the number of filters and their width. More importantly, ELGs can only be detected in certain redshift intervals, which makes contamination from other sources more likely because the emission lines may be confused; for example, \oiii emitters may be detected as \Ha emission line objects. 
\par The special design of the Javalambre Physics of the Accelerating Universe Astrophysical Survey \citep[J-PAS,][]{2014arXiv1403.5237B} enables overcoming some of the caveats for spectroscopic and traditional photometric surveys. J-PAS will play a crucial role in the upcoming years, which will be very competitive compared to the new generations of spectroscopic surveys such as DESI \citep{2016arXiv161100036D},  {\it Euclid} \citep{2011arXiv1110.3193L}, or the WHT Enhanced Area Velocity Explorer-Stellar Population at intermediate redshift Survey \citep[WEAVE-StePS;][]{2019A&A...632A...9C}.
\par The unprecedented area that J-PAS will cover ($\sim 8000$ deg$^2$ of the northern sky) is perhaps one of the main advantages compared to previous and current surveys. J-PAS will observe the sky with 56 bands: 54 narrowband filters in the optical range, plus two medium-band filters, one in the UV and another in the near-infrared. Separated by $100$ \AA,  each narrowband filter has a width of $ \sim 145 $ \AA, which provides a resolving power of $ R \sim 60 $ (J-spectrum hereafter). These unique characteristics make J-PAS an ideal survey for galaxy evolution studies \citep{2021A&A...653A..31B}, superseding the scientific impact achieved by other previous medium-band imaging surveys, such as ALHAMBRA ($ R \sim 20 $). The narrowband setup of J-PAS allows the detection and measurement of galaxies with emission lines in a continuous range in redshift within a nonsegregated area \citep[hereafter MS21]{MS21}. J-PAS observations will be carried out with the 2.55 m telescope (T250) at the Observatorio Astrof\'isico de Javalambre, a facility developed and operated by the Centro the Estudios de F\'isica del Cosmos de Arag\'on (CEFCA, in Teruel, Spain) using JPCam, a wide-field 14 CCD-mosaic camera with a pixel scale of 0.2267 arcsec/px and an effective field of view (FoV) of $\sim  4.7$ deg$^2$ \citep{2014JAI.....350010T,2015IAUGA..2257381M,2021A&A...653A..31B}.
\par The pathfinder camera of J-PAS started its observations using 60 optical bands in four fields of the sky that overlap with the All-wavelength Extended Groth Strip International survey \citep[AEGIS;][]{2007ApJ...660L...1D} , amounting to $1$deg$^2$ with more than $60\,000$ objects\footnote{http://www.j-pas.org/}; hereafter, this is referred to as the miniJPAS survey \citep{2021A&A...653A..31B}. The pathfinder instrument used by the J-PAS collaboration is a single CCD direct imager ($9.2k \times 9.2k$, $10\mu m$ pixel) located at the center of the T250 FoV with a pixel scale of 0.23 arcsec pix$^{-1}$, vignetted on its periphery. This provides an effective FoV of 0.27 deg$^2$.
\par The goal of this paper is to identify the ELG population in the AEGIS field and characterize them through their SFR and the stellar population properties. This work shows the potential of J-PAS data in this regard. We apply a method based on artificial neural networks (ANN) developed in MS21 to obtain the EW of the main emission lines in the optical range: \Ha, \Hb, \oiii, and \nii. Afterward, we analyze the main ionization mechanisms in galaxies through WHAN and BPT diagrams, and we compare the nebular properties of the gas with the properties of the stellar populations of their host galaxies derived in \cite{2021arXiv210213121G}. We characterize the SFR-M$_*$ relation derived from the flux of \Ha, and we compute the cosmic evolution of $\rho_{\text{SFR}}$ up to $z=0.35$.
\par This paper is organized as follows. In Sect.~\ref{sec:datasample} we present the galaxy sample taken from miniJPAS, which is the subject of this study. In Sect.~\ref{sec:method} we summarize the method we employed, which is based on previous works of MS21 and \cite{2021arXiv210213121G}. In Sect.~\ref{sec:main_results} we identify the ELG population by means of the EWs of the emission lines and their relations with the stellar population properties: stellar mass, intrinsic colors, luminosity-age, and so on. We derive the fraction of AGN, quiescent, and star-forming galaxies in miniJPAS. In Sect. \ref{sec:SFR} we characterize the star-forming galaxy population. We derive their SFR through \Ha emission, and we fit the SFMS. In Sect.~\ref{sec:discussion} we discuss the implications of our results in detail and compare them with previous works. We derive the $\rho_{\text{SFR}}$ up to $z=0.35$. Finally, we provide the outlook for J-PAS in Sect.~\ref{sec:outlook}, and we summarize in Sect.~\ref{sec:conclusions}. Throughout this work, we adopt a $\Lambda$CDM cosmology with $H_0 = 70 ~\mathrm{km}$ ~s$^{-1}$ ~$\mathrm{Mpc}^{-1}$, $\Omega_\mathrm{M} = 0.3,$ and $\Omega_\mathrm{\Lambda}= 0.7$. All magnitudes are presented in the AB system \citep{1983ApJ...266..713O}, and a \cite{chabrier2003} initial mass function (IMF) was employed.
\section{Sample and data}\label{sec:datasample}
The galaxy sample studied in this paper is a subsample of the galaxies analyzed in  \citet[][see Sect.~2.3]{2021arXiv210213121G}. We selected all the objects detected in miniJPAS with a photometric redshift (\photoz ) lower than $0.35$, which is the highest redshift at which \Ha can be observed in miniJPAS. The \photoz  $ $ was estimated with the \texttt{JPHOTOZ} package developed by the \photoz $ $ team at CEFCA. This package is a customized version of the \texttt{LePhare} code \citep{arnouts2011}, which has a new set of stellar population synthesis galaxy templates that were optimized for the miniJPAS filter system \citep{2021A&A...654A.101H}. At the depth of miniJPAS ($5 \sigma$ limits between $\sim 21.5$ and $22.5$ mag for the narrowband filters and $\sim 24$ mag for the broadband filters in a $3''$ aperture), there are $17\,500$ galaxies per deg$^2$ with valid \photoz $ $ estimates (rSDSS < 23), of which $\sim 4\,200$ have $|\Delta z| < 0.003$. The typical error for rSDSS $< 23$ galaxies is $\sigma_{\text{NMAD}} = 0.013$ with an outlier rate of $\eta=0.39$. The target \photoz $ $ accuracy $\sigma_{\text{NMAD}} = 0.003$ is achieved after imposing $odds > 0.82$ \citep[see][for details]{2021A&A...654A.101H}.
\par We imposed a maximum class-star probability of $0.1$, as defined in \texttt{SExtractor}, in order to select only extended sources. We discarded galaxies with an S/N lower than 1.8 in the filters to capture the flux of the emission lines. The estimates of the EWs with the ANN for galaxies with a very low S/N yield large errors. Therefore, these errors indicate the limit to which galaxies can be analyzed. For this reason, we favor a more conservative approach by setting a very low constraint on the S/N of the filters with which the flux of the emission lines is captured. Thus, we can exclude galaxies a posteriori when their EW measurements are not reliable. The magnitude limit cut of the sources was set at $22.5$~mag in the rSDSS band. This is the completeness limit for miniJPAS extended sources \citep{2021A&A...653A..31B}.  Finally, the sample is composed of 2154 galaxies in total. 
\par
\begin{figure}
  \centering
  \includegraphics[width=\hsize]{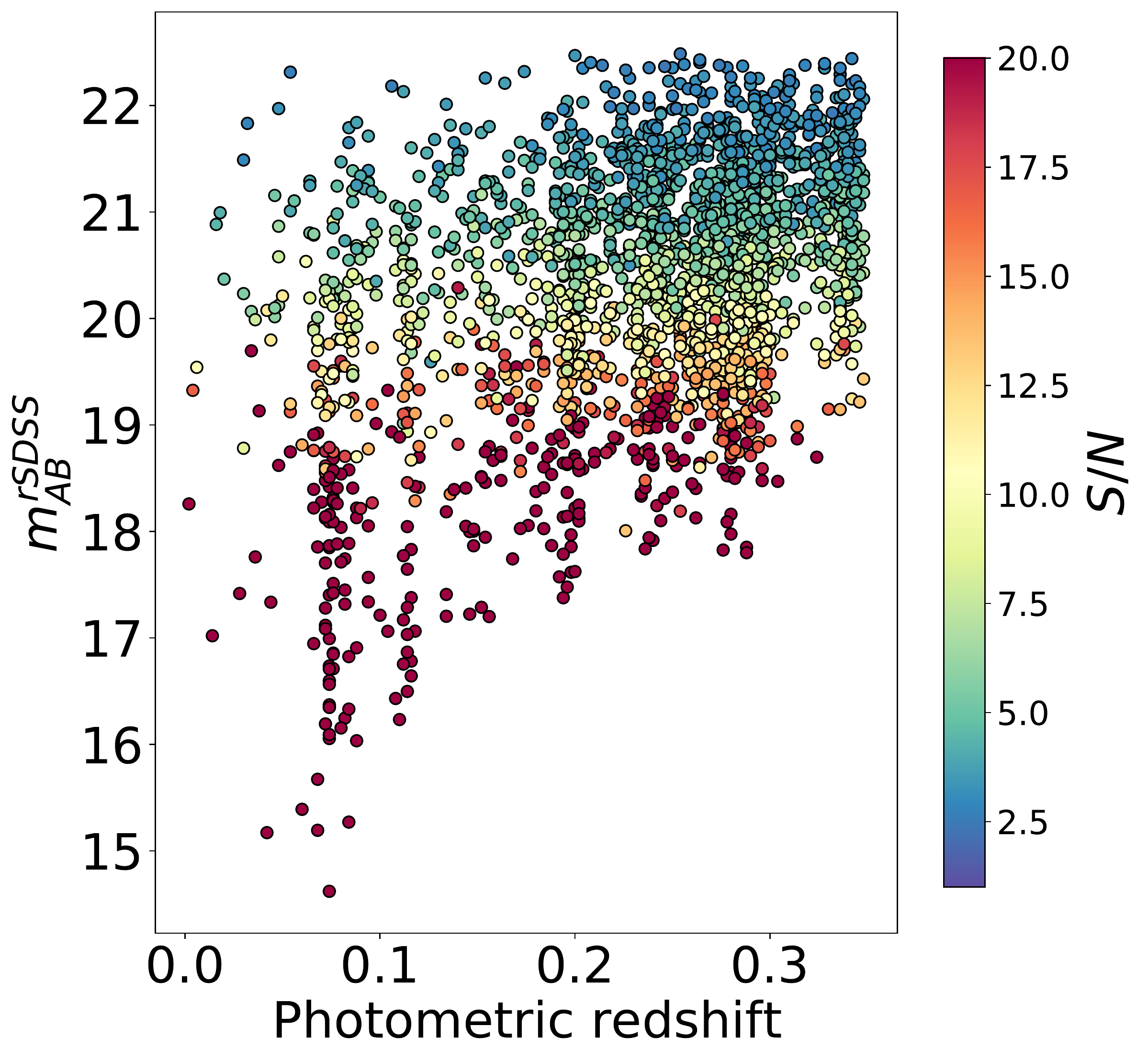}
  \caption{\tiny{Relation between the apparent magnitude in the rSDSS band and redshift for all galaxies in the parent sample. We used the \texttt{MAG\_AUTO} photometry. Dots are color-coded according to the median S/N of the J-PAS narrowband filters.}}
  \label{fig:sample}
  \end{figure}

 In Fig. \ref{fig:sample} we show the relation between the apparent magnitude in the rSDSS band and the redshift for the galaxies in the parent sample. The color bar indicates the median S/N measured in the J-PAS narrowband filters. In this work, we made use of the \texttt{MAG\_AUTO} photometry from the miniJPAS dual-mode catalog because it captures the entire light from the galaxy. Most of the galaxies in this sample ($ \sim $ 68 $\%$) are higher than 0.205 in redshift and have an S/N lower than 10.
\par
\begin{figure*}
        \includegraphics[width=6.4cm,height=5.7cm]{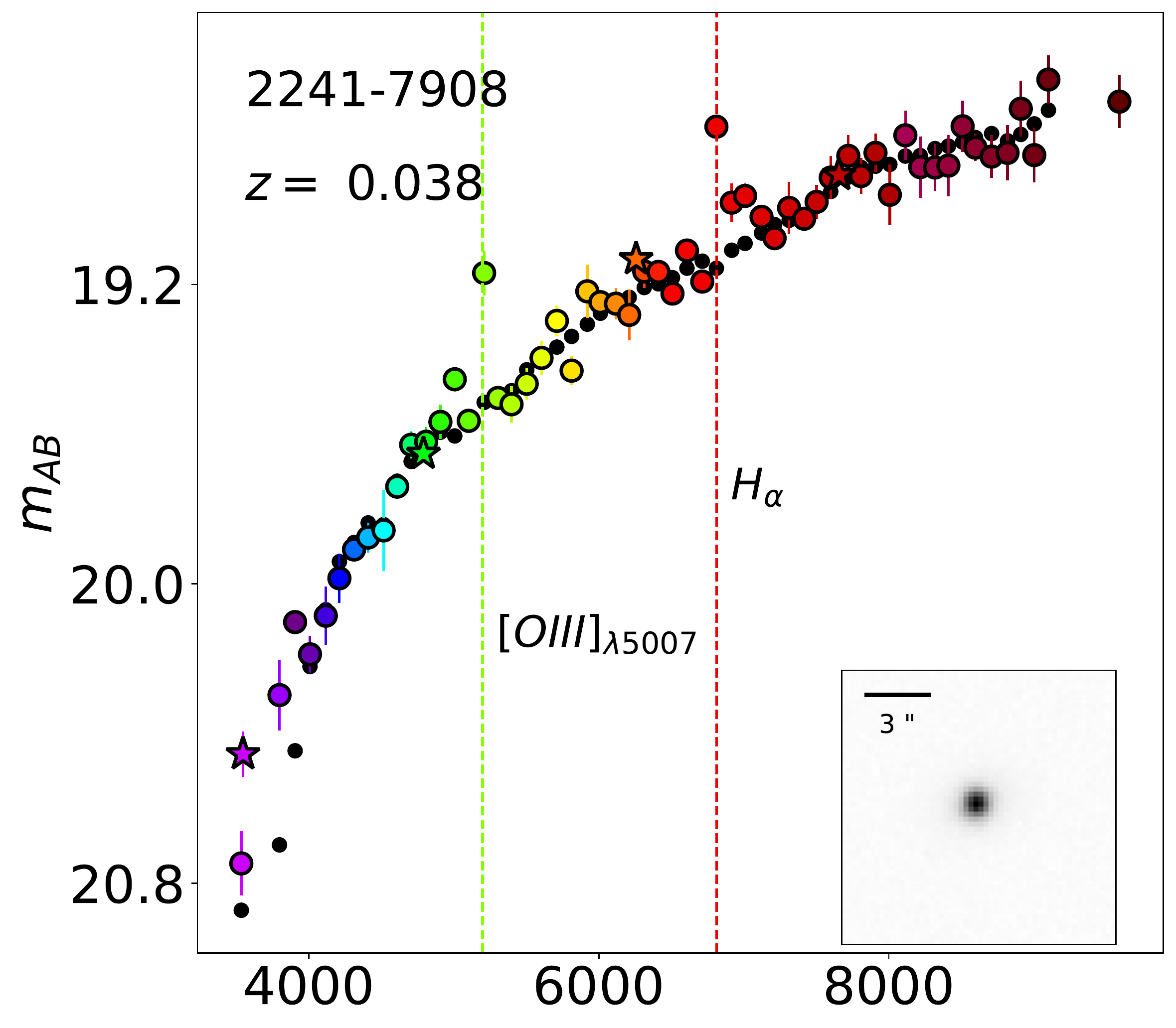} 
        \includegraphics[width=6.4cm,height=5.7cm]{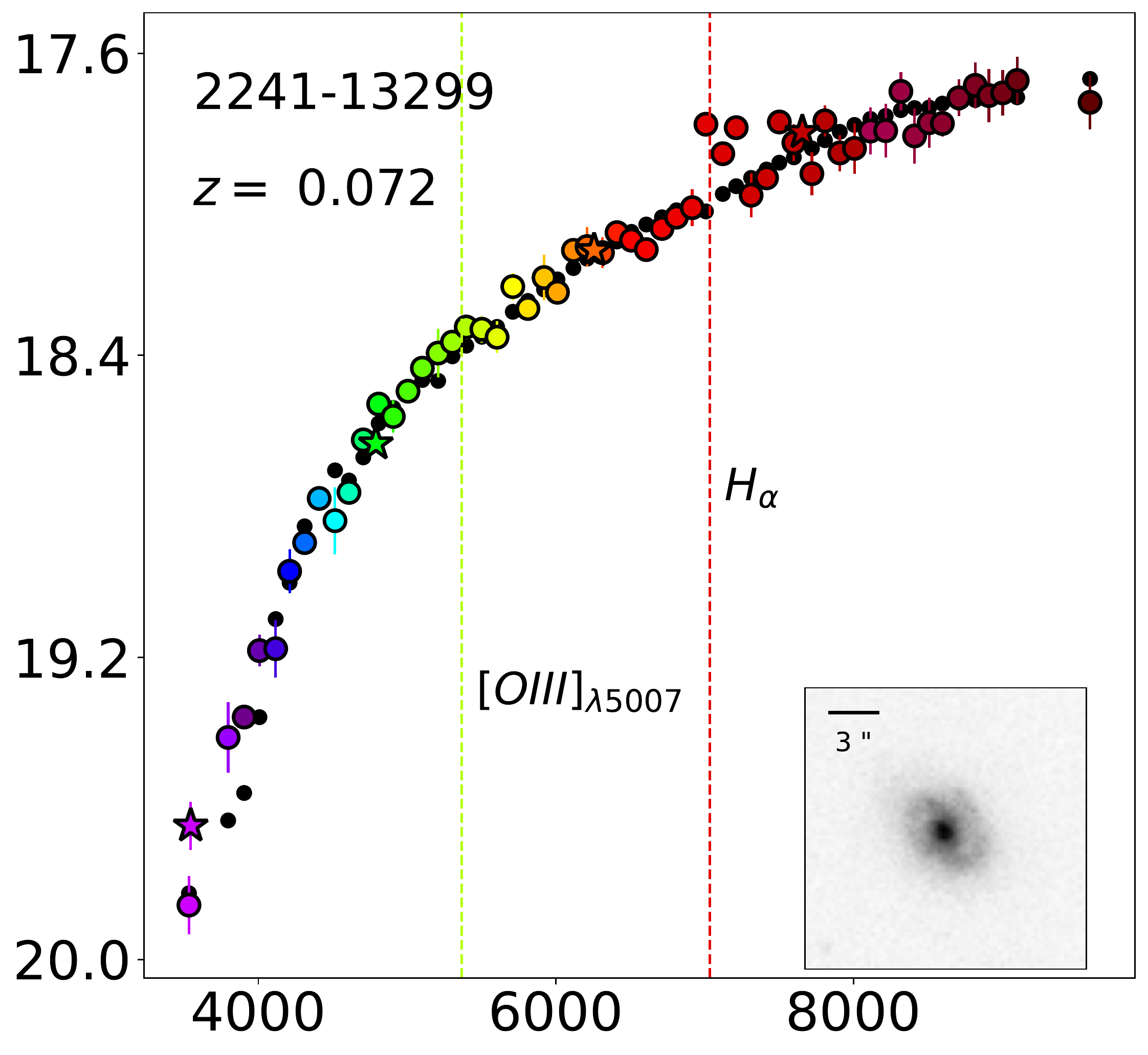}
        \includegraphics[width=6.4cm,height=5.7cm]{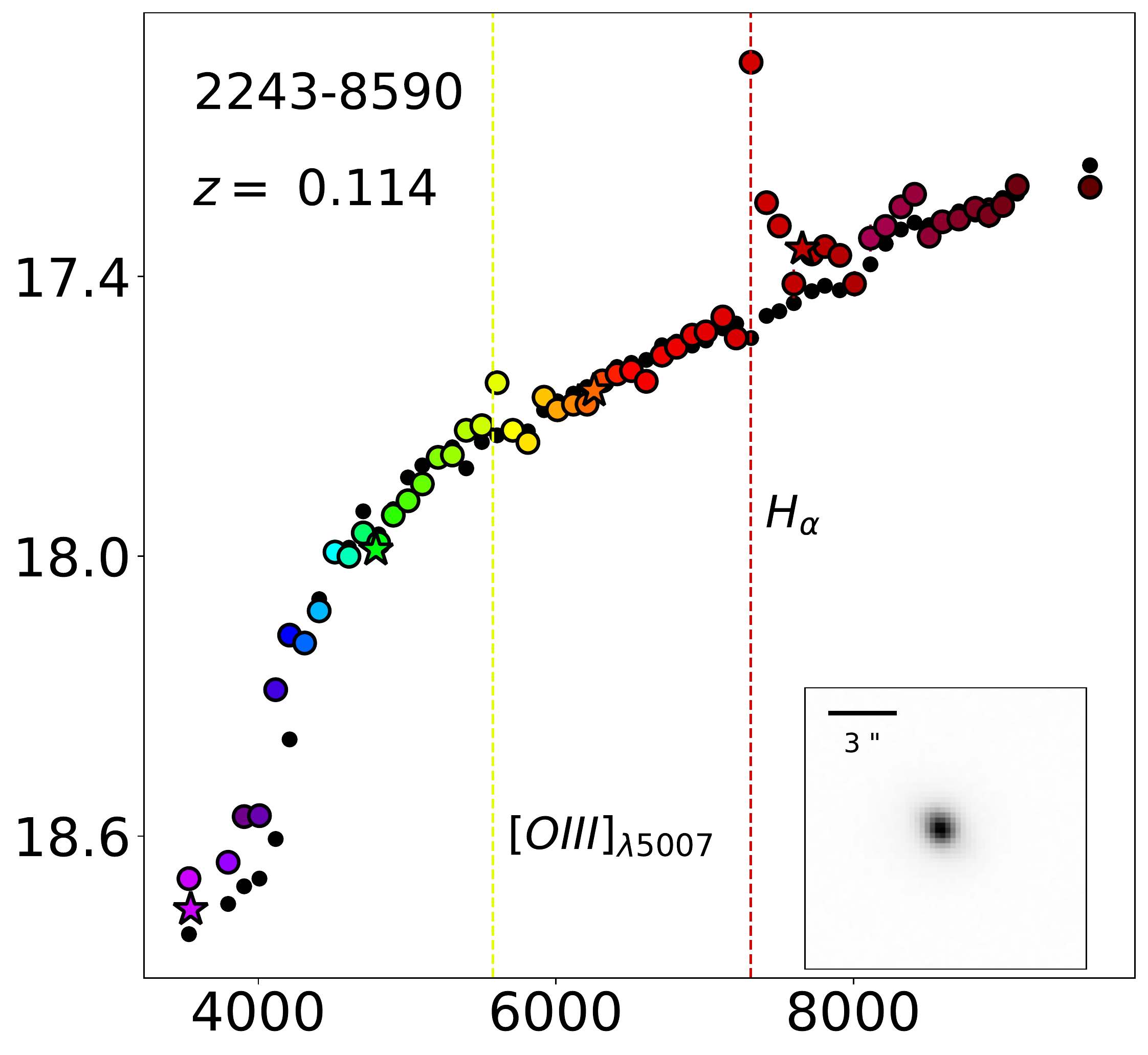}
        
        \includegraphics[width=6.4cm,height=6cm]{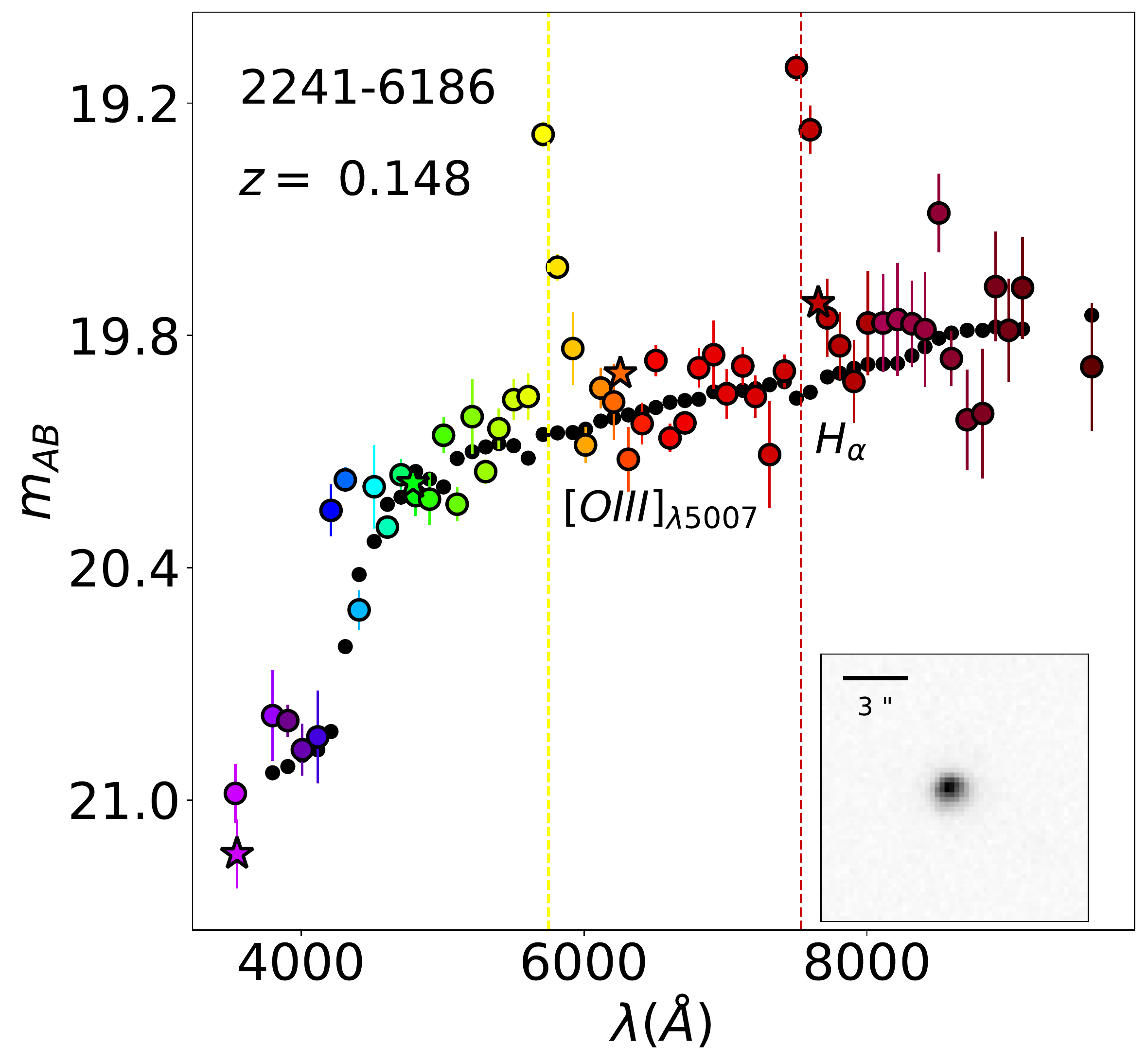}
        \includegraphics[width=6.4cm,height=6cm]{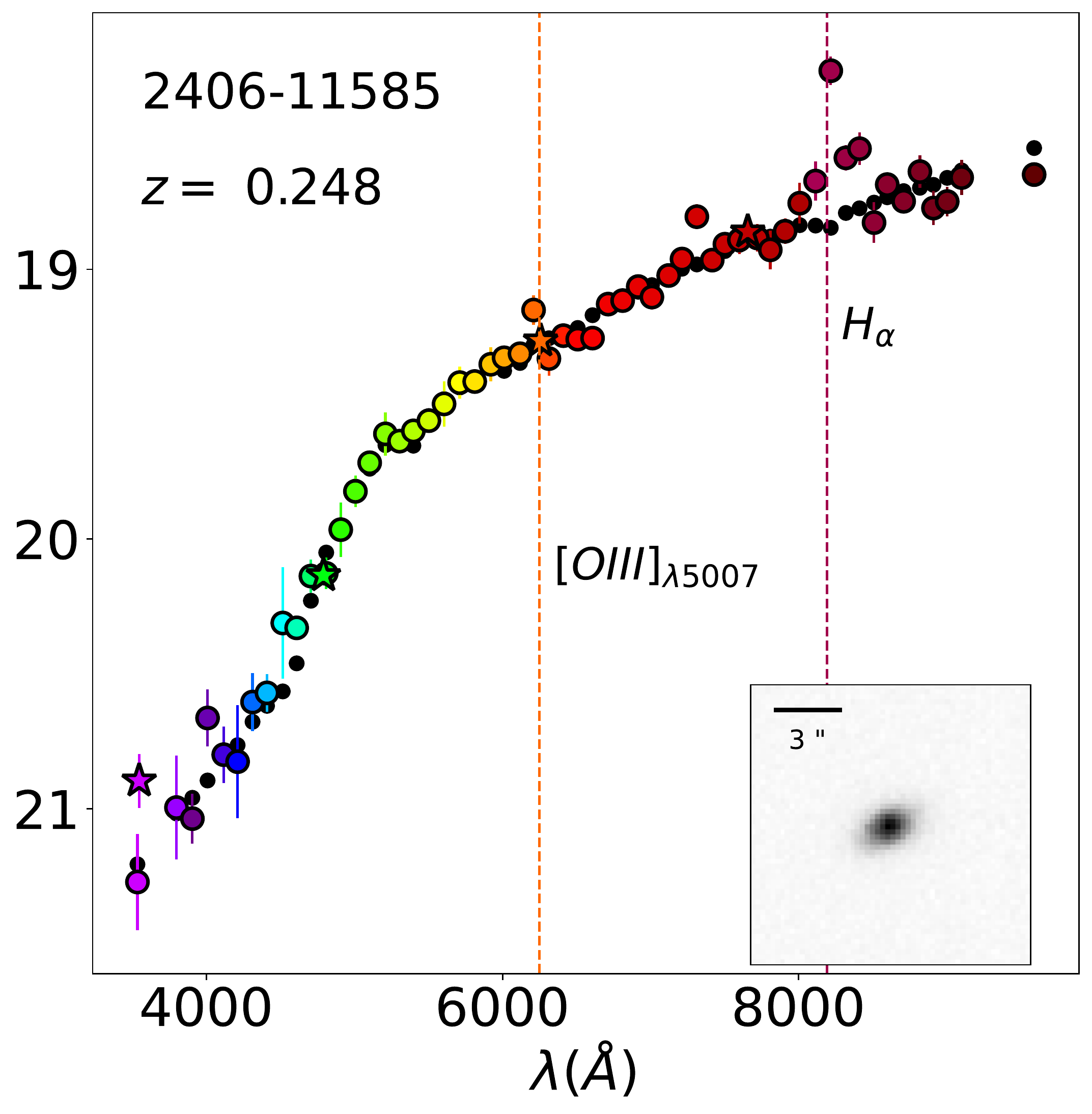}
        \includegraphics[width=6.4cm,height=6cm]{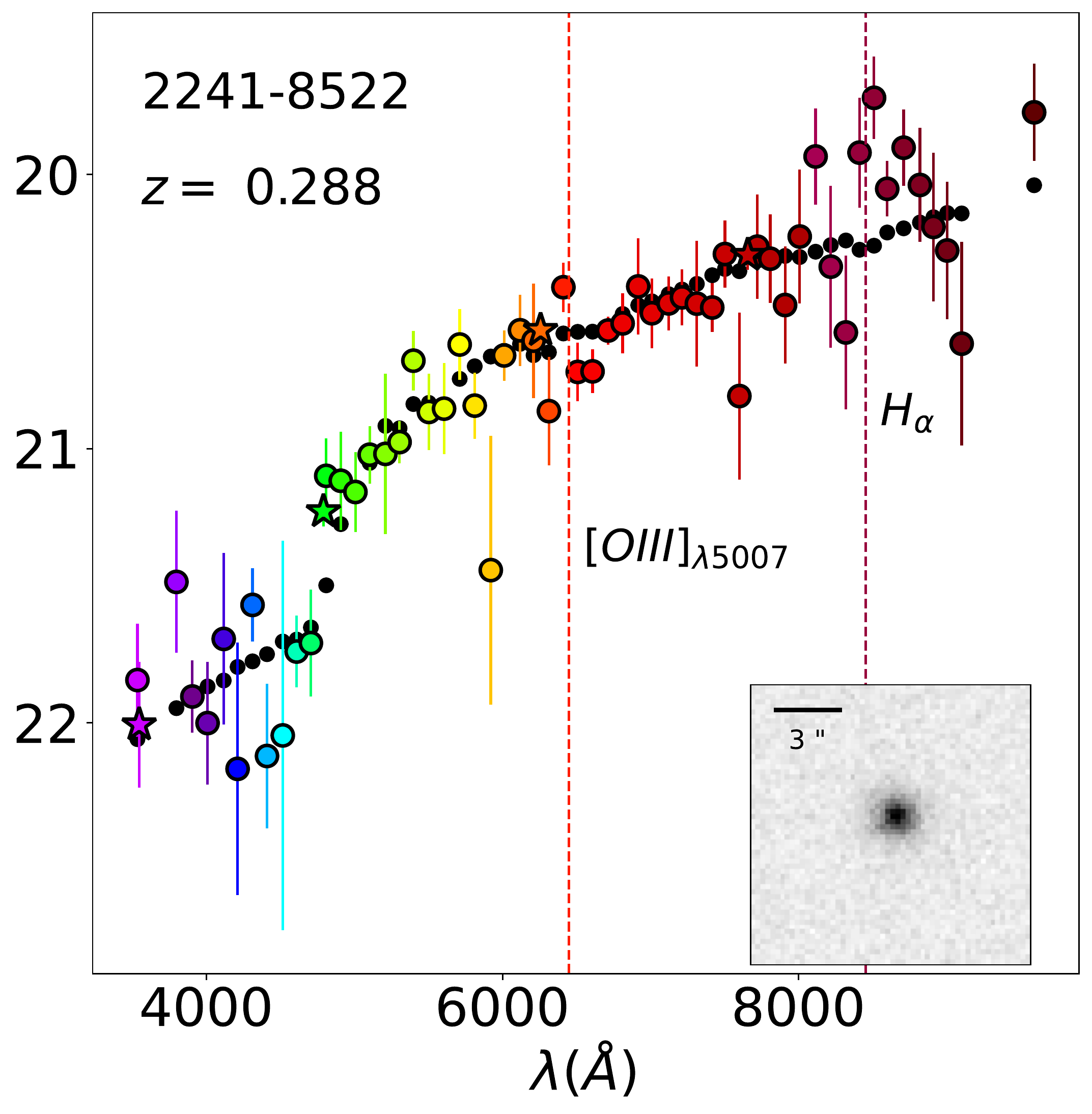}

 \caption{\tiny{J-spectra in magnitudes (\texttt{MAG\_AUTO} photometry) for a set of galaxies within the AEGIS field observed by miniJPAS. Stars correspond to broadband filters ($u_{JPAS}$, and SDSS g, r, and i). Black dots are the best fit obtained with \texttt{BaySeAGal} to the stellar continuum. Filters including the wavelength of \Ha and \oiii emission lines within their bandpass are marked with dashed vertical lines. The images of these galaxies in the rSDSS band are attached in the lower left inset. The miniJPAS ID and the \photoz $ $ are shown in black in the left corner of each figure. }}
 \label{fig:jpas-examples}
 \end{figure*}

 In Fig. \ref{fig:jpas-examples} we show some examples of galaxies in this sample at different redshift and magnitude bins. Emission lines such as \Ha or \oiii are clearly visible in most of them. Some lines are captured by more than one filter (see, e.g., 2241-6186). This is caused by the overlapping adjacent filters, whose separation ($100$ \AA) is smaller that their width ($ \sim 145 $ \AA).  

\section{Method}\label{sec:method}
\subsection{Artificial neural networks}\label{subsec:ANN}
The analysis of the emission lines was carried out with a machine-learning code based on ANN and described in MS21. Different ANNs were trained with the J-PAS synthetic photometry extracted from CALIFA, MaNGA, and SDSS galaxies after convolving the spectra with the J-PAS photometric system. The ANNs learned to perform different tasks. First, an ANN was trained to estimate the EW values for the main emission lines in the optical range: \Ha, \Hb, \oiii, and  \nii. This ANN is referred to as ANN$_\mathrm{R}$. As inputs, the ANNs used photometry colors measured with respect to the J-PAS filter, in which the \Ha flux dominates. As outputs, the ANNs received the values of the EWs that were measured directly in the spectrum. We estimated the uncertainty in the EWs with a Monte Carlo approach. We considered the error in the \photoz $ $ and the error in photometric fluxes (see MS21 for further details). Second, another ANN was trained to distinguish galaxies with emission lines from those without them. This classifier (ANN$_\mathrm{C}$) also relies on the EWs, but it is independent of the prediction from the ANN$_\mathrm{R}$. Galaxies were previously classified as class 1 or class 2 depending on whether they exceeded a preselected EW threshold in any of the emission lines. Several ANN$_\mathrm{C}$ with different thresholds ($\mathrm{EW}_{\mathrm{min}} = 3,5,8,11,\text{and }14$ \AA) were trained in order to better study the regime of low emission, in which the ANN$_\mathrm{R}$ is less sensitive.
\par As we discussed in MS21, there are many ways of combining the CALIFA, MaNGA, and SDSS surveys to build up a training set. Each survey has its own observational biases, and the emission lines were measured with different approaches. In this work, we made use of the CALMa training set for the ANN$_\mathrm{R}$, which performs better in unseen data (SDSS test sample). The CALMa training set employs both CALIFA and MaNGA spectra from spatially resolved regions over many diverse physical states, including AGN emission and SF regions. With the CALMa training set, we are able to fully reproduce the position of SF galaxies in the BPT diagram. We reached a precision of 0.092 and 0.078 dex for $\log$~(\nii/\Ha) and $\log$~(\oiii/\Hb), respectively, assuming an average S/N in the photometry of 10. We can measure an EW of 10 \AA $ $ in the \Ha, \Hb, \nii, and \oiii lines with a median S/N of 5, 1.5, 3.5, and 10, respectively.
\\ \\For the the ANN$_\mathrm{C}$ classifier, we employed the CALIFA set, which is a subset of the CALMa set, but only includes CALIFA galaxies. The two training sets performed very similarly in the SDSS test sample. For the sake of simplicity, we therefore employed the CALIFA set.
\subsection{Stellar population analysis}\label{subsec:stellar}
The stellar population properties of the galaxies in this sample were analyzed with \texttt{BaySeAGal} \citep[Amorim et al. in prep.,][]{2021arXiv210213121G}. This is a Bayesian parametric code that fits stellar metallicity ($Z_{\star}$), dust attenuation ($\tau_{V}$), and the parameters related to the star formation history of galaxies. We assumed a delayed-$\tau$ model of the form
\begin{equation}
\Psi (t) = \phi \frac{t_0-t}{\tau} \exp \left[-(t_0-t)/\tau\right],
\label{eq:delta_delay}
\end{equation}
where $t$ is the lookback-time, $t_0$ is the starting point of star formation in lookback-time, $\tau$ is the SFR e-folding time, and $\phi$ is the normalization constant related to the total mass formed in stars. $t_0$ and $\tau$ are sampled uniformly in logarithmic scale, which can vary between 1.4 and the maximum age at the redshift of the galaxy (13.7~Gyr at $z=0$), and between 0.1 and 10~Gyr, respectively. For the present work, we chose the attenuation law proposed by \citet{calzetti2000}, which adds a unique foreground screen with a fixed ratio of $R_V=4.05$ (the average value for the Milky Way).
\par The code used the 2017 version of the \cite{2003MNRAS.344.1000B} stellar population (SSP) synthesis models (hereafter CB17). The SSP covers the  metallicity range $\log Z_{\star}/Z_\mathrm{\odot}=$ -2.3, -1.7, -0.7, -0.4, 0, and +0.4, and the ages span from 0 to 14~Gyr. The CB17 models follow the PARSEC evolutionary tracks \citep{2013MNRAS.434..488M,2015MNRAS.452.1068C} and use the Miles \citep{2006MNRAS.371..703S,2011A&A...532A..95F,2011A&A...531A.165P} and IndoUS \citep{2004ApJS..152..251V,2016A&A...585A..64S} stellar libraries in the spectral range observed by J-PAS.
\par It is important to emphasize that filters capturing the nebular emission lines are masked and were not used in the SED fitting. Therefore, the galaxy properties are only based on the stellar continuum, and it does not include the emission of nebular regions or the result of the AGN activity. The stellar continuum is derived from the ensemble of best fits and allows us to determine stellar masses ($M_{*}$) , metallicities ($Z_*$), the amount of dust attenuation ($A_V$), or the luminosity-weighted age ($<\log t>_L$)  of galaxies. Furthermore, it is also used to extrapolate the photometry in the filters that lack a measurement or have a very low S/N (lower than $1.8$). Because the ANN (as we designed it) cannot work with missing data, these extrapolations allow the ANN to access all the inputs needed (photometric fluxes). This does not apply to the filters containing emission lines at each redshift and the filters that are immediately next to them. For instance, the \Ha emission line is captured by the J0660 filter for a galaxy in the local Universe ($z=0$). Therefore, the fluxes in filters J0650, J0660, and J0670 are never extrapolated. When problems in the photometry with these filters occurred, we did not include the corresponding galaxies in our sample.
\par The use of alternative SED fitting codes to derive stellar population properties of miniJPAS galaxies does not affect the main results in this paper.  \cite{2021arXiv210213121G} analyzed in detail how the main properties derived for galaxies might change with different SED fitting approaches. The results are consistent between each other: nonparametric codes such as \texttt{MUFFIT} \citep{2015A&A...582A..14D}, \texttt{Alstar} \citep[the algebraic version of starlight][]{2005MNRAS.358..363C}, or \texttt{TGASPEX} \citep{2015PASP..127...16M} and \texttt{BaySeAGal} all obtained similar distributions of rest-frame $(u-r)$ color, stellar mass, age, and metallicity up to $z=1$. 
\par
\begin{figure}
        \includegraphics[width=\hsize]{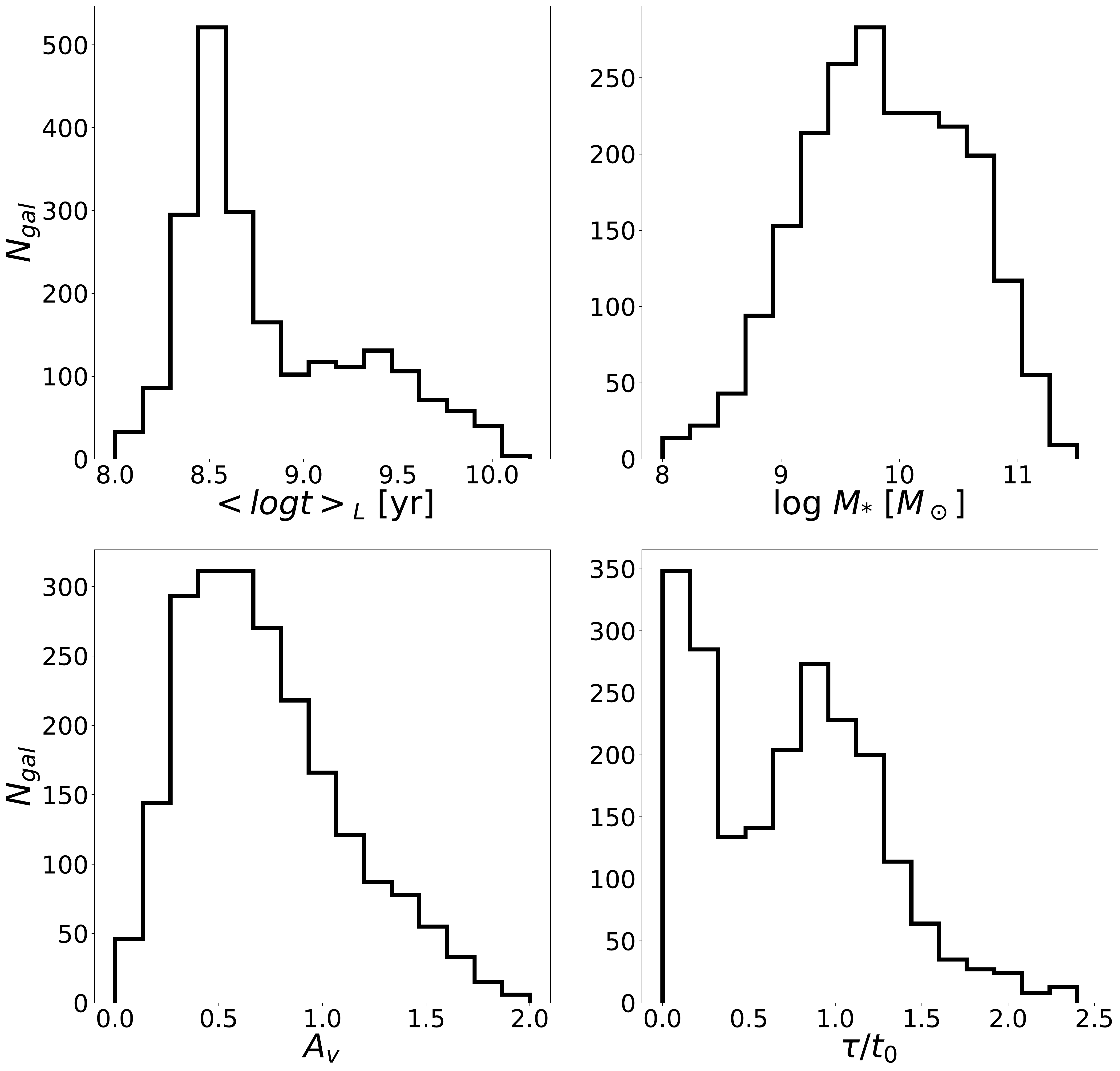}
        \caption{\tiny{Distributions of mean stellar luminosity-weighted age (top left panel), galaxy stellar mass   (lower right panel), extinction (lower left panel), and  $\tau/t_0$ ratio (bottom right panel) obtained by \texttt{BaySeAGal} for the sample of galaxies described in Sect.~\ref{sec:datasample}.}}
 \label{fig:hist_stellar}
 \end{figure}

 A summary of the stellar population properties of the galaxies we analyzed is shown in Fig. \ref{fig:hist_stellar}. The distributions of the galaxy ages and the $\tau/t_0$ ratio are bimodal.  \texttt{BaySeAGal} provides rest-frame colors and extinction-corrected colors. In particular,  $(u-r)_{int}$ is very useful for distinguishing between red and blue galaxies. We followed the criterion of \citet[][hereafter the color criterion,]{2019A&A...631A.156D} in order to distinguish them. This criterion was adapted to match the miniJPAS photometric system (DÃ­az-GarcÃ­a et. al in prep.). For a galaxy to be part of the red sequence, this criterion establishes a limit in $(u-r)_{int}$ from the galaxy stellar mass and redshift,
\begin{equation}
\label{eq:redbluegalaxies} 
(u - r)_\mathrm{int}^\mathrm{lim} =  0.16 \times (\logMt  \ - 10) - 0.3 \times (z - 0.1) + 1.7.
\end{equation}
Galaxies with $(u-r)_{\mathrm{int}}$ above $(u - r)_\mathrm{int}^\mathrm{lim}$ are classified as red galaxies, otherwise, they are considered to be blue. Furthermore, \texttt{BaySeAGal} provides the probability distribution function (PDF) for the model parameters.
The uncertainty on the derived stellar population properties is defined as the standard deviation. As expected, the uncertainty depends on the S/N of the photometry. The median errors are lower in the red sequence than in the blue cloud.
That is, $\langle \sigma (\logMt ) \rangle = 0.16 \pm 0.03$ dex, $\langle \sigma (\ageMt ) \rangle  = 0.19 \pm 0.05$ dex, $\langle \sigma (A_V) \rangle = 0.19\pm 0.07$ mag, and $\langle \sigma (\tau/t_0) \rangle = 0.10 \pm 0.04$ for galaxies in the red sequence, and  $\langle \sigma (\logMt ) \rangle = 0.28 \pm 0.04$ dex, $\langle \sigma (\ageMt ) \rangle  = 0.25 \pm 0.05$ dex, $\langle \sigma (A_V) \rangle = 0.33\pm 0.05$ mag, and $\langle \sigma (\tau/t_0) \rangle = 0.5 \pm 0.19$ for those in the blue cloud. 

\section{Identification of ELGs}\label{sec:main_results}
In this section, we show the potential of our methods to identify ELG in the AEGIS field and determine their main ionization mechanism. The EW of \Ha, \Hb, \oiii, and \nii and their relative strengths allow us to distinguish between different types of ELGs and derive the fraction of star-forming, Seyfert, and quiescent galaxies in miniJPAS.
\subsection{Identification with ANN$_R$: EW distributions}
\begin{figure}
        \includegraphics[width=\hsize]{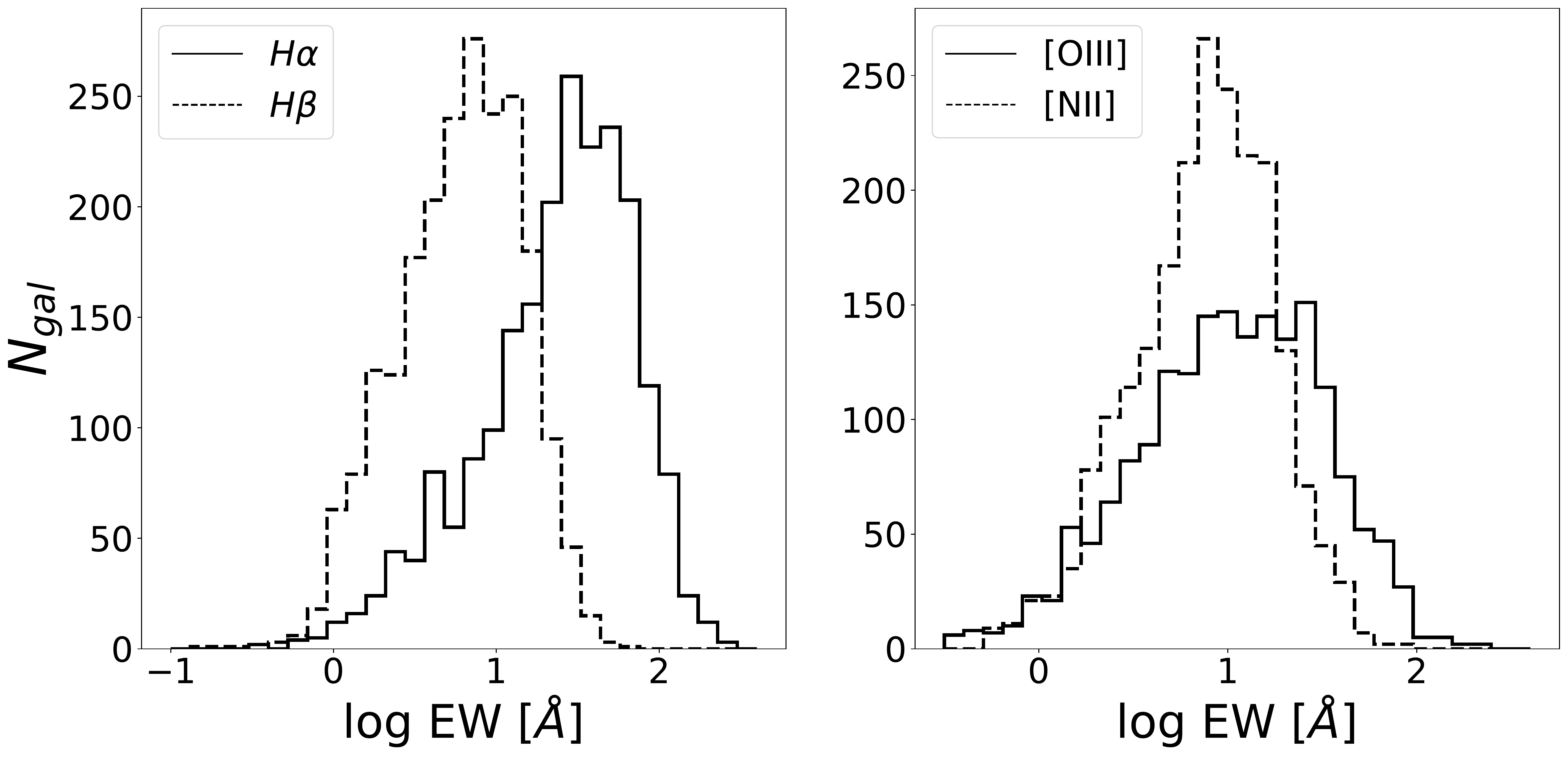}
        \caption{\tiny{Distribution of the EW of \Ha and \Hb (left), \oiii, and \nii (right) in log scale as obtained with the ANN$_{R}$.}}
 \label{fig:EW_distribution}
 \end{figure}

First, we show the EW distribution of the \Ha, \Hb, \oiii, and \nii lines in Fig. \ref{fig:EW_distribution} derived with the ANN$_\mathrm{R}$. We excluded from the histograms galaxies where the EWs are below zero. Even though the ANN$_\mathrm{R}$ was not trained with absorption lines, certain configurations can indeed lead to negative values of the EWs. If the fluxes in the filters in which the emission lines are expected to appear are suppressed or are highly uncertain, or if they mimic the shape of an absorption line, the ANN$_\mathrm{R}$ might predict EWs that are below zero. We find 20, 2, 299, and 23 galaxies with negative EWs in \Ha, \Hb, \oiii, and \nii , respectively. The median S/N in the EWs for these galaxies is below one, which indicates that these values are compatible with positive and null values.
 \par Generally, blue galaxies are star-forming galaxies, while red galaxies are quiescent. However, a galaxy might appear to be part of the red sequence due to the presence of dust, which absorbs a fraction of the total radiation more efficiently on the blue side of the spectrum. Therefore it is important to correct for dust extinction in order to distinguish between red and dust-reddened star-forming galaxies.
\par
\begin{figure*}
        \includegraphics[width=1.\linewidth]{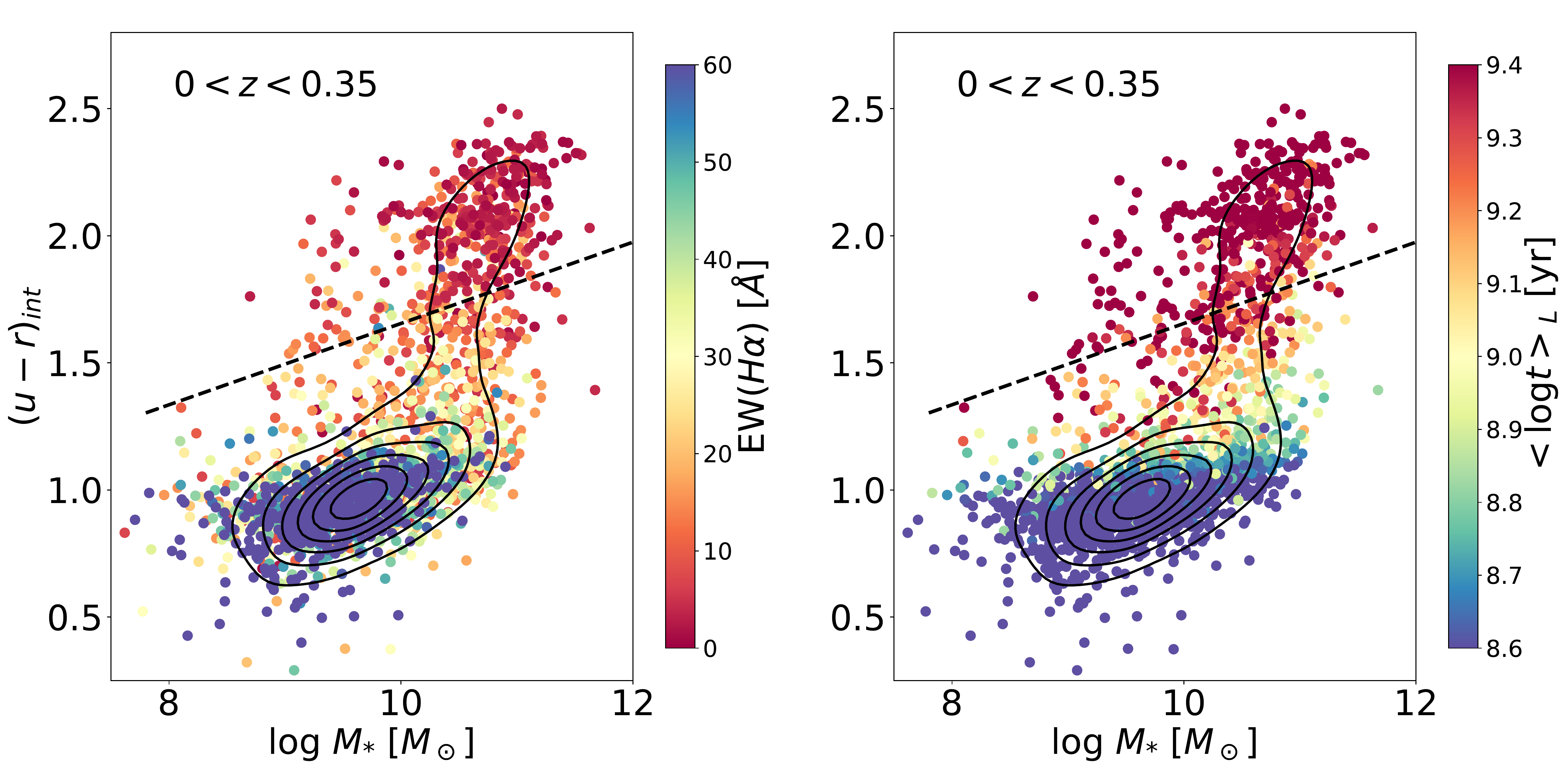}
        \caption{\tiny{Color--mass diagram for our sample of galaxies. The (u -- r) color-corrected for dust extintion vs. stellar mass. Galaxies are color-coded with the EW of \Ha (the luminosity-weighted stellar age) on the left side (right side). The intrinsic color, stellar mass, and luminosity-weighted age are obtained via \texttt{BaySeAGal}. Dashed black lines separate blue and red galaxies following Eq. \ref{eq:redbluegalaxies}, where we considered the median redshift of the sample ($z = 0.25$). Density contours are drawn in black at the top.}}
 \label{fig:colour-mass}
 \end{figure*}

 Figure \ref{fig:colour-mass} shows as expected that blue galaxies contain young populations of stars with high values of EW(\Ha), while red galaxies are older and lack \Ha emission or have very low values of EW(\Ha). Between the red sequence and the blue cloud, we observe galaxies in the GV with intermediate ages and moderate values in the EWs of \Ha. 

\subsection{Identification with the ANN$_C$: Strong and weak ELGs}\label{subsec:galclass}
\begin{figure*}
    \centering
 \includegraphics[width=9.5cm,height=8.7cm]{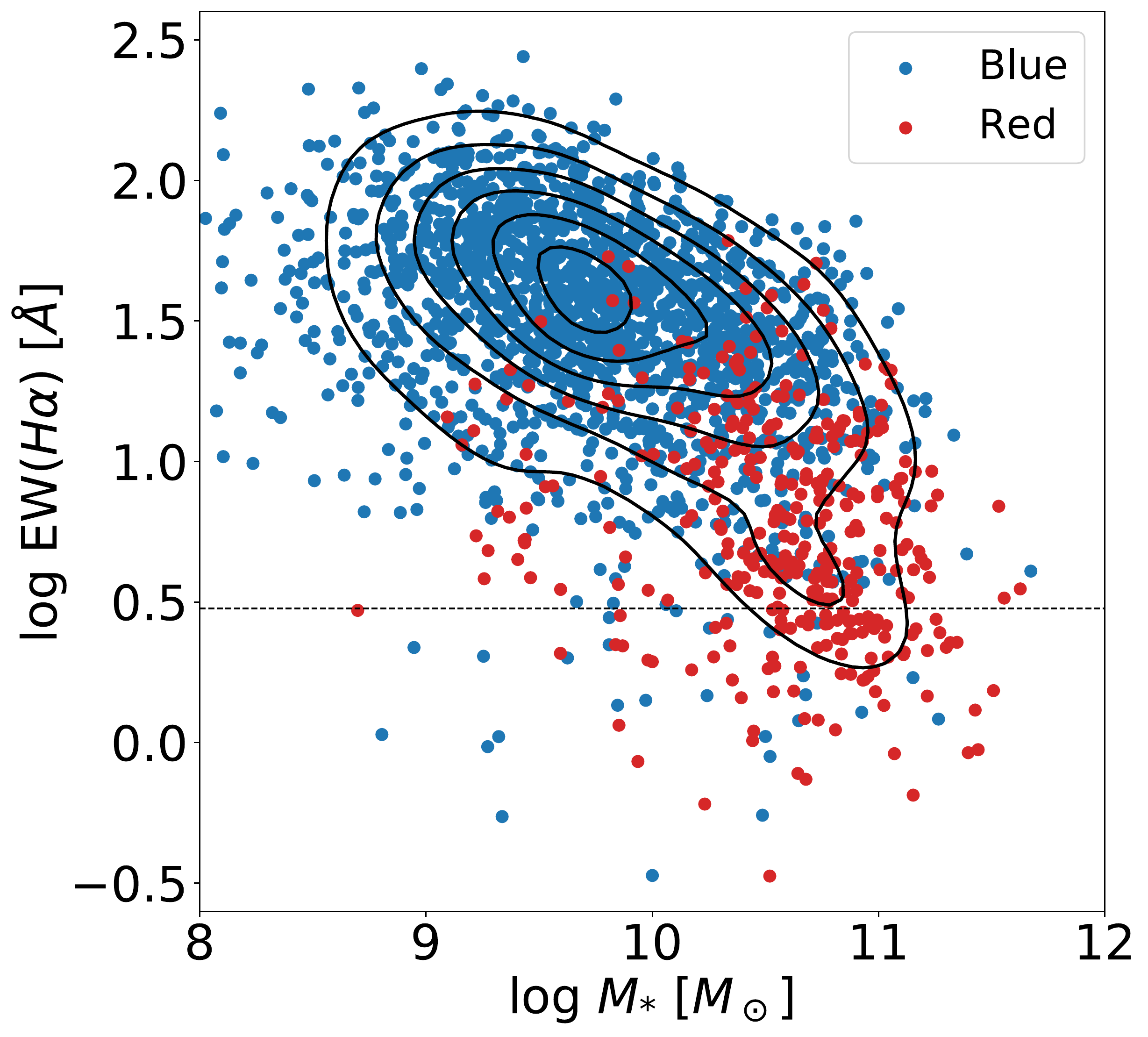}
 \includegraphics[width=8.3cm,height=8.7cm]{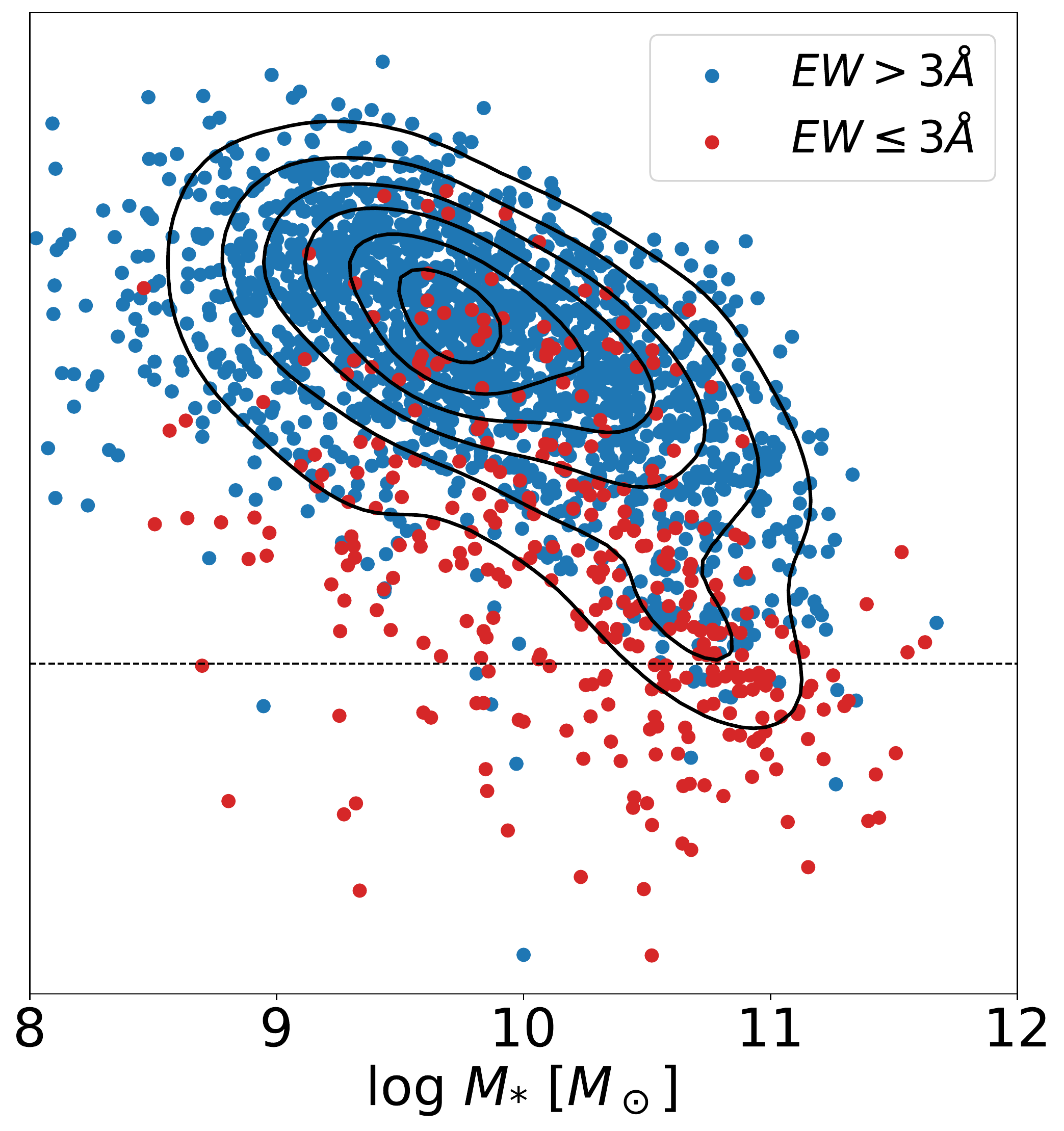}
        \caption{\tiny{Equivalent width of \Ha as a function of the stellar mass of the galaxy. In the left panel, we used Eq. \ref{eq:redbluegalaxies} to distinguish between red and blue galaxies. In the right panel, we relied on the classification performed with a machine-learning code trained with strong EL and weak EL galaxies. Strong ELs were defined as those with EWs greater than 3 \AA $ $ in any of the following emission lines: \Ha, \Hb, \oiii, or \nii, and weak ELs are all others. The dashed horizontal lines mark the 3 \AA\ limit in the EW(\Ha). Density contours are drawn in black at the top.}}
 \label{fig:EW_class}
 \end{figure*}

In addition to the color-criterion, we can also make use of the predictions of the ANN$_\mathrm{C}$ to distinguish between galaxies above and below a certain threshold limit in the EW. The EW of \Ha quantifies the relative intensity of the emission line flux with respect to the stellar continuum, and therefore it is a good indicator of the sSFR in the galaxy \citep{2016MNRAS.460.3587M,2021MNRAS.503.5115K}. In Fig. \ref{fig:EW_class} we plot the $\log$ EW(\Ha) as a function of the stellar mass. In the left panel, we indicate in blue (red) the galaxies that belong to the blue cloud (red sequence) following the color criterion. On the right panel we show a similar scheme but galaxies are separated according to the class defined by the ANN$_\text{C}$ with $\mathrm{EW}_{\text{min}} = 3 $ \AA. In other words, galaxies are considered strong ELs if any of the emission lines present an EW greater than 3~\AA $ $ and weak ELs if all lines are below this limit. For a threshold of $0.1$ in the ANN$_\mathrm{C}$ probability, strong ELs represent $83 \%$ of the sample, while weak ELs are the remaining $17 \%$. With the color criterion, $82 \%$ of the galaxies in the parent sample are classified as red and the remaining $18 \%$ are blue.
\par The dashed line in Fig. \ref{fig:EW_class} illustrates the $\mathrm{EW}(\Ha) = 3$ \AA $ $ limit. As expected, most of the galaxies below this limit are classified as weak ELs. However, we detect a non-negligible number of weak ELs or red galaxies above this limit in both panels. We have to take into account that the ANN$_\mathrm{R}$ is less accurate at low EWs and has a tendency to overestimate their values. Moreover, the relative errors in this regime are higher (see MS21). Therefore, it is not surprising to find a fraction of weak EL galaxies above this limit. Moreover, although \Ha leads the ANN$_\mathrm{C}$ classification, the algorithm includes other emission lines in addition to \Ha, which might occasionally overcome this limit. At high EWs, the number of weak EL galaxies decreases significantly, and the discrepancy between the ANN$_\mathrm{R}$ and the ANN$_\mathrm{C}$ can be explained by the high uncertainty found in the \photoz\ or a low S/N in the photometric fluxes.
\par In two panels in Fig. \ref{fig:EW_class}, the two methods of classifying galaxies present a consistent picture. Most of the blue galaxies are strong ELs, and red galaxies are weak ELs. Nevertheless, we found some disagreement between the last two populations. While the ANN$_\mathrm{C}$ is trained to separate galaxies as a function of the EW, Eq. \ref{eq:redbluegalaxies} depends mainly on the global color and the mass of the galaxy. Thus, it is expected to find some galaxies with red intrinsic colors and a low level of star formation reflected on the nebular emission with EWs greater than 3 \AA.
\par Finally, it is clear from these diagrams that galaxies are less efficient at forming new stars as the mass of the galaxy increases at $z < 0.35$. At some point around $M_* = 10^{11} M_\mathrm{\odot} $, the EW of \Ha falls sharply, with most galaxies above this mass showing red colors and low values in the EW(\Ha), suggesting that the main sequence of star-forming galaxies has already ended.
\subsection{Identification of star-forming galaxies and AGNs: BPT and WHAN diagrams}\label{subsec:BPT_WHAN}
The BPT diagram ($\log $\oiii/\Hb versus $\log $\nii/\Ha) provides a means to unveil the main ionization mechanism of galaxies. It involves four emission lines, and galaxies are classified into four groups by three dividing lines: star-forming, composite, Seyfert, and LINERs. The \citet[hereafter Ka03,]{2003MNRAS.346.1055K} curve is derived empirically using the SDSS galaxies and defines the region populated by SF galaxies. Usually referred to as the SF wing, galaxies evolve from high (low) to low (high) \oiii/\Hb (\nii/\Ha) ratios, increasing their mass \citep{2019A&ARv..27....3M}. The \citet[hereafter Ke01,]{2001ApJ...556..121K} curve is determined using both stellar population synthesis models and photoionization. It defines the AGN wing that is dominated by AGN (including LINER or LINER-like emission, and shocks). Between these two lines lies the composite region, which might be populated by galaxies with a composite spectrum, that is, the ionization mechanism is a mix of star-formation processes and AGN activity or galaxies with very weak emission lines that are leaving the SFMS. Finally, the \citet[hereafter S07,]{2007MNRAS.382.1415S} line is an empirical division that distinguishes between Seyfert and LINER galaxies.
\par
\begin{figure*}
    \centering
 \includegraphics[width=8.7cm,height=8.7cm]{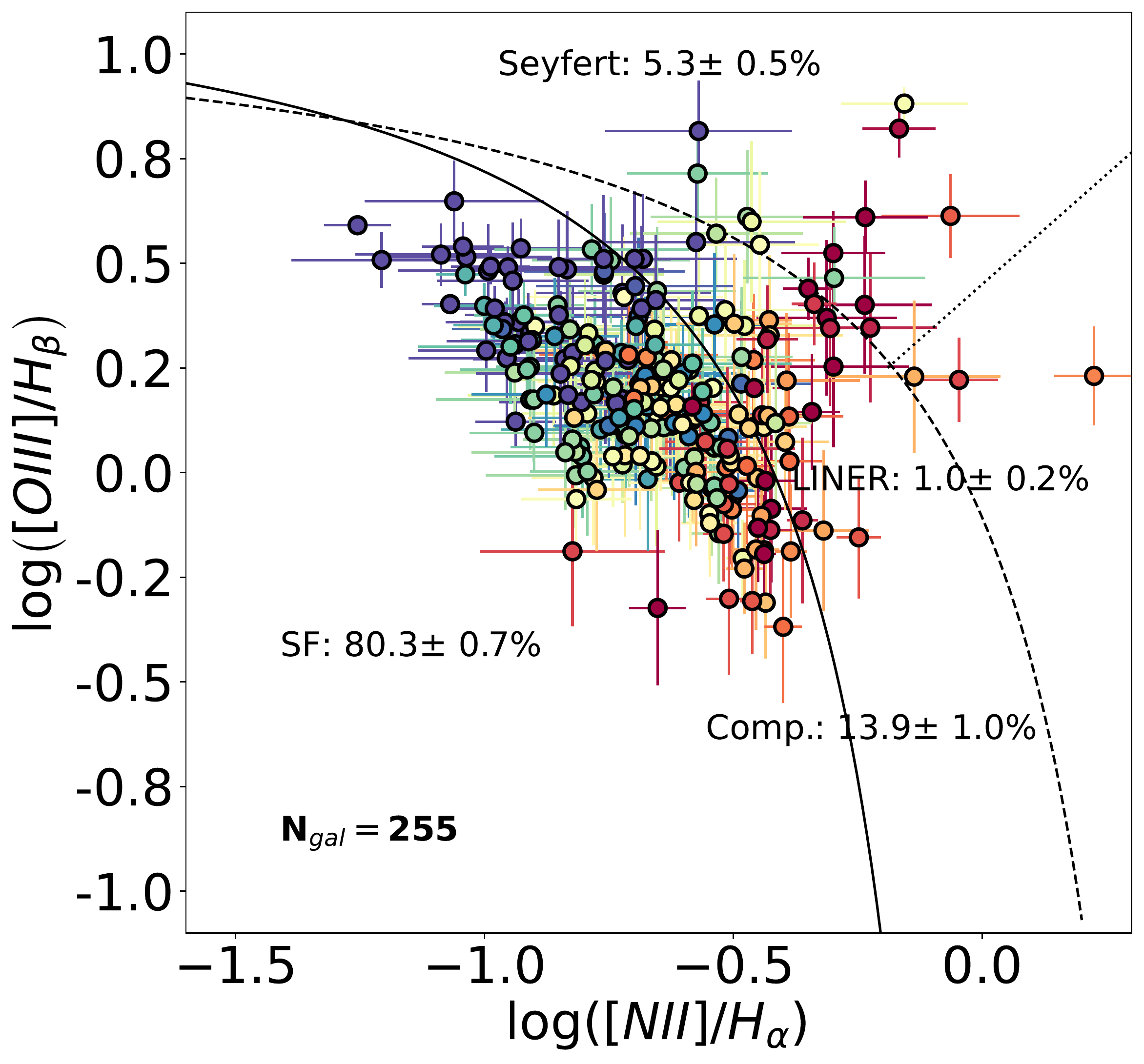}
 \includegraphics[width=9.1cm,height=8.7cm]{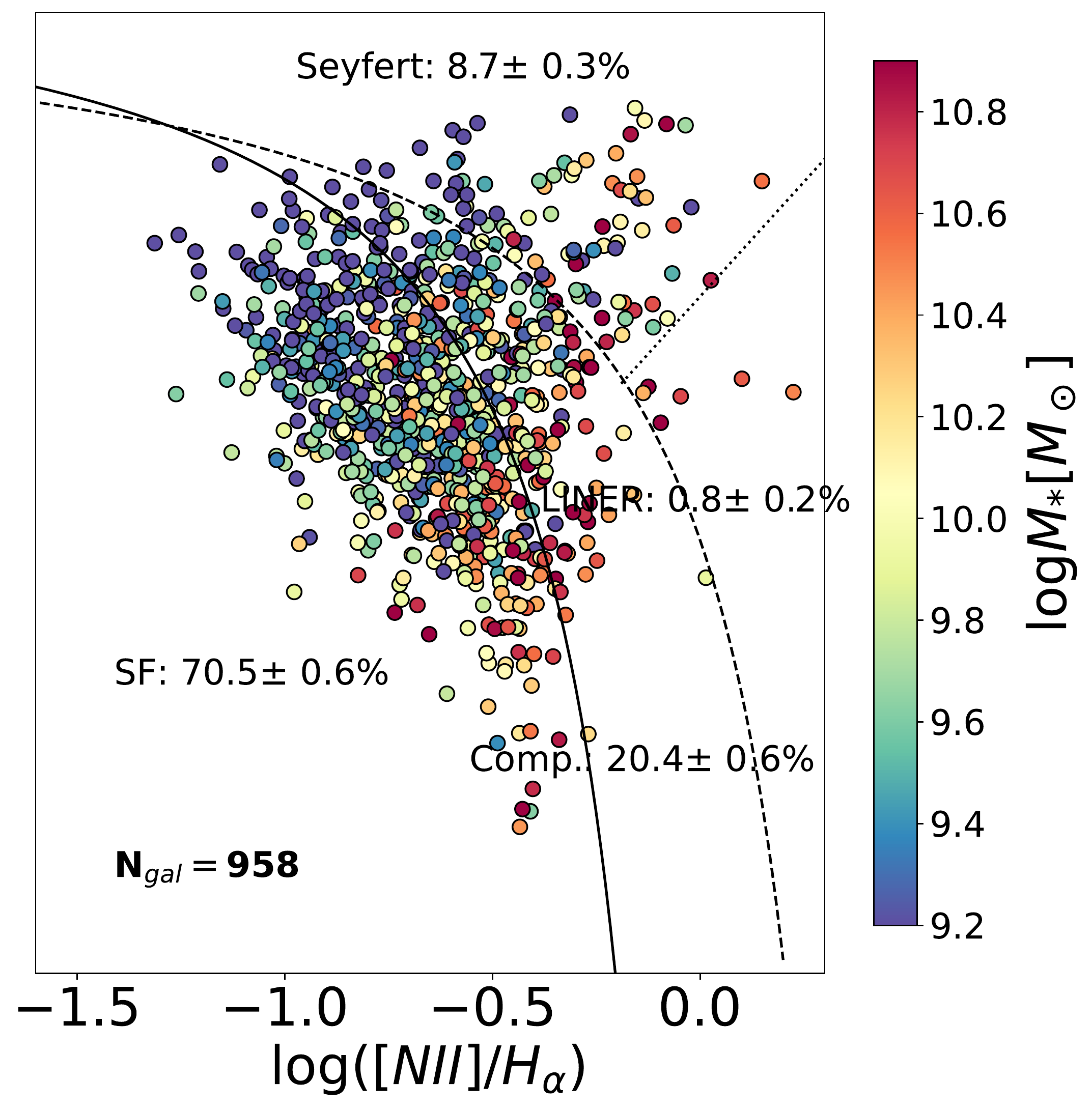}
        \caption{\tiny{BPT diagram for the galaxies in the sample with an error of $0.2$ dex ($0.5$ dex) in the \oiii/\Hb and \nii/\Ha ratios in the left (right) panel. The errors are not plotted in the right panel for clarity. The color bar indicates the stellar mass of the galaxy. The solid (Ka03), dashed (Ke01), and dotted lines (S07) define the regions for the four main spectral classes. The relative percentage of each galaxy type in each subsample is indicated in the figure. In each panel, the number of galaxies is specified in the lower left corner. The parent sample contains 2154 galaxies.}}
 \label{fig:BPT}
 \end{figure*}

 We show the BPT diagram for
the galaxies in the parent sample with error lower than $0.2$ dex in \oiii/\Hb
and \nii/\Ha in the left panel of Fig. \ref{fig:BPT}. In the right panel, we relax this threshold to $0.5$ dex. These thresholds are arbitrary, and they have been chosen to show how the BPT diagram changes when galaxies with a high uncertainty in the measurement of the emission lines are included. However, they are not used for the final selection of SF galaxy sample. The stellar mass distribution of galaxies in the BPT is consistent with expectations: galaxies grow in mass while they evolve through the SF wing. However, as the error increases (right panel), some galaxies populate regions that are less likely to be occupied (the narrowest wedge at the top left within the composite region).
\par Galaxies with very faint emission lines may be misclassified as LINERs from a BPT diagnostic. Sometimes called fake AGN \citep{2011MNRAS.413.1687C}, one of the advantages of the WHAN ($\log$ EW(\Ha) versus $\log$ (\nii)/\Ha)) diagram is that it can identify these galaxies. Even more important is the fact that the WHAN diagram provides a simpler way of determining the main ionization mechanism of galaxies. 
\par
\begin{figure*}
    \centering
 \includegraphics[width=8.7cm,height=8.85cm]{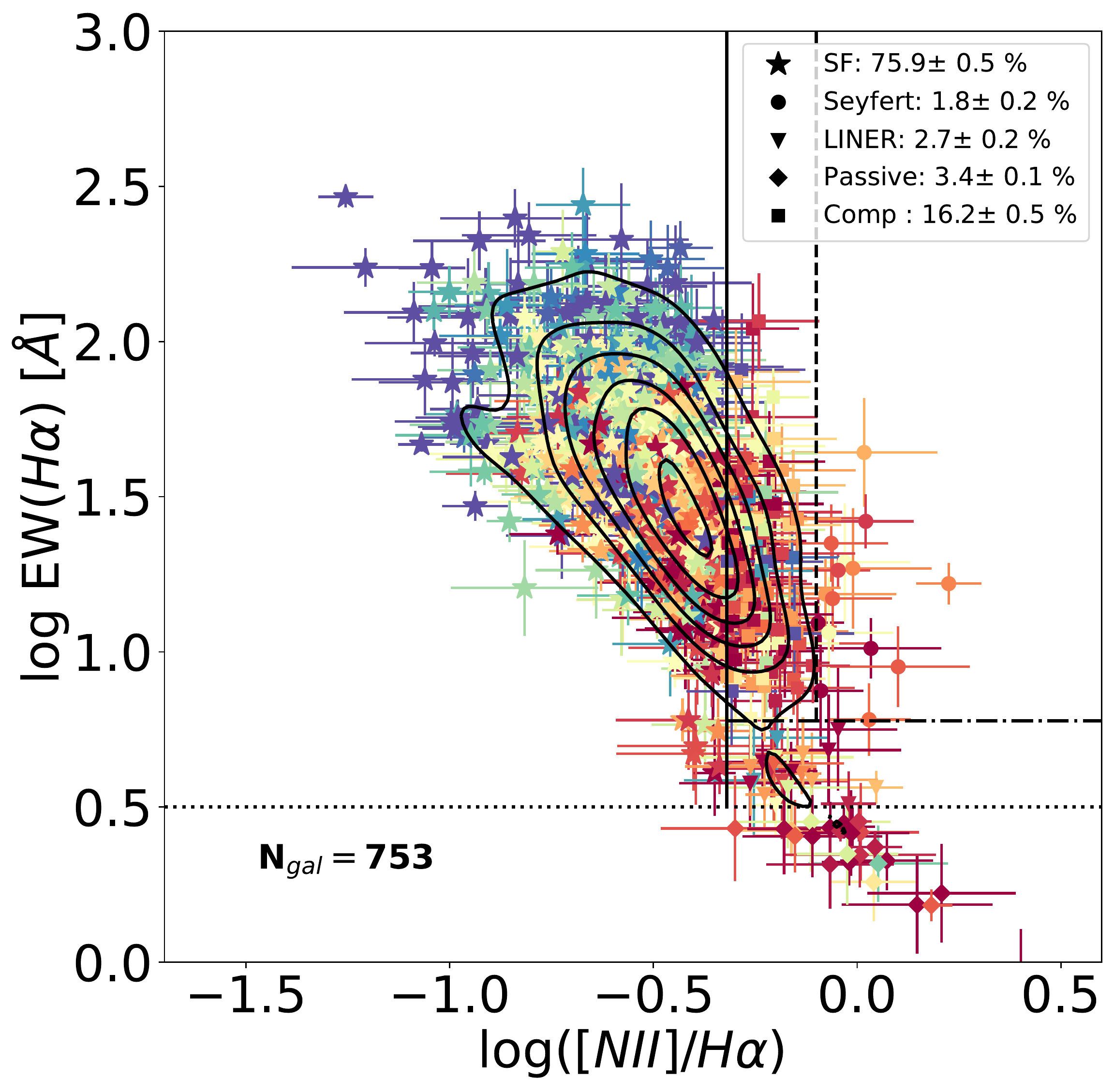}
 \includegraphics[width=9.1cm,height=8.7cm]{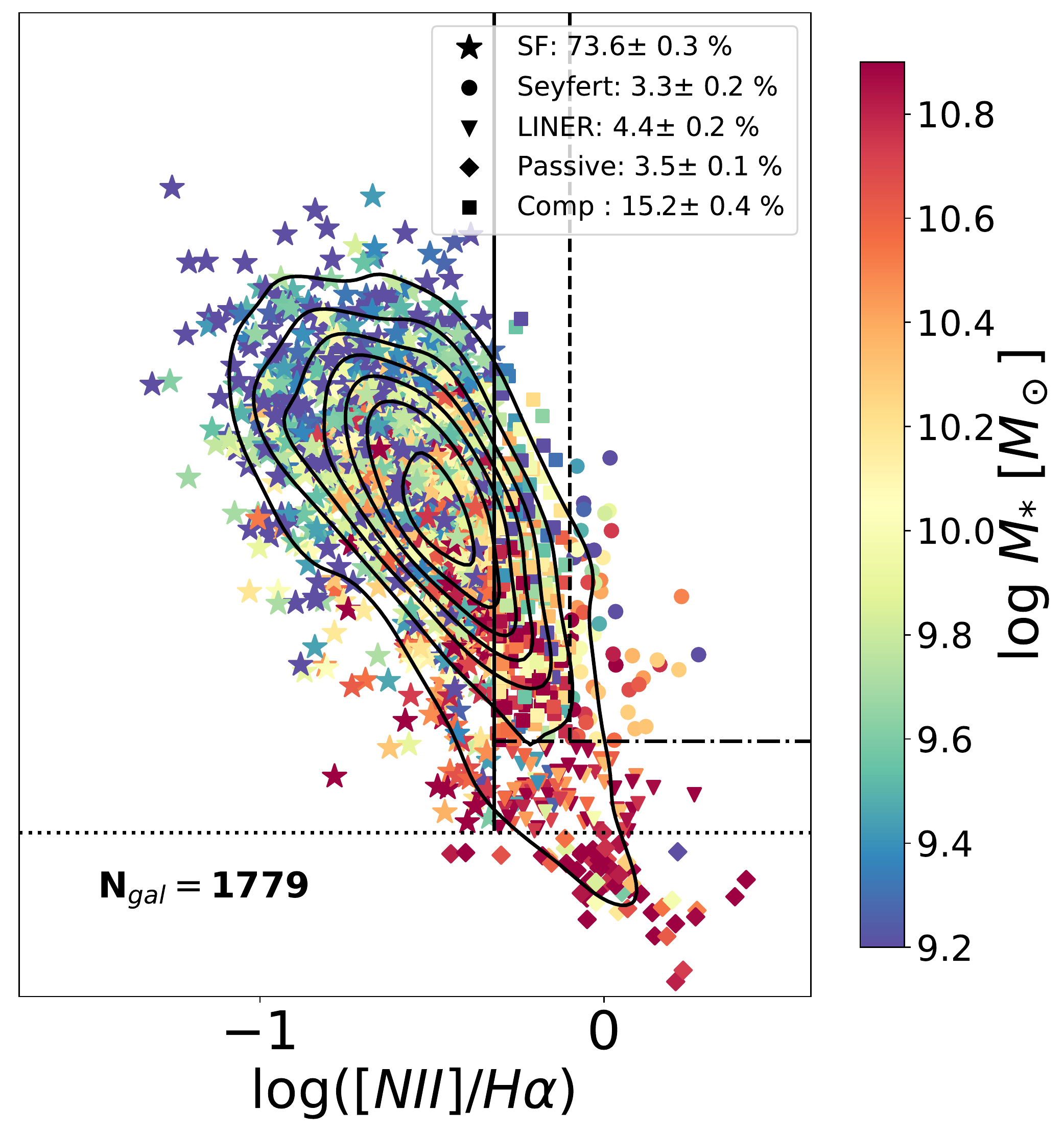}
        \caption{\tiny{WHAN diagram for galaxies with an error smaller than $0.2$ dex ($0.5$ dex) in both the EW(\Ha) and the \nii / \Ha ratio in the left (right) panel. The errors are not shown in the right panel for clarity. The color bar indicates the stellar mass of the galaxy. The inset shows the relative percentage of each galaxy type in each subsample. Dashed and solid vertical lines define the optimal projections of the Ke01 and the Ka03 lines in the WHAN diagram \citep{2010MNRAS.403.1036C,2011MNRAS.413.1687C}. Similarly, the dash-dotted horizontal line at $\mathrm{EW}(\Ha) = 6$  \AA  $ $  is the optimal transposition of the S07, and the dotted line at $\log$ $\mathrm{EW}(\Ha) = 0.5$ \AA $ $ defines the limit of ELGs.  In each panel, the galaxy counts are specified in the lower left corner. The parent sample contains 2154 galaxies}. Density contours are drawn in black at the top.}
 \label{fig:WHAN}
 \end{figure*}

  Fig. \ref{fig:WHAN} we show the WHAN diagram for the galaxies in the parent sample. The solid and dashed vertical lines represent the optimal projection of Ka03 and Ke01 onto the $\log$ EW(\Ha) versus $\log$ (\nii)/\Ha) space, that is, the dividing lines that better distinguish galaxy types in the WHAN diagram as they are defined in the BPT \citep{2010MNRAS.403.1036C,2011MNRAS.413.1687C}. Similarly, the division between Seyferts and LINERs at $\mathrm{EW}(\Ha) = 6$  \AA  $ $ corresponds to the optimal projection of S07.  Finally, the area below the dashed horizontal line at $\mathrm{EW}(\Ha) = 3.16$  \AA $ $ is composed of galaxies with highly uncertain line measurements that are therefore compatible with quiescent galaxies. We did not distinguish between retired and passive galaxies as in \cite{2011MNRAS.413.1687C} because our precision is not high enough to measure values of the EWs in the range of a few \AA. 
\par In the left panel of Fig.~\ref{fig:WHAN} we show galaxies with an error smaller than $0.2$ dex in both the EW(\Ha) and the \nii / \Ha ratio, while in the right panel, we relax this requirement to $0.5$ dex. The percentage of each galaxy type is indicated in the legend. Galaxies with lower EW(\Ha) have higher relative errors. Furthermore, many red galaxies do not appear in this diagram.
\par The color gradient in Fig. \ref{fig:WHAN} indicates that galaxies are more massive as the EW(\Ha) decreases and the \nii/\Ha ratio increases. Therefore, star-forming galaxies are on average less massive than Seyferts, while LINERs and passive galaxies are the most massive galaxies.
\par By comparing the position of each galaxy (i.e., their values with errors) in both diagrams, it is noticeable that the values of a given galaxy in the BPT convey more uncertainties than in their counterpart spot in WHAN. The reason is that the error in the y-axis of the BPT diagram stems from two sources: the error in the \oiii and \Hb emission lines. However, in the WHAN diagram, the only error source is the \Ha emission line. As a consequence, with a maximum error of $0.2$ dex, we can estimate the position of only 255 galaxies of the sample in the BPT and 753 galaxies in the WHAN. The median S/Ns of these subsamples in the narrowband filters are $10.7$ and $11.4$.
\subsection{Fraction of galaxy types in miniJPAS}
$\text{We identify }83$\% of the galaxies (1787) from the parent sample (2154 galaxies) in the AEGIS field as strong ELGs, and the remaining $17$\% (367 galaxies) are weak ELGs. In Table~\ref{Tab:galaxytype} we show the percentages of each galaxy type according to the WHAN diagram for all galaxies with an error smaller than 1 dex in the EW(\Ha) and \nii / \Ha ratio. This criterion is fulfilled by 2000 galaxies, which leaves 154 galaxies from the parent sample unclassified. We eliminate the composite population, but we indicate the percentage of SF and Seyfert galaxies in the different separation curves: Ka03, Ke01, or \citet[hereafter S08]{2008MNRAS.391L..29S}. Although we showed in Fig. \ref{fig:WHAN} the percentages for LINERs and passive galaxies, we grouped both classes together in this table. The emission lines for LINER galaxies are at the limit of what we can detect with the ANN given the S/N in the photometry. Hence, it is more challenging to distinguish them in the low S/N regime. We estimated the percentages and the errors of each galaxy type with a Monte Carlo (MC) method using the position of each galaxy in the diagram and its errors. Then, we computed the median and the standard deviation.

\begin{table}
\caption{\tiny{Percentage of each galaxy type according to the WHAN diagram. Quiescent galaxies include LINERs and passives. }}
\begin{tabular}{llll}
\hline
\hline
 \nii / \Ha &  Star-forming & Seyfert &  Quiescent  \\
  &  \hspace{0.6cm}$[\%]$ & \hspace{0.2cm}$[\%]$ &  \hspace{0.35cm}$[\%]$  \\

 \hline \\ 
 $ \le 0.79$ (S08) & $89.8  \pm 0.2$  & $3.5 \pm 0.2$ & $6.7 \pm 0.2$  \\\\
 $\le 0.48$ (Ka03) & $72.8 \pm 0.4$  & $17.7 \pm 0.4$  & $9.4 \pm 0.2$  \\\\
 $ \le 0.40$ (Ke01) & $62.4 \pm 0.3$  & $27.5 \pm 0.4$  & $10.1 \pm 0.2$  \\\\\hline
\end{tabular}
\label{Tab:galaxytype}
\end{table}
\par
\begin{figure}
        \includegraphics[width=\hsize]{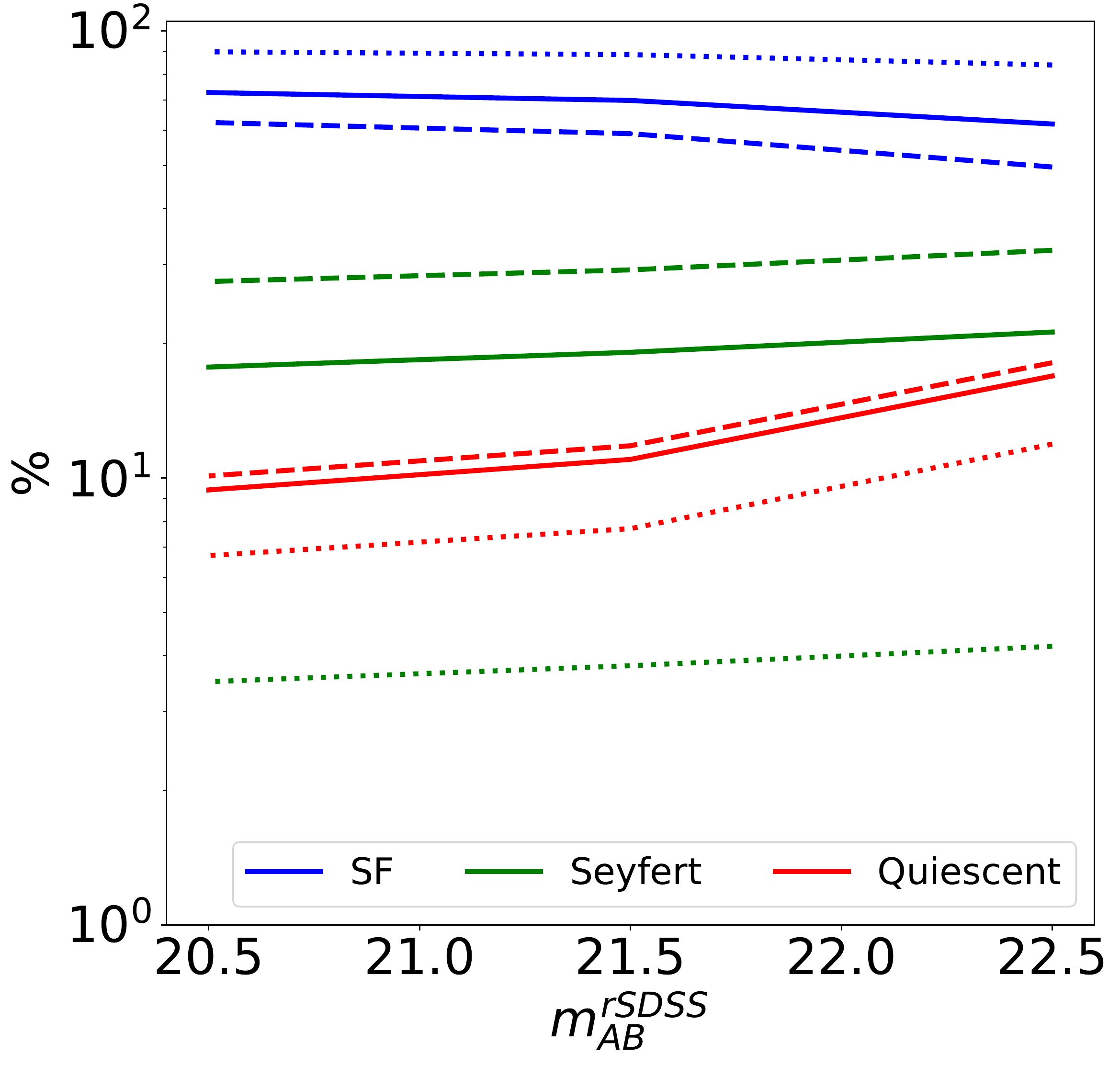}
        \caption{\tiny{Fraction of SF, Seyfert, and quiescent (passive or LINER) galaxies as a function of the maximum rSDSS apparent magnitude of each subsample. Solid, dashed, and dotted lines represent the fraction of each galaxy type according to the Ka03, Ke01, and S08 curves, respectively.}}
 \label{fig:frac_mag}
 \end{figure}

 Finally, we studied how the fractions of SF, Seyfert, and quiescent (passive or LINER) galaxies varied when we imposed brighter flux limit constraints. For this purpose, we generated new samples of galaxies that are below 20.5, 21.5, and 22.5 mag in the rSDSS band and computed the fraction of each galaxy type. The results are shown in Fig. \ref{fig:frac_mag}. We do not observe a strong correlation with the rSDSS apparent magnitude. The fraction of each galaxy type is more uncertain when one or another of the separation curves is chosen.
 
 \section{Characterization of star-forming galaxies}\label{sec:SFR}
In this section, we characterize the star-forming galaxy population in miniJPAS. We traced the SFR through the \Ha emission line. First, we selected a suitable sample of star-forming galaxies with the identification tools we presented in the previous section. Then, we corrected the \Ha flux from nebular extinction and derived the position of SF galaxies in the SFMS. We also analyzed the correlation between nebular and stellar extinction and the relation between the star formation history (SFH) of galaxies obtained with the SED fitting and their position in the SFMS.
\subsection{Selection of star-forming galaxies}\label{subsec:SFselection}
Our sample of star-forming galaxies was obtained from the parent sample (Sect.~\ref{sec:datasample}) by imposing different constraints. We relied on the WHAN diagram to exclude the galaxies in which the main ionization mechanism is not driven by star formation (AGN-like galaxies). We chose the Ka03 curve. In order to consider a galaxy as a member of the main sequence, we therefore imposed a maximum \nii/\Ha of $0.48$. We also discarded galaxies with very low emission in the diagram (LINER and passive galaxies). Finally, galaxies must be classified as blue with the color criterion and the ANN$_\mathrm{C}$ to be part of our sample. We found 1178 galaxies in total (SF sample hereafter).
\par
\begin{figure}
        \includegraphics[width=\hsize]{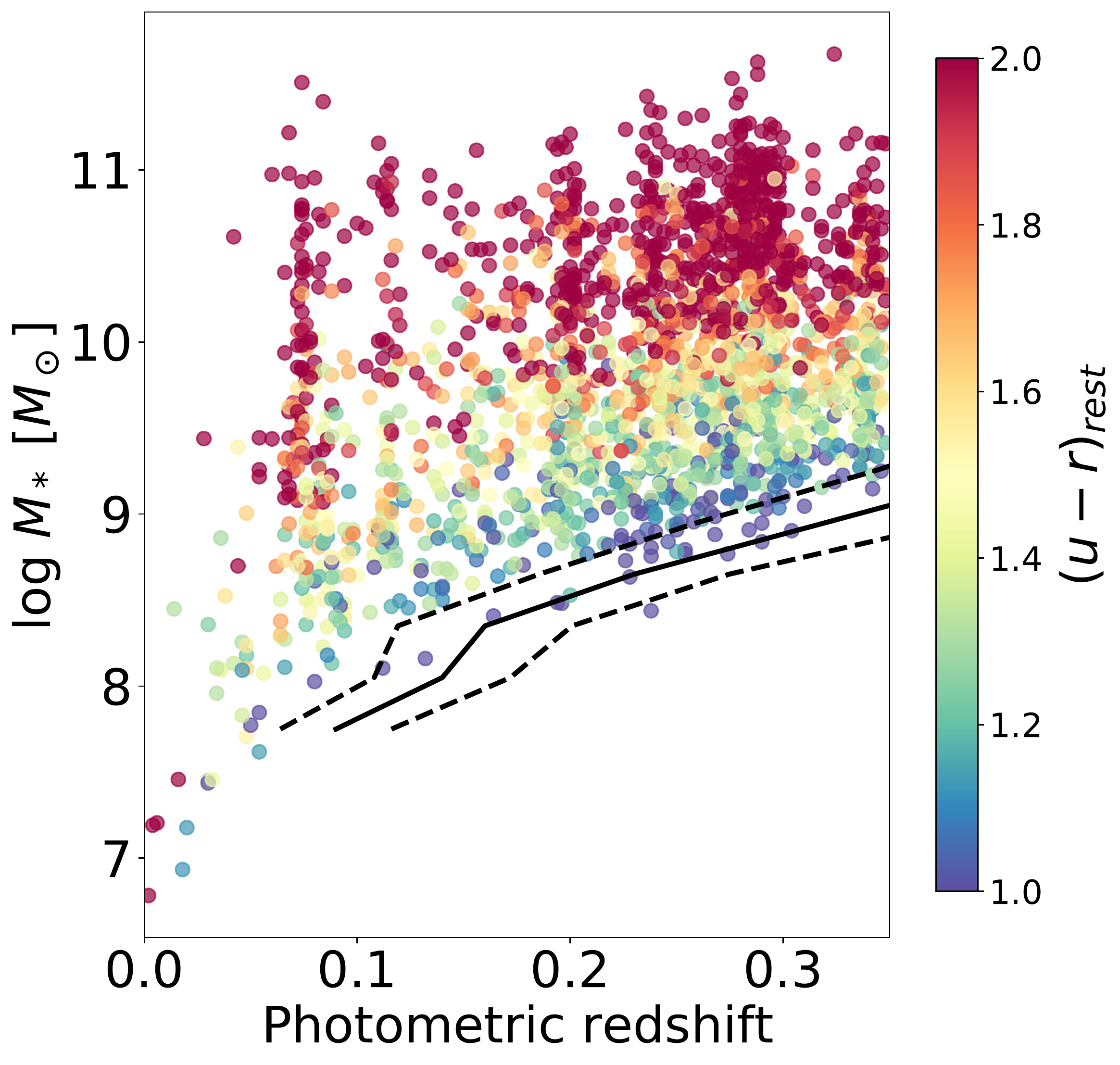}
        \caption{\tiny{Relation between galaxy stellar mass and redshift for all galaxies in the parent sample. The solid black line is the limit at which galaxies can no longer be observed with the criteria we used to select the sample (see Sect.~ \ref{sec:datasample}). Dashed black lines represent the uncertainty limit ($\pm \sigma$.) Galaxies are color-coded according to their (u--r) rest-frame color.}}
 \label{fig:mass_redshift}
 \end{figure}

 In Fig. \ref{fig:mass_redshift} we show the relation between the total stellar mass and the redshift for all galaxies in the parent sample. The solid black line indicates the limit at which galaxies cannot be observed in our flux-limited sample (see Sect.~\ref{sec:datasample}). In order to be complete in mass, we would need to discard a large fraction of galaxies and risk to loose statistical reliability. Furthermore, the mass dynamical range would be significantly reduced at high redshift. Therefore, we fit the SFMS in two cases: using the whole SF sample, or using only galaxies in the SF sample that are above the stellar mass detection limit  (see Sect.~\ref{subsec:Bayesianroutine}). We will also study how stronger flux limit constraints affect the shape of the SFMS. As soon as J-PAS observes larger areas of the sky, we will be able to be more conservative in the mass limit of the selected sample. 
 
\subsection{Dust correction}\label{subsec:dust}
In order to account for the extinction of dust, we followed the empirical extinction relation described in \cite{1994ApJ...429..582C}. The intrinsic luminosity of galaxies ($L_{\mathrm{int}}$) is attenuated by interstellar dust through the following equation:
\begin{equation}
L_{\mathrm{int}}(\lambda) = L_{\mathrm{obs}} (\lambda) 10^{0.4A_\lambda} =  L_{\mathrm{obs}} (\lambda) 10^{0.4k(\lambda)E(B-V)}
,\end{equation}
where $L_{\mathrm{obs}}$ is the observed luminosity, $A_{\lambda}$ is the extinction at wavelength $\lambda$, and $k(\lambda)$ is the reddening curve. We considered the reddening curve of \cite{2000ApJ...533..682C} with $R_{\mathrm{V}} = 4.05$. The nebular color excess $E(B-V)$ can be obtained from the Balmer decrement assuming regular gas conditions in star-forming galaxies \citep[for a detailed description, see, e.g.,][and references therein]{2013ApJ...763..145D} as follows:
\begin{equation}
E(B-V) = 1.97 \log_{10} \left[\frac{(\Ha/\Hb)_{\mathrm{obs}}}{2.86}\right],
\end{equation}
where \Ha and \Hb stand for the emission line fluxes. As the ANN$_\mathrm{R}$ provides the values of the EWs, we used the stellar continuum derived from \texttt{BaySeAGal} at the \Ha and \Hb wavelengths to compute the total flux of the emission lines. 
\par
\begin{figure*}
        \includegraphics[width=\hsize]{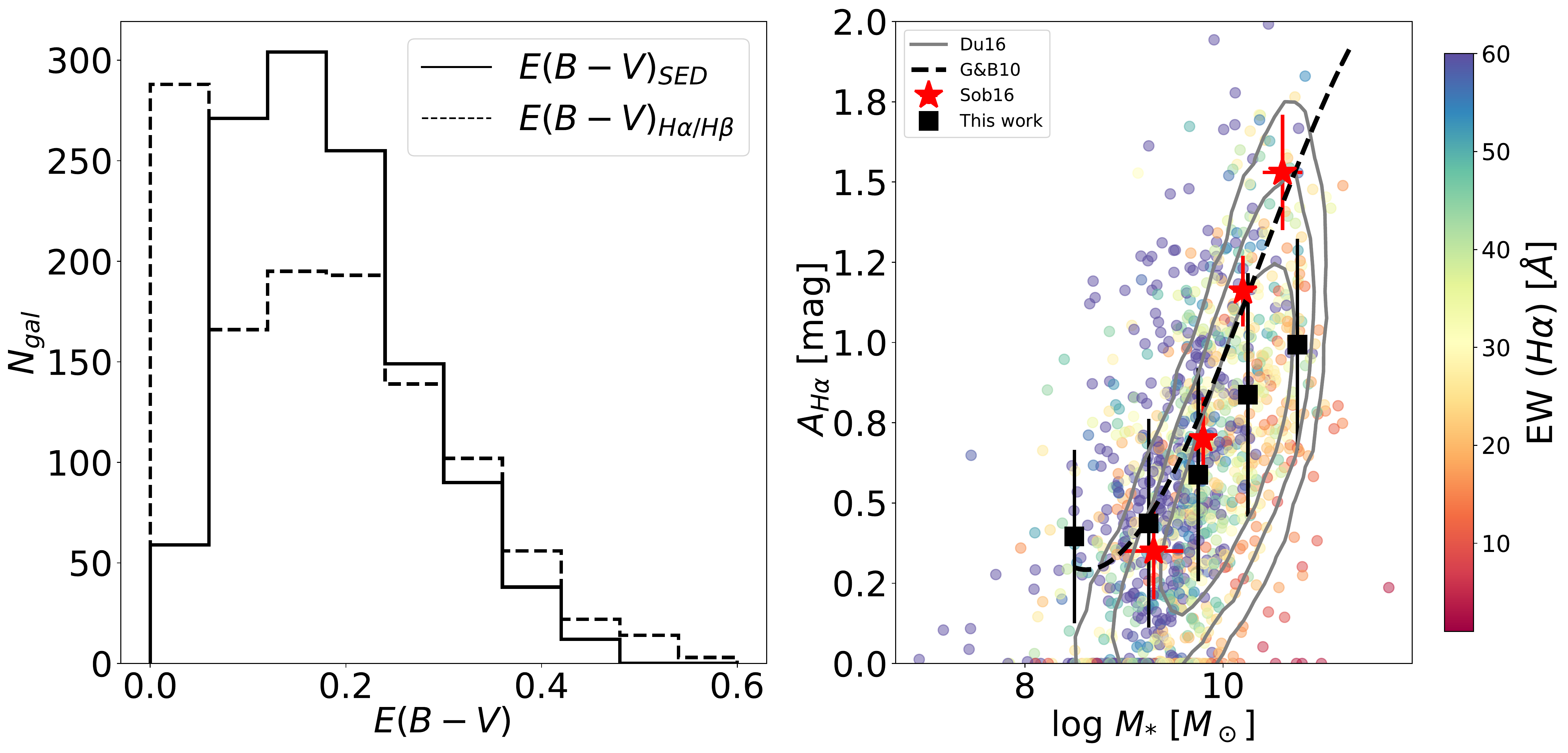}
        \caption{\tiny{Distribution of the nebular $(E(B-V)_{H\alpha/H\beta})$ and stellar $(E(B-V)_{SED})$ color excess (left). Nebular extinction at the \Ha wavelength as a function of stellar mass (right). Galaxies are color-coded with the EW of \Ha and belong the SF sample described in Sect.~\ref{subsec:SFselection}. Black squares are the median obtained in the following stellar mass bins: $8 < \log M_* \leq 9$, $9 < \log M_* \leq  9.5$,  $9.5 < \log M_* \leq  10$, $10 < \log M_* \leq  10.5$, and $10.5 < \log M_* \leq 11$. The error bars on the y-axis represent the standard deviation, gray contours represent the density of sources for 1 $\sigma$, 2 $\sigma,$ and 3 $\sigma$ derived from SDSS galaxies in \cite{2017A&A...599A..71D}. Red stars are the values obtained by \cite{2016MNRAS.458.3443S} by means of spectroscopy measurements in SF galaxies within the cluster Cl0939+4713 at $z=0.41$. The dashed blakc line is the best polynomial fit obtained by \cite{2010MNRAS.409..421G} in a sample of SDSS galaxies.}}
 \label{fig:AvOB}
 \end{figure*}

 In the left panel of Fig. \ref{fig:AvOB} we show the distribution of the nebular $(E(B-V)_{H\alpha/H\beta})$ and stellar $(E(B-V)_{SED})$ color excess. A fraction of galaxies in the SF sample ($\sim 15 \%$) have a Balmer decrement below the theoretical value (2.86), but very close to it. Furthermore, its errors indicate that nebular extinction for these galaxies is compatible with null or very low values. Either way, we set the $E(B-V)_{H\alpha/H\beta}$ to zero for these galaxies. $E(B-V)_{SED}$ is $0.017$ mag higher on average than $E(B-V)_{H\alpha/H\beta}$ with a dispersion of $0.072$ mag. The median error on the $E(B-V)_{H\alpha/H\beta}$ and  $E(B-V)_{SED}$ is 0.089 and 0.015, respectively. Some authors reported that $E(B-V)_{H\alpha/H\beta}$ is twice $E(B-V)_{SED}$ on average \citep{2000ApJ...533..682C,2019ApJ...886...28Q,2019PASJ...71....8K}. However, other studies found similar levels of nebular and stellar extinction \citep{2013ApJ...777L...8K,2016A&A...586A..83P}. In particular, we found agreement with the results of \citet[see Fig. 8 and Table 1,]{2021arXiv210700974K}  who argued that nebular extinction is much more pronounced in the nuclear regions, affecting the relations found by single spectroscopic surveys such as the SDSS, which cannot capture the whole light produced in galaxies.  
\par The right panel of Fig. \ref{fig:AvOB} shows the nebular extinction at the \Ha wavelength ($A_{H\alpha}$) as a function of the galaxy stellar mass. We found a similar trend as in other studies. Red stars are the values obtained by \cite{2016MNRAS.458.3443S} by means of spectroscopy measurements in SF galaxies within the cluster Cl0939+4713 at $z=0.41$. Gray contours represent the density of sources for 1 $\sigma$, 2 $\sigma,$ and 3 $\sigma$ derived from all SDSS SF galaxies in \cite{2017A&A...599A..71D}. Finally, the dashed black line is the best polynomial fit obtained by \cite{2010MNRAS.409..421G} in a sample of SDSS galaxies. Applying aperture correction to the $\Ha/\Hb$ ratio as in \cite{2017A&A...599A..71D} lowers the extinction 0.2 mag in average.

\subsection{Fitting the star formation main sequence}\label{subsec:Bayesianroutine}
The SFR was obtained from the \Ha luminosity using the \cite{1994ApJ...435...22K} relation converted to employ a Chabrier IMF \citep{chabrier2003} and assuming case B recombination,
\begin{equation}
\text{SFR} [M_{\mathrm{\odot}} \hspace{0.2cm}\text{yr}^{-1}] = 4.9\times 10^{-42}L_{\Ha} [\mathrm{erg/s}]
.\end{equation}
We used this relation to derive the SFR from the corrected \Ha luminosity. Then, we fit the SFMS for the galaxies in the SF sample assuming a power-law relation between the stellar mass ($M_{*}$) and the SFR,
\begin{equation}
\log \text{SFR} = \alpha \times \log M_{*} + \beta. 
\label{eq:linear_SFR}
\end{equation}
We assumed that galaxies deviate from this relation with a scatter perpendicular to the line that we parameterized in terms of the scatter along the y-axis ($\sigma_y$), often called $\sigma_{\text{int}}$. We employed a Bayesian approach to derive the posterior distribution of $\sigma_y$, $\alpha$, and $\beta$. We followed \cite{2015PASA...32...33R} in order to construct the likelihood function,
\begin{align}
\ln L = - \frac{1}{2} \sum_{i=0}^{N_{\text{gal}}} \frac{(\log \text{SFR}_i -\alpha \log M_{*,i} - \beta)^2}{\sigma_i^2} \hspace{0.2cm} +  \hspace{0.2cm} \ln \sigma_i^2 \nonumber \\ 
- \ln (\alpha^2 + 1),
\label{eq:likelihood}
\end{align}
where $\sigma_i^2$ reads
\begin{equation}
\label{eq:sigmay}
\sigma_i^2= \sigma_y^2 + \sigma_{\log \text{SFR}_i}^2 +  \alpha^2 \sigma_{\log \text{M}_i}^2.
\end{equation}
We assumed that the errors in the SFR and stellar mass of the galaxies are not correlated. This hypothesis is justified because both quantities are derived independently from each other. Although the flux of stellar continuum at \Ha wavelength is used to estimate the total \Ha flux, its error is negligible compared to the error in the EW. The errors are considered Gaussian and heteroscedastic, that is, each data point is drawn from a different Gaussian distribution. The last term in Eq. \ref{eq:likelihood} ensures that the data are rotationally invariant. In other words, data have no defined predictor or response variable, and therefore we can predict the SFR from the stellar mass of the galaxy and vice versa. 
\par The posterior distribution was sampled with the Markov chain Monte Carlo (MCMC) method, using the \texttt{emcee} Python implementation \citep{2013PASP..125..306F}, with 250 walkers and 5000 steps per walker. 
We used a burn-in phase of 3500 steps. 
\par
\begin{figure}
        \includegraphics[width=\hsize]{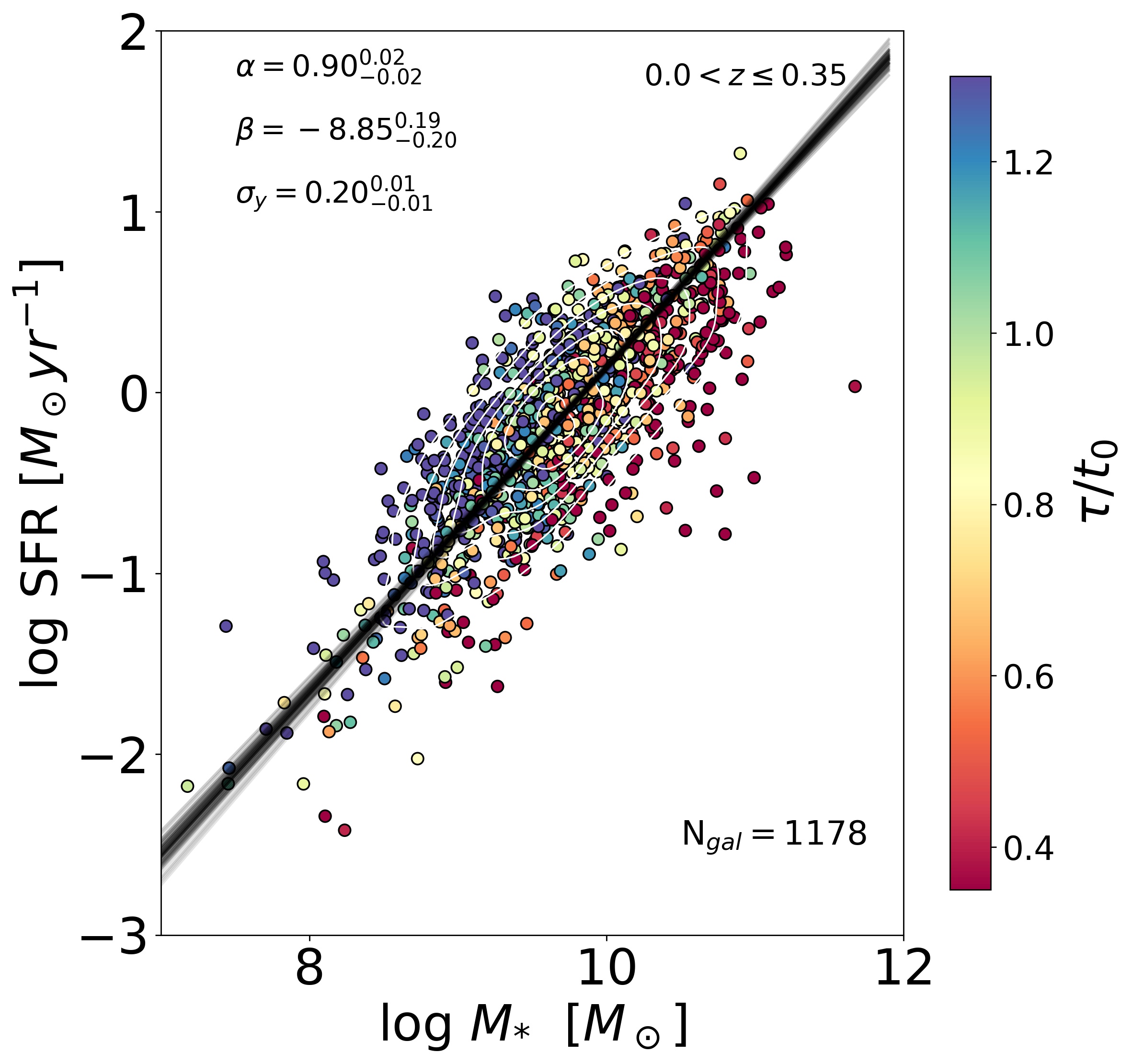}
        \caption{\tiny{SFR vs. stellar mass for the galaxy sample described in Sect.~ \ref{subsec:SFselection}. Galaxies are color-coded with the $\tau/t_0$ ratio (see Sect.~  \ref{subsec:stellar}). Black lines are the best fits obtained with the Bayesian routine. The median posterior value and $1 \sigma$ confidence interval are shown for each of the parameters.}}
 \label{fig:SFRvsMass}
 \end{figure}

 Figure \ref{fig:SFRvsMass} shows the SFMS for the galaxies in the SF sample; in black we plot the ensemble of best fits obtained with the Bayesian routine. Galaxies are color-coded with the $\tau/t_0$ ratio, which is an indicator of the SFH (see Eq. \ref{eq:delta_delay}). High values of $\tau/t_0$ indicate an SFH with almost constant SFR throughout cosmic time, while low values are related to galaxies with a burst of star formation long ago with a decreasing SFR ever since. 
\par On the one hand, the color gradient observed in Fig. \ref{fig:SFRvsMass} suggests that galaxies with higher values of $\tau/t_0$ are more likely to be found above the SFMS and preferentially have stellar masses below  $10^{10} M_\mathrm{\odot}  $. On the other hand, lower values of $\tau/t_0$ are associated with  massive galaxies that lie below the SFMS. 
\par We investigated how the parameters of the SFMS are affected when we included only the galaxies in the SF sample that lie above a certain flux limit. Additionally, we generated a new sample of galaxies that were selected from the SF sample with stellar masses above $10^9 M_\mathrm{\odot} $ (SF0 sample). This is the stellar mass detection limit for the redshift between 0 and 0.35 (black line in Fig. \ref{fig:mass_redshift}). Subsequently, we studied again how the flux limit cut affects the parameters of the SFMS. The results are summarized in Table \ref{tab:maglimSFRM}. We conclude that the selection function that depopulates the SFMS below $m_{AB} = 22.5$ in the rSDSS band does not affect the shape of the SFMS. The results for the SF and SF0 sample are consistent (compatible within the errors).

\begin{table}
\caption{\tiny{Parameters of the SFMS with different selection cuts in the rSDSS band for the SF (SF0) sample at the top (bottom). }}
\begin{tabular}{llll}
\hline
\hline
\\
rSDSS &  $ \alpha $  & $\beta$ &  $\sigma_y$ \\\\ \hline \\ 
 $ \le 22.5$ &  $0.90^{+ 0.02}_{-0.02}$ &  $-8.85^{+ 0.19}_{-0.20}$ &   $0.20^{+ 0.01}_{-0.01}$ \\\\
  $ \le 21.5$ &  $0.93^{+ 0.02}_{-0.02}$ &  $-9.15^{+ 0.21}_{-0.21}$ &   $0.21^{+ 0.01}_{-0.01}$ \\\\
 $ \le 20.5$ &  $0.93^{+ 0.03}_{-0.03}$ &  $-9.27^{+ 0.26}_{-0.27}$ &   $0.22^{+ 0.01}_{-0.01}$ \\\\
\hline \\
 $ \le 22.5$ &  $0.93^{+ 0.03}_{-0.03}$ &  $-9.17^{+ 0.29}_{-0.29}$ &   $0.21^{+ 0.01}_{-0.01}$ \\\\
  $ \le 21.5$ &  $0.95^{+ 0.03}_{-0.03}$ &  $-9.37^{+ 0.30}_{-0.33}$ &   $0.21^{+ 0.01}_{-0.01}$\\\\
 $ \le 20.5$ &  $0.97^{+ 0.04}_{-0.04}$ &  $-9.66^{+ 0.30}_{-0.30}$ &   $0.23^{+ 0.02}_{-0.01}$\\\\

\end{tabular}

\label{tab:maglimSFRM}
\end{table}

\subsection{SFR at different redshift}\label{sub:SFR-bins}
The relation of the SFR and the stellar mass is expected to change as a function of the redshift due to  changes in the cosmic gas accretion rates and the gas depletion timescales. Some authors modeled this relation with a power law  \citep[SFR $\propto (1 + z)^a$,][]{2018A&A...619A..27B,2015A&A...575A..74S}, others assumed that the evolution takes place in the zeropoint \citep[$\log$ SFR $\propto \beta z $,][]{2021MNRAS.501.2231S}. Another common approach is to split the sample into redshift bins and fit them independently \citep[e.g.,][]{2016MNRAS.461..458D,2020arXiv201113605T}. Because the redshfit range of the SF sample is limited, we decided to employ the latter approach and fit the SFMS in three different redshift bins: $0 < z \leq 0.15$, $0.15 < z \leq 0.25$, and $0.25 < z \leq 0.35$. We removed all galaxies in each sample that lay below the stellar mass limiting value (solid black line in Fig. \ref{fig:mass_redshift}). 
\par
\begin{figure*}
    \centering
 \includegraphics[width=6.0cm,height=6.1cm]{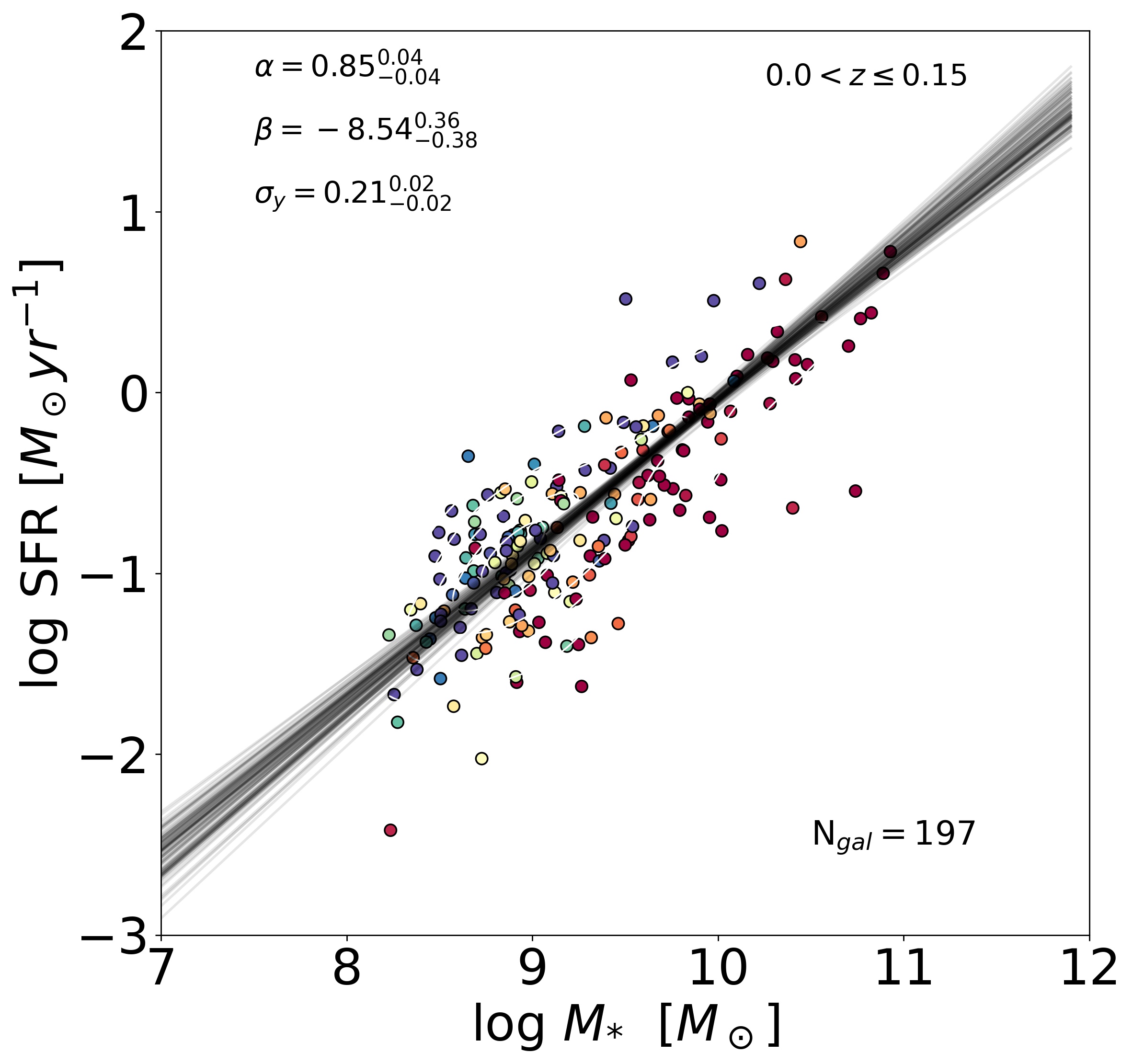}
 \includegraphics[width=5.6cm,height=6cm]{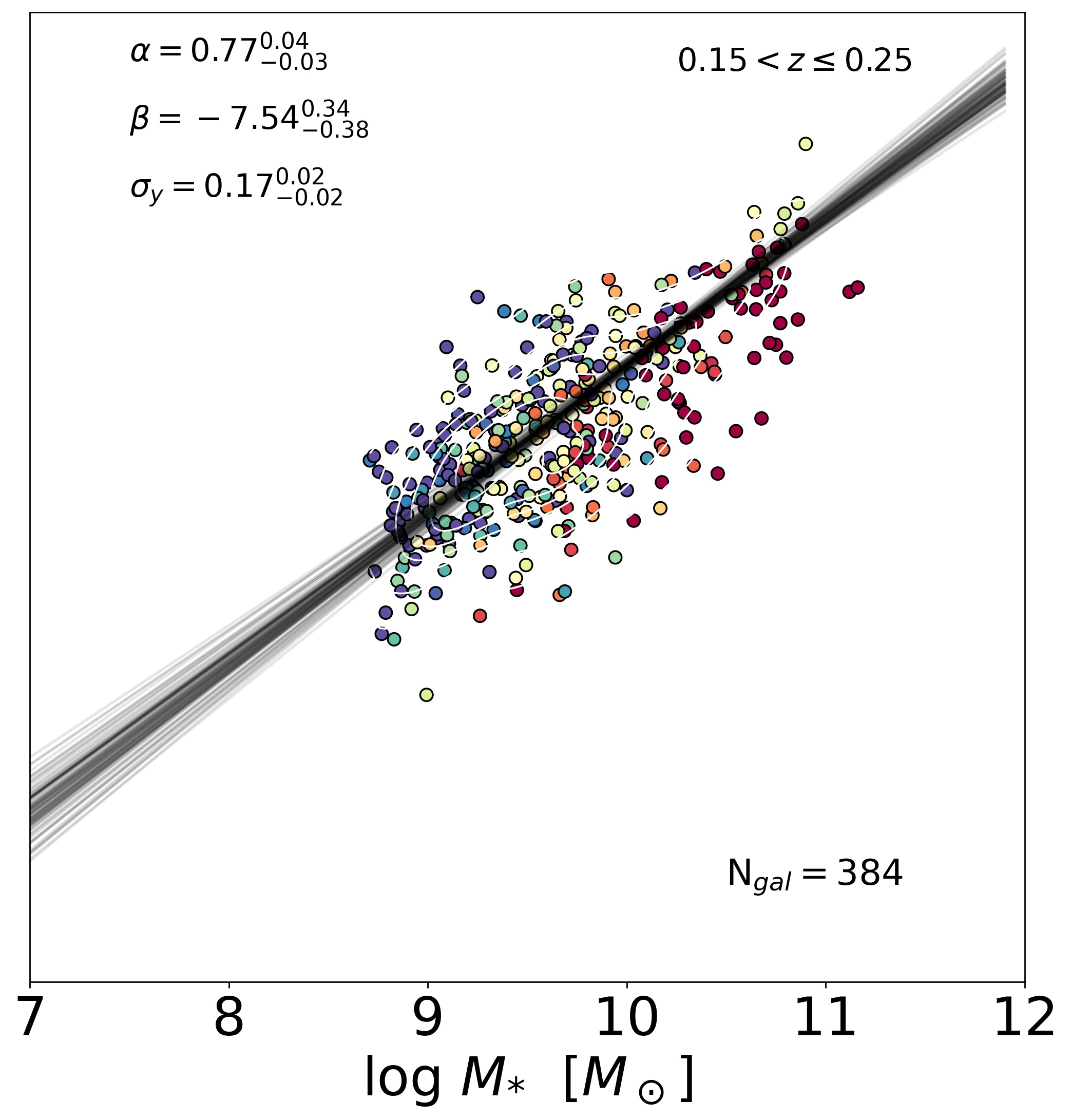}
 \includegraphics[width=6.5cm,height=6cm]{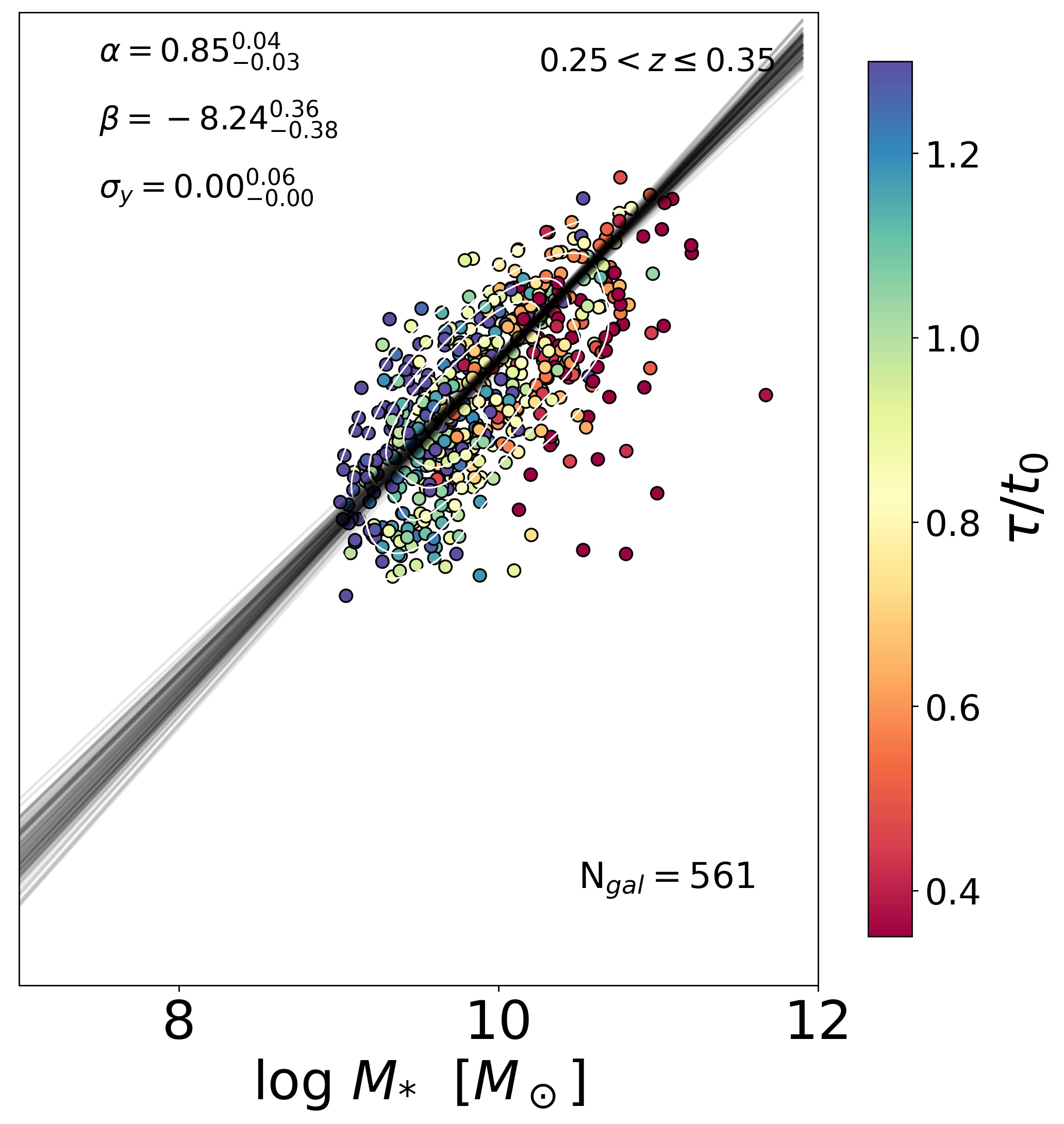}
        \caption{\tiny{SFR vs. stellar mass for galaxies in different redshift bins color-coded with their the $\tau/t_0$ ratio (see Sect.~  \ref{subsec:stellar}.) Black lines are the best fits obtained with the Bayesian routine. The median posterior value and $1 \sigma$ confidence interval are shown for each of the parameters. The number of galaxies within each redshift bin is also indicated.}}
 \label{fig:SFR_bins}
 \end{figure*}

 We show the results in Fig.~\ref{fig:SFR_bins}. A small flattening of the relation is seen at intermediate redshifts, but it may not be significant. As expected due to the anticorrelation between the slope and the zeropoint, the latter becomes higher in the $0.15 < z \leq 0.25$ bin. Most likely, these discrepancies are caused by the effect of fitting the SFMS within a smaller dynamical range of mass and by the lower statistics. The intrinsic scatter of galaxies along the SFMS decreases at higher redshifts. This may be caused by a dependence on stellar mass rather than on redshift. Galaxies below $1.6 \times 10^8 M_\mathrm{\odot}  $, $5 \times 10^8 M_\mathrm{\odot}  $, and  $10^9 M_\mathrm{\odot}$ for $0 < z \leq 0.15$, $0.15 < z \leq 0.25$, and $0.25 < z \leq 0.35$, respectively, cannot be detected with fluxes brighter than 22.5 in the rSDSS band. We discuss the implication of this result in more detail in Sect.~ \ref{subsubsec:intrinsic_scatter}. 

\subsection{Turnover mass hypothesis}\label{subsec:turn_over}
Several studies have shown evidence that the relation between the SFR and the stellar mass turns over at a mass of $M_{*}\sim~10^{10}~M_{\mathrm{\odot}}$ \citep{2014ApJ...795..104W,2015ApJ...801...80L,2015A&A...575A..74S,2016ApJ...817..118T}. In this section, we investigate this scenario by fitting a quadratic power law (Eq. \ref{eq:quadratic_law}) and a broken power law (Eq.\ref{eq:broken_power_law}) to the SF sample,
\begin{equation}
\log \text{SFR} = \alpha \times \log M_{*} + \gamma \times (\log M_{*})^2 + \beta
\label{eq:quadratic_law}
\end{equation}
\begin{equation}
\log \text{SFR} = \beta  - \log \left [1 + (M_*/M_0)^{-\alpha}\right]
\label{eq:broken_power_law}
.\end{equation}
We obtained a turnover mass ($\log M_0 = 10.93^{+ 0.22}_{-0.17}$) that is very close to the highest mass that we have in the SF sample ($\log M^{\mathrm{max}}_*=11.2$). Furthermore, only 14 out of 1178 galaxies have a mass higher than $M_0$. For the quadratic model, we obtained a quadratic term near zero ($\gamma = -0.08^{+ 0.02}_{-0.02} $). In Table \ref{Tab:comparison_SFMS} (see next section) we show the best-fitting  parameters for different separation curves. We employed the Bayesian information criterion (BIC) to determine the model that better describes the observed SMFS. The BIC is defined as $\text{BIC} = n_{\text{param}} \ln N_{\text{gal}}-2 \ln L$, where $n_{\text{param}}$ is the number of parameters in the model, $N_{\text{gal}}$ is the number of galaxies, and $L$ is the likelihood function. The linear model (Eq. \ref{eq:linear_SFR}) obtained the lowest value. Therefore, it is the most likely model. 
\subsection{AGN selection criteria}\label{sub:AGN_criteria}
The exclusion of AGN-like galaxies from the SF sample is based on the $\nii /\Ha$ ratio and the EW of \Ha. We chose the curve of Ka03 to select SF galaxies, but we could have relied on other separation curves, such as Ke01 or S08. In this section we study how these choices can impact our result. 
\par In Table \ref{Tab:comparison_SFMS} we show the best-fit parameter values as a function of the separation curves, the redshift bin, and the fitting equation used to model the SFMS. The results are marginally consistent, meaning that the retrieved parameter does not change the main conclusion of the previous sections. Nevertheless, we observed a trend in the slope, the quadratic term, and in the turnover mass as we relaxed the maximum $\nii /\Ha$ ratio allowed to be part of the SFMS. Galaxies at the border of the dividing lines populate the high-mass end. As a consequence, the quadratic terms and the turnover mass increase as the slope of the SFMS flattens. Nonetheless, the intrinsic scatter exhibits little variation, except for the highest redshift bin, where  higher-mass galaxies increase the scatter. This exercise demonstrates that the SFMS can be affected by AGN contamination, which is only one ingredient in the definition of the SFMS. Other criteria based on color cuts or sSFR thresholds are also important and can have a non-negligible impact on the derived parameters of the SFMS \citep{2018MNRAS.477.3014B,2019MNRAS.482.1557S,2021MNRAS.503.5115K}.
\section{Discussion}\label{sec:discussion}
In the following sections, we compare the results of the SFMS with the literature. We derive the cosmic evolution of the star formation rate density up to $z=0.35$, and we discuss the differences we found with respect to other studies that did not trace the SFR with \Ha emission line.
\subsection{SFMS: Comparison with the literature}\label{subsec:literature}
We have modeled the SFMS in the mass range from $10^8$ up to $10^{11} M_\mathrm{\odot} $ in the redshift range $0 < z < 0.35$. We employed a Bayesian approach (Sect.~\ref{subsec:Bayesianroutine}) that considers the intrinsic scatter of the SFMS and the heteroscedastic errors on the stellar masses and the SFRs. We derived the SFRs from the \Ha emission line, and we corrected for dust extinction through the Balmer decrement. We relied on the \nii/\Ha ratio to remove from the sample galaxies hosting an AGN. The linear model explains the relation between the $\log \text{SFR}$ and $\log M_*$ for the sample of SF galaxies better. Our selection criteria combine color-cut and emission line diagnostics and consequently favour a pure rather than a complete sample of SF galaxies. Most probably, we also excluded most of the GV population, and this might explain why the turnover-mass scenario is not compatible with our results. We compare our results with the literature below. We focus our attention on the slope of the SFMS and on the intrinsic scatter.
\subsubsection{Slope}
\begin{figure}
    \centering
\includegraphics[width=\hsize]{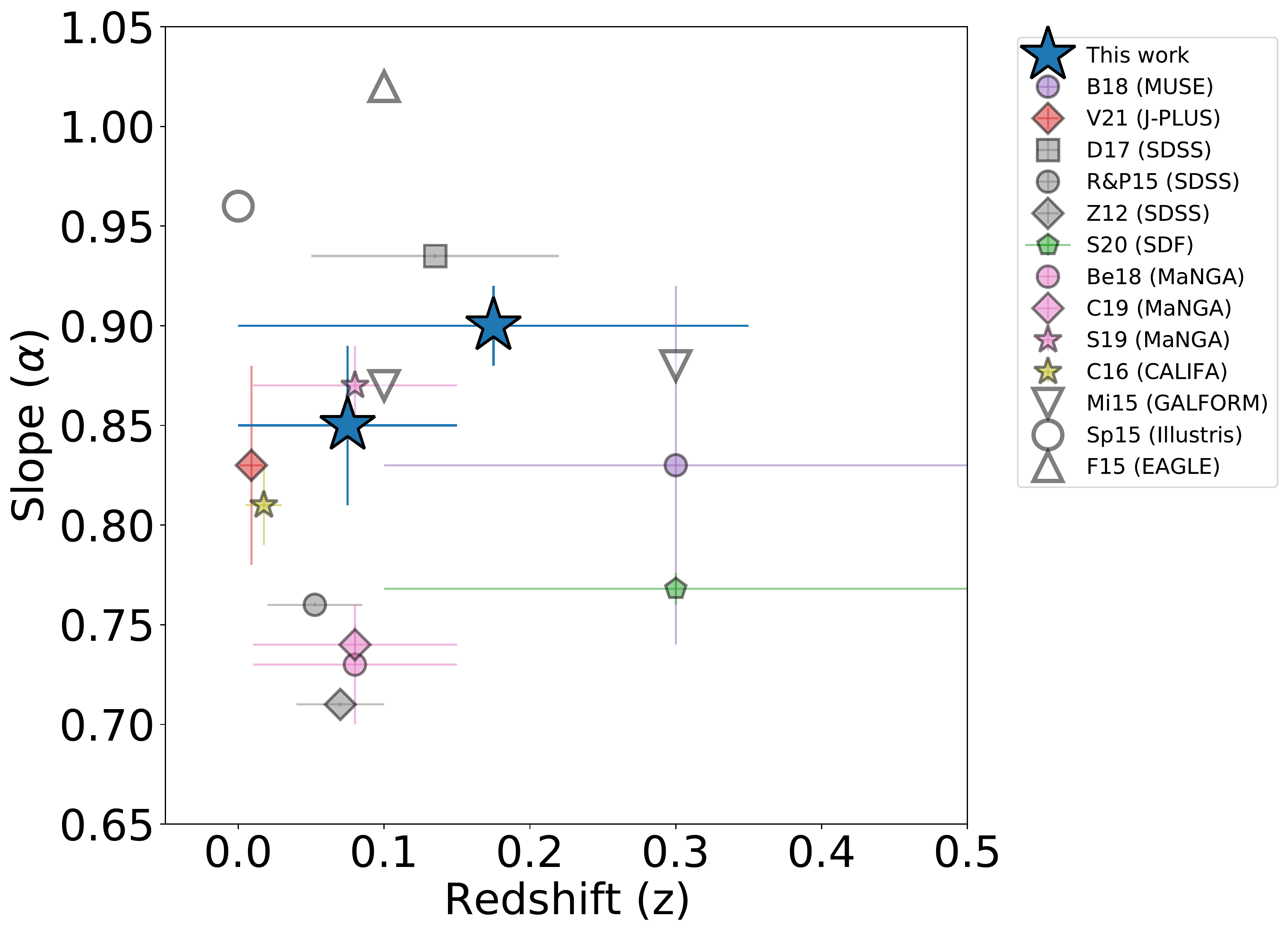}
        \caption{\tiny{Slope of the SFMS derived from the \Ha emission line by different works as a function of the redshift.  The bars on the x-axis represent the redshift range of the galaxies involved in each study. Our best fit of the SFMS is shown with large blue stars for the lowest redshift range ($0 < z \le 0.15$) and the SF sample ($0 < z \le 0.35$). The results of the literature are from \cite{2018A&A...619A..27B} (B18), \cite{2021arXiv210104062V} (V21), \cite{2017A&A...599A..71D} (D17), \cite{2015ApJ...801L..29R} (R$ \& $P15), \cite{2012ApJ...757...54Z} (Z12), \cite{2021MNRAS.501.2231S} (S20),
        \cite{2018MNRAS.477.3014B} (Be18),
        \cite{2019MNRAS.488.3929C} (C19),
        \cite{2019MNRAS.482.1557S} (S19), and
        \cite{2016ApJ...821L..26C} (C16). We also include the results derived by GALFROM (a semianalytical model) \citep{2014MNRAS.444.2637M} (Mi15), and from hydrodynamical simulations, \cite{2015MNRAS.447.3548S} (Sp15) and \cite{2015MNRAS.450.4486F} (F15).}}
\label{fig:slope_literature}
\end{figure}

We find a result very similar to those of \cite{2019MNRAS.482.1557S} (MaNGA) and \cite{2016ApJ...821L..26C} (CALIFA), but our slope is steeper than those of \cite{2018MNRAS.477.3014B} and \cite{2019MNRAS.488.3929C}, who used MaNGA data. 
Our results are also consistent with the recent work of \cite{2021arXiv210104062V}, who studies the SFR of galaxies in the nearby Universe with J-PLUS data. SDSS galaxies have also been used to analyze the SFMS. The slopes found by \cite{2012ApJ...757...54Z} and \cite{2015ApJ...801L..29R} are flatter than our results. Nevertheless, \cite{2017A&A...599A..71D} applied aperture correction based on CALIFA data \citep{2016ApJ...826...71I} to recover the total flux from SDSS fiber spectroscopy and found a slope of 0.935, which is very close to our slope, which we obtained with the SF sample in the $0 < z \le 0.35$ redshift range (see Fig. \ref{fig:slope_literature}).  \cite{2021MNRAS.501.2231S} obtained a flatter slope than we did based on galaxies from the Subaru Deep Field at intermediate redshift ($0.1 < z \le 0.5$). However, we recovered a slope that is marginally consistent with the one found by \cite{2018A&A...619A..27B}, who used data from the Multi Unit Spectroscopic Explorer (MUSE) and employed the same method as we used to fit the SFMS. 

\subsubsection{Intrinsic scatter}\label{subsubsec:intrinsic_scatter}
The amount of intrinsic scatter is hard to  constrain because the scatter caused by the measurements errors in both the stellar masses and the SFRs needs to be accounted for. As pointed out by \cite{2018A&A...619A..27B}, this is one of the advantages of using the fitting model of \cite{2015PASA...32...33R}. We obtained an intrinsic scatter of $0.20$ dex for the SF sample ($0 < z \leq 0.35$). This is consistent with previous works, which found values ranging from 0.15 up to 0.5 dex \citep[see, e.g.,][]{2012ApJ...754L..29W,2012ApJ...754L..14S,2014ApJS..214...15S,2015A&A...575A..74S,2015A&A...579A...2I}. 
\par Many factors than can impact the amount of intrinsic scatter. First of all, different SFR indicators account for variations in the SFH on different timescales \citep[see, e.g.,][and references therein]{2016MNRAS.461..458D}. For instance, while \Ha provides a direct measure of the current SFR in galaxies ( < $10-20$ Myr), UV-like tracers can detect changes in the SFH in only the last $100$ Myr and are therefore less sensitive to recent episodes in the SFH that enhance or suppressed the star formation in the galaxy. Secondly, the selection criteria that defined the SFMS can boost or decrease artificially the scatter by excluding or including a fraction of galaxies that `belong' or not to the SFMS. 
\par The results obtained in each redshift bin show a decrease in intrinsic scatter for galaxies with higher redshift. 
The MC approach predicts $\sigma_y$ to be compatible with zero in the last redshift bin. This might be the effect of the method. When we averaged over all galaxies in Eq. \ref{eq:sigmay} and solved for $\sigma_y$  , we found $\sigma_y = $ 0.19, 0.09, and 0.17 dex for $0 < z \leq 0.15$, $0.15 < z \leq 0.25$, and $0.25 < z \leq 0.35,$ respectively. However, we found a very similar value for the SF sample of galaxies (0.22 dex). As we pointed out in Sect.~ \ref{sub:SFR-bins}, the selection function in the SF sample together with the low statistics in each redshift bin might affect the results. 
\subsubsection{SFMS with \texttt{BaySeAGal} }\label{subsubsec:SFM_Bay}
The SED fitting performed by \texttt{BaySeGal} yields the SFH of galaxies, and therefore we can estimate the current SFR in each galaxy by summing all the  mass that formed stars in the last 30 Myr. Since tau-delayed models cannot account for a bursty SFH, any value between 10 to 200 Myr provides essentially the same SFR. A comparison of the results of the SFMS derived from the flux of \Ha with a different and independent technique provides valuable information about the potential inaccuracies and strengths of our method. 
\par
\begin{figure}
        \includegraphics[width=\hsize]{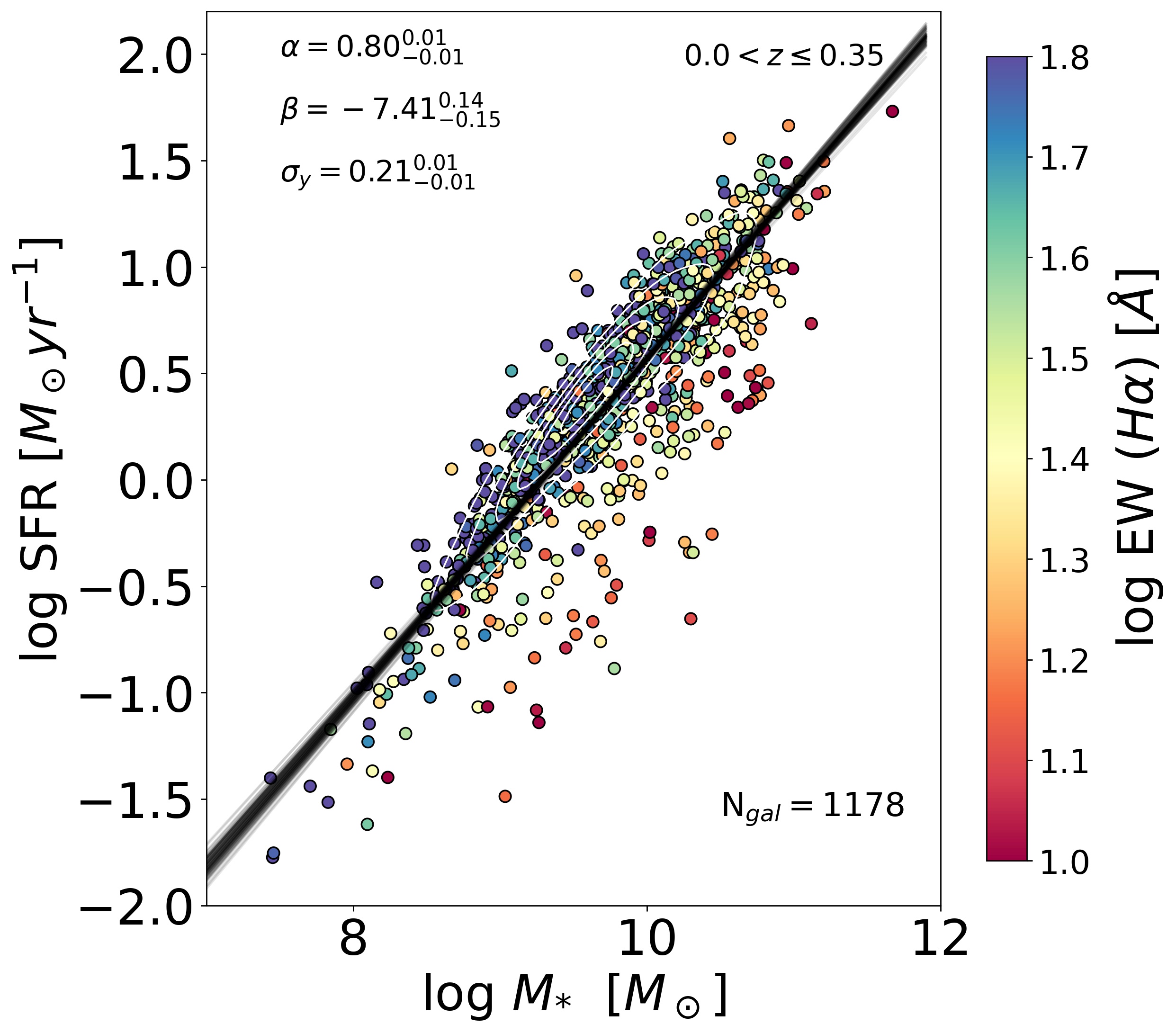}
        \caption{\tiny{SFR vs. stellar mass for the galaxy sample described in Sect.~ \ref{subsec:SFselection}. SFRs are derived from \texttt{BaySeGal}. Galaxies are color-coded with the EW of \Ha. Black lines are the best fits obtained with the Bayesian routine. The median posterior value and the $1 \sigma$ confidence interval are shown for each of the parameters.}}
 \label{fig:SFRBayvsMass}
 \end{figure}

 In Fig. \ref{fig:SFRBayvsMass} we show the SFMS for the same sample of galaxies described in Sect. \ref{subsec:SFselection} that is plotted in Fig. \ref{fig:SFRvsMass}. The color code now represents the EW of \Ha. As expected, galaxies with higher values in the EW of \Ha are placed above the main sequence. This suggests that the two methods are consistent overall. Nevertheless, we obtained a zeropoint that is higher, meaning that the SFR derived from the analysis of the stellar populations gives higher values on average. This discrepancy later translates into the cosmic SFR density and the number of ionizing photons. In Sect. \ref{subsec:diss_CESR}, we discuss the possible origin of this difference in detail.
\par We obtain a slope that is slightly flatter, but still closer to what we retrieved with \Ha. The different assumptions made by each method mean that this difference is expected. While the \Ha flux is very sensitive to recent changes in the star formation activity of a galaxy, the SFR derived from the SED fitting traces the SFR on longer timescales. As a consequence, recent episodes that  enhance or suppress the SFR might result in a global change in slope with respect to an SFMS derived from the average SFR over the last 200 Myr. 

\subsection{Cosmic evolution of the star formation rate density}\label{sec:CSFR}
The star formation rate density of the universe has been estimated by different means. Galaxy redshift surveys found that $\rho_{\text{SFR}}$ peaks at $\sim$ 3.5 Gyr after the Big Bang ($z \sim 2$) and has decreased ever since \citep[e.g.,][]{2013MNRAS.433.2764G,2013MNRAS.428.1128S,2014ARA&A..52..415M,2018MNRAS.475.2891D}. A similar trend was confirmed  with galaxies in the nearby Universe using the so-called fossil record method \citep{2018A&A...615A..27L,2019MNRAS.482.1557S,2020MNRAS.498.5581B}. Very recently,  \cite{2021arXiv210213121G} employed this method to derived the $\rho_{\text{SFR}}$ from a subsample of galaxies in miniJPAS ($0.05 \leq z \leq  0.15$). The agreement with cosmological surveys is remarkable, even though different SED-fitting codes were used. In this section, we estimate the $\rho_{\text{SFR}}$ from the SFR derived with the flux of \Ha at the same redshift bins as described in Sect.~ \ref{sub:SFR-bins}. 
\par The miniJPAS area comprises only 0.895 deg$^2$ of the central regions of the AEGIS field. Therefore, our cosmological volume is somewhat limited, especially at low redshift. In this regard, a study of the $\rho_{\text{SFR}}$ using miniJPAS data maybe affected by cosmic variance effects \citep{2010MNRAS.407.2131D,2011ApJ...731..113M}. The main source of uncertainty of $\rho_{\text{SFR}}$ comes from this effect. We followed eq. 4 in \cite{2010MNRAS.407.2131D} to quantify the cosmic variance of miniJPAS at different redshift bins,
\begin{equation}
\begin{split}
\zeta_{\text{Cos. Var.}} (\text{per cent}) = [1.00 - 0.03 \sqrt{A/B -1}]  \\ 
\times \hspace{0.1cm} [219.7 - 52.4 \log (AB \times 291.0 )] \\
+ 3.21 \log (AB \times 291.0 )^2/\sqrt{NC/291.0},
\end{split}
\end{equation}
where $N$ is the number of fields observed by miniJPAS (simply one), A and B are the median transverse lengths, and C is the radial depth. We obtained a cosmic variance for the comoving number density of galaxies of 37$\%$ (0.16 dex), 27$\%$ (0.12 dex), and 21$\%$ (0.09 dex) for the volumes within $0 < z \leq 0.15$, $0.15 < z \leq 0.25$, and $0.25 < z \leq 0.35$, respectively. In the future, J-PAS will scan $\sim8000$ deg$^2$ in the northen sky, and the effect of cosmic variance will be negligible (less than 1\%).
\par In order to estimate $\rho_{\text{SFR}}$ , we computed the total sum of the SFR for the galaxies in our sample and divided it by the volume contained in each redshift bin ($V_{int}$). We selected them from the parent sample with the same criteria as we used in Sect.~\ref{subsec:SFselection} to generate the SF sample. However, we relied on the Ke01 curve to exclude AGNs. We found a total of 1361 galaxies. In this way, we ensured that we did not underestimate $\rho_{\text{SFR}}$ by excluding objects that lie between the Ke01 and Ka03 lines, which might contribute much to the flux of \Ha through ionized interstellar gas. In any case, the difference between selecting SF galaxies with the Ka03 or the Ke01 line is only 0.05 dex in $\log \rho_{\text{SFR}}$. 
\par The photometric depth of miniJPAS prevents us from detecting a fraction of galaxies below a certain mass limit. This effect becomes stronger for galaxies at higher redshift. Therefore, we have to apply volume corrections to reduce the impact of the lack of low-mass galaxies in the highest resdshift bins in this work. We used the classical $V_{int}/V_{max}$ technique described originally in \cite{1968ApJ...151..393S} and \cite{1973ApJ...186..433H}, \cite[see Appendix C in][for a detailed discussion of this correction]{2021arXiv210104062V}. This is formally expressed as:
\begin{equation}
    \rho^{int}_{\text{SFR}} = \sum_{i \in j} \frac{\text{SFR}_i}{V_{int}} w_i,
    \label{eq:p_SFR}
\end{equation}
where $w_i = V_{int}/V^{max}_i$  is the weight that each galaxy has in the total $ \rho^{int}_{\text{SFR}}$ , and $V_{max}$ is the maximum volume occupied by a galaxy assuming that it cannot be observed at a magnitude fainter than $22.7$. For galaxies with $V_{int} \leq V^{max}_i$ , the weight is simply one, but galaxies with $V_{int} > V^{max}_i$  will contribute more.
\par
\begin{figure}
    \centering
        \includegraphics[width=\hsize]{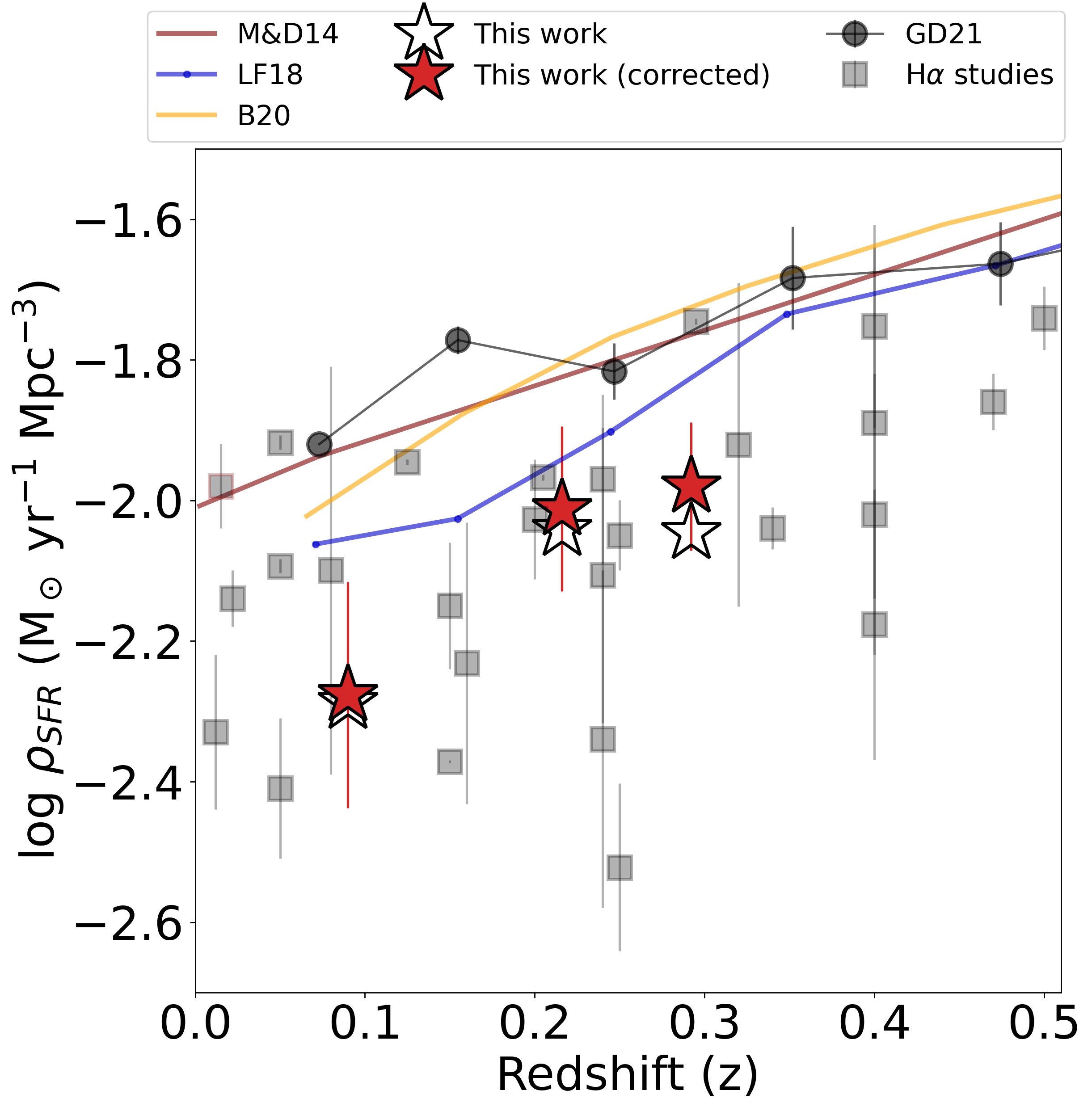}
        \caption{\tiny{Star formation rate density at $z < 0.35$. Red stars show the values obtained in this work from the luminosity of \Ha. Empty stars are uncorrected values that do not take galaxies with undetectable nebular emission lines or with very low S/N (see text in Sect.~\ref{sec:CSFR}) into account. Black circles are the values obtained by \cite{2021arXiv210213121G} applying the fossil record method to a sample of miniJPAS galaxies in the range $0.05 < z \leq 0.15$. Squares are studies based on \Ha (see references in Table \ref{Tab:HalphaSFR}). Solid lines represents the trends obtained by different studies based on the stellar continuum: \citet[][M$\&$D14]{2014ARA&A..52..415M} , \citet[][LF18]{2018A&A...615A..27L}, and \citet[][B20]{2020MNRAS.498.5581B}. All values are scaled to the \cite{chabrier2003} IMF.}}
\label{fig:CSFR}
\end{figure}

 A direct comparison of $\rho_{\text{SFR}}$ with the results obtained in \cite{2021arXiv210213121G} also requires applying a correction to account for the galaxies that are detectable in the rSDSS band and are consequently fitted by the SED-fitting codes, but their emission lines cannot be measured because of the low S/N ratio. From the galaxies that belong to this group, we took those that were classified as blue by the color criterion and used their mass to place them in SFMS derived in Sect. \ref{subsec:Bayesianroutine}. In this  way, we can estimate their SFR with Eq. \ref{eq:linear_SFR} and add their contribution to $\rho_{\text{SFR}}$. These corrections are indeed minor, as shown in Fig. \ref{fig:CSFR} (red stars are the corrected values, and empty stars represent the uncorrected stars), but become slightly stronger at higher redshift.
\par In Fig. \ref{fig:CSFR} we also show the values obtained by several studies that used the \Ha flux to estimate the $\rho_{\text{SFR}}$ at different redshift bins (squares, see references in Table \ref{Tab:HalphaSFR}). It is remarkable that most of them predict lower values of $\rho_{\text{SFR}}$ than works that used the stellar continuum (solid line). Finally, black circles show the values obtained with the fossil record method by  \cite{2021arXiv210213121G} for miniJPAS galaxies in the range $0.05 < z \leq 0.15$. 
\par Our results reproduce the $\rho_{\text{SFR}}$ well that was found with other studies using \Ha as a tracer to measure the SFR. Nevertheless, we found a non-negligible difference with respect to the results found by studies based on the stellar populations \citep{2014ARA&A..52..415M,2018MNRAS.475.2891D,2018A&A...615A..27L,2019MNRAS.482.1557S,2019ApJ...876....3L,2020MNRAS.498.5581B,2021arXiv210213121G}. Our estimation of $\rho_{\text{SFR}}$ does not take the SFR into account that is ongoing in galaxies hosting an AGN.
\begin{table}
\caption{\small{Compilation of star formation rate densities derived from \Ha. All values are scaled to \cite{chabrier2003} IMF.  $\log \rho_\star$ is in units of $M_\mathrm{\odot} \text{yr}^{-1} \text{Mpc}^{-3}$. }}
\begin{tabular}{lll}
\hline
\hline \\
\textbf{References} & \textbf{Redshift} &  \boldmath $\log \rho_\star$
\\ \\ \hline \\
\cite{1995ApJ...455L...1G} & 0.022 & -2.14 $\pm$ \hspace{0.1cm} 0.04 \\ \\
\cite{2007ApJ...657..738L} & 0.08 & -2.01 $\pm$ \hspace{0.1cm} 0.29 \\
 & 0.24 & -2.34 $\pm$ \hspace{0.1cm} 0.24 \\
 & 0.4 & -2.02 $\pm$ \hspace{0.1cm} 0.20 \\ \\
 \cite{2008ApJS..175..128S} & 0.24 & -1.97 $\pm$ \hspace{0.1cm} 0.12 \\ \\
 
 \cite{2010ApJ...712L.189D} & 0.16 & -2.23 $\pm$ \hspace{0.1cm} 0.20 \\
 & 0.24 & -2.11 $\pm$ \hspace{0.1cm} 0.21 \\
 & 0.32 & -1.92 $\pm$ \hspace{0.1cm} 0.23\\ 
  & 0.40 & -1.89 $\pm$ \hspace{0.1cm} 0.25 \\ \\
 
  \cite{2010ApJ...708..534W} & 0.05 & -2.41 $\pm$ \hspace{0.1cm} 0.10 \\
 & 0.15 & -2.15 $\pm$ \hspace{0.1cm} 0.09 \\
 & 0.25 & -2.05 $\pm$ \hspace{0.1cm} 0.05\\ 
  & 0.34 & -2.04 $\pm$ \hspace{0.1cm} 0.03 \\ \\

\cite{2013MNRAS.433..796D} & 0.25 & -2.52 $\pm$ \hspace{0.1cm} 0.12 \\
 & 0.4 & -2.18 $\pm$ \hspace{0.1cm} 0.19 \\
 & 0.5 & -1.74 $\pm$ \hspace{0.1cm} 0.05 \\ \\

\cite{2013MNRAS.428.1128S} & 0.40 & -1.75 $\pm$ \hspace{0.1cm} 0.15 \\ \\
\cite{2013MNRAS.433.2764G}  & 0.05 & -1.92 $\pm$ \hspace{0.1cm} 0.06 \\ 
 (GAMA) & 0.125 & -1.95 $\pm$ \hspace{0.1cm} 0.06 \\
 & 0.205 & -1.97 $\pm$ \hspace{0.1cm} 0.09 \\
 & 0.295 & -1.75 $\pm$ \hspace{0.1cm} 0.09 \\ \\
 \cite{2013MNRAS.433.2764G} & 0.05 & -2.01 $\pm$ \hspace{0.1cm} 0.06 \\ 
 (SDSS)  & 0.15 & -2.37 $\pm$ \hspace{0.1cm} 0.09 \\\\

 \cite{2015MNRAS.453..242S} & 0.2 & -2.03 $\pm$ \hspace{0.1cm} 0.09 \\ \\
 \cite{2016ApJ...824...25V} & 0.015 & -1.98 $\pm$ \hspace{0.1cm} 0.06 \\ \\
 \cite{2020MNRAS.493.3966K} & 0.47 & -1.86 $\pm$ \hspace{0.1cm} 0.04 \\ \\
 \cite{2021arXiv210104062V} & 0.012 & -2.34 $\pm$ \hspace{0.1cm} 0.11 \\ \\

\textbf{This work}  & 0.09 & -2.28 $\pm$ \hspace{0.1cm} 0.16 \\ 
& 0.216 & -2.02 $\pm$ \hspace{0.1cm} 0.11 \\
& 0.292 & -1.98 $\pm$ \hspace{0.1cm} 0.09 \\ \\

\end{tabular}

\label{Tab:HalphaSFR}
\end{table}

\subsection{Differences between the SFR derived through \Ha and the SED fitting}\label{subsec:diss_CESR}
\begin{figure*}
    \centering
        \includegraphics[width=\hsize]{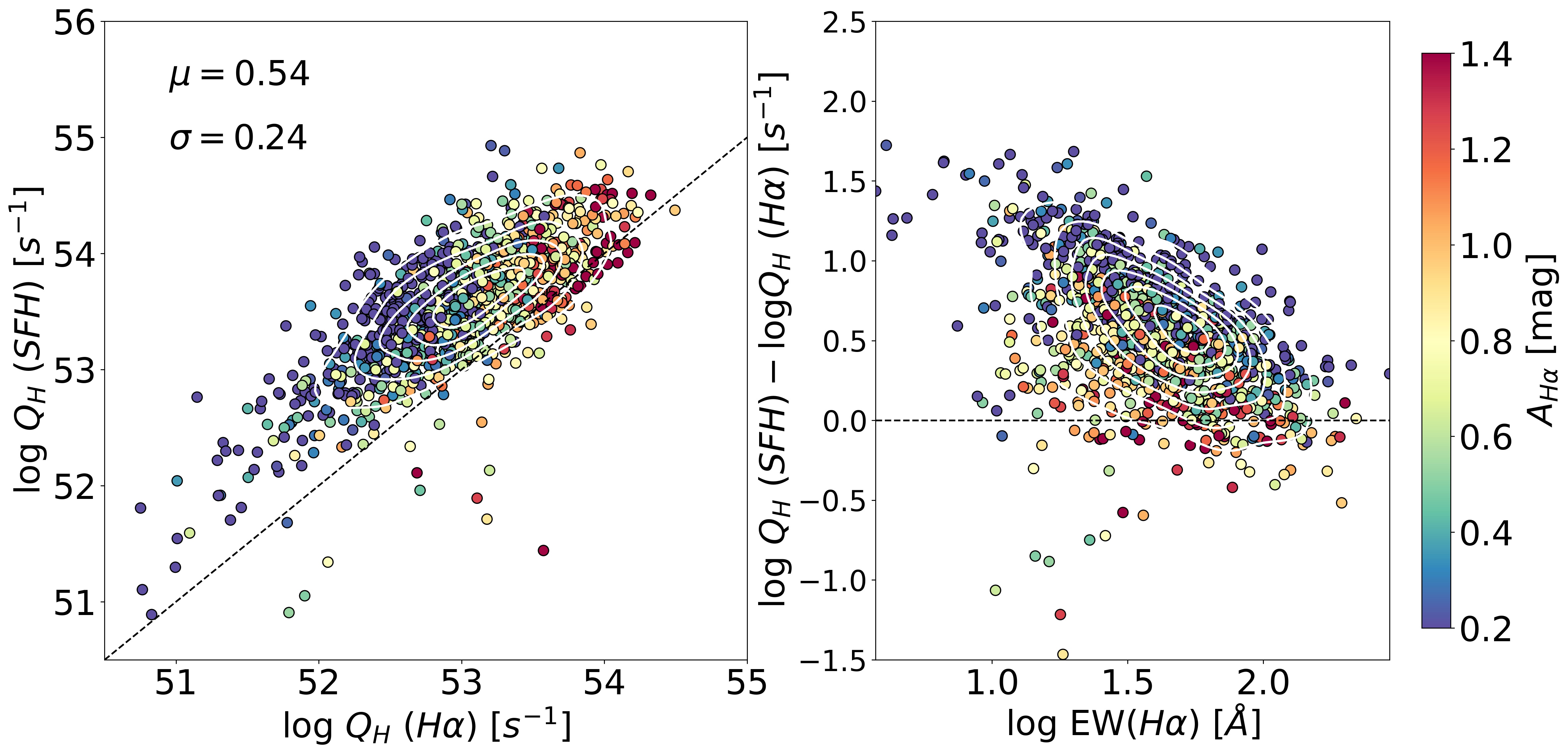}
        \caption{\tiny{Comparison of the ionizing photon rates computed from \Ha emission line and from the fit obtained with the analysis of the stellar populations with \texttt{BaySeAGal} (left; see text in Sect. \ref{subsec:diss_CESR}). The dashed black line represents the 1:1 relation. $\mu$ and $\sigma$ are the bias and the standard deviation. The right panel shows the difference between these quantities as a function of the EW of \Ha. Density contours are drawn in black. In both cases, the galaxies are color-coded with the extinction of the interstellar gas calculated from the Balmer decrement.}}
\label{fig:QH_Bay}
\end{figure*}

The star formation rate density derived in this work is compatible with previous studies that used the \Ha luminosity to determine its evolution with cosmic time in the nearby Universe. Nevertheless, our predictions are lower than those obtained with other methods based on the SED fitting of the stellar continuum. Even though $\rho_{\text{SFR}}$ might be lower in the miniJPAS field, meaning we are affected by the large cosmic variance, our results differ from those derived with the analysis of the stellar populations in \cite{2021arXiv210213121G}. 
\\ \\ In order to shed light on this difference, we compared the ionizing photon rates expected from \Ha luminosity and from the SED fitting. When we assume that no photons escape from \Hii regions, the relation between the dust-corrected luminosity of \Ha and the ionizing photon rates is
\begin{equation}
    Q_H^{H\alpha}  = x_{H\alpha} \frac{L_{H\alpha}}{h \nu_{H\alpha}}
\label{Eq:QH_halpha}
,\end{equation}
where $ x_{H\alpha} = 2.206 $ for case B hydrogen recombination. 
\\\\ In the case of the SED fitting, \texttt{BaySeAGal} provides the mass fraction ($\mu_j$) of each SSP that better describes the observed spectrum. In other words, for each galaxy, we can reproduce the SFH. Therefore, we can retrieve the ionizing photon rates by weighting the number of H ionizing photons emitted per unit time and initial mass for the jth SSP ($q_{H,j} = q_{H}(t_j,Z_j)$),
\begin{equation}
    Q_H^{SFH}  = M_\star \sum_{j=1}^{221} \mu_j q_{H,j} 
.\end{equation}
We compare the two quantifies in Fig. \ref{fig:QH_Bay}. $Q_H^{SFH}$ is 0.54 dex higher than $Q_H^{H\alpha}$ on average. We observe a clear trend with the nebular extinction (color bar) and the EW of \Ha. Galaxies where we estimated low values of the nebular extinction lie farther away from the 1:1 line. On the same line, the differences between $Q_H^{SFH}$ and $Q_H^{H\alpha}$ become smaller as the EW of \Ha increases. 
\par Interestingly, $Q_H^{SFH}$ and $Q_H^{H\alpha}$ are closer at higher values. This trend has also been found in comparisons between the SFR derived from \Ha and from the UV both in the integrated spectrum and in spatially resolved galaxies \citep{2009ApJ...706..599L,2016ApJ...817..177L,2021arXiv210614363B}. Specifically, \cite{2021arXiv210614363B} concluded that deficient \Ha fluxes in the extended disks of galaxies are tightly correlated with recent starbursts, which are being rapidly suppressed over the last 10 Myr. This phenomenon can explain the difference found in the slope of the SFMS in Sect.~ \ref{subsubsec:SFM_Bay}. Because galaxies with a low \Ha luminosity have higher SFRs according to the SED fitting, the slope becomes flatter. 
\par $Q_H^{SFH}$ might also be overestimated if the mass fraction attributed to young stellar populations (YSP) were higher than it should be. This might happen if the SFH in the last 20 Myr were different from the global SFH that accounts for the formation and growth of mass in galaxies on scales of billion years and/or because our parametric code overestimated the fraction of mass that formed in recent epochs with respect to nonparametric codes that are more flexible to varying the fraction of the young stellar population on a shorter timescale.  In order to determine how our result might be affected by different assumptions of the SFH, we used the SFH from ALSTAR and computed $Q_H^{SFH}$. We found that there is a bias of 0.81 dex, which is even higher than the results found with \texttt{BaySeAGal}. 
\par Studies that retrieved the stellar population properties of a sample of galaxies based on optical spectra (either form SDSS or CALIFA) and based on photometry from the GALEX survey showed that when the UV part of the spectrum is not included in the SED fitting, a brighter YSP contribution is found \citep{2016MNRAS.458..184L,2019MNRAS.483.2382W}. However, this excess of light in the UV does not have a strong impact on the mass content of YSP because the mass is dominated by older stars. 
\par \texttt{BaySeAGal} does not yet include a model of nebular emission lines. Therefore, the SED fitting only accounts for the emission of the stellar continuum and masks the filters in which the emission lines peak. We do not know how this might affect the shape of the SFH and the mass fraction attributed to the young stellar population. Moreover, a delta-delayed model might not be sufficient to describe SFHs with a recent burst of SFR. In the future, we expect to explore this aspect further. 
\par Furthermore, other hypotheses need to be taken into account to explain this discrepancy. First, we should consider whether we underestimate the nebular extinction. Certainly, we would expect that galaxies with very low S/N show this effect more. When we rebuild Fig. \ref{fig:QH_Bay} and include only galaxies with an error in \Ha luminosity smaller than 0.25 dex, the bias decreases by 0.17 dex. Additionally, when we assume for the SF sample that the nebular extinction is underestimated by a factor of two, which would mean $E(B-V)_{H\alpha/H\beta} \sim 2 E(B-V)_{SED}$ , as some studies reported \citep{2019ApJ...886...28Q,2019PASJ...71....8K}, the difference would only be reduced by 0.22 dex. In other words, it is plausible that we did not properly estimate the nebular extinction for a fraction of galaxies in the SF sample, but in the worst scenario, this effect alone cannot explain the difference between $Q_H^{SFH}$ and $Q_H^{H\alpha}$. 
\par Another effect that might also contribute to this difference is the  ionizing radiation that leaks from the \Hii regions. In this case, Eq. \ref{Eq:QH_halpha} would underestimate the \Ha ionizing photon rates. Several studies have shown precisely that there is a fraction of ionizing photons that escapes, and they are therefore unable to ionize the interstellar gas \citep{2005A&A...438..599G,2010A&A...511A..61O,2012ApJ...755...40P,2015ApJ...800..101A}. Nevertheless, the average fraction is still debated and can vary from galaxy to galaxy and from region to region within the same galaxy. Unfortunately, there is no means to quantify this effect with the data employed in this work. Nonetheless, when we assume that 30 $\%$ of the ionizing radiation leaks from \Hii regions, the difference could be reduced 0.16 dex.
\par Most probably, the difference that we observe between $Q_H^{SFH}$ and $Q_H^{H\alpha}$ is a combination of all these factors. Certainly, fitting the SED of miniJPAS galaxies with information from the UV from GALEX or HST-UV observations and/or the IR from \textit{SPITZER} would be very useful to unveil the origin of the discrepancy and test some of the previous hypotheses. However, this analysis is not the main goal of this work. 
\section{Outlook for J-PAS}\label{sec:outlook}
The results presented in this paper prove that the main properties of ELGs can be studied with J-PAS data. The miniJPAS Pathfinder instrument allowed us to test and combine different methods of analysis to fully exploit the scientific potential of the data and draw the baseline for the prospect of J-PAS.
\par The vast amount of data to be collected by J-PAS will allow us to perform a more comprehensive research, exploring other aspects that remained elusive or were limited within the area covered by miniJPAS. For instance, we will be able to derive the properties of blue and SF galaxies in groups and clusters, the fraction of AGN, and their role in the quench of SF galaxies within dense and very low density environments.
\par For instance, if in 1 deg$^2$ we were able to estimate the position of 255 galaxies in the BPT with an error smaller than 0.15 dex, the ionization mechanism of about two million galaxies in the Universe ($z < 0.35$) could be studied at the end of the J-PAS survey. With this amount of data, we will be able to determine the SFMS parameters better and place constraints on the evolution of $\rho_{\text{SFR}}$ at least up to 0.35 in redshift. Thus, it will be possible to further explore the discrepancies found in Sect.~ \ref{sec:CSFR}.
\par
\begin{figure}
    \centering
        \includegraphics[width=\hsize]{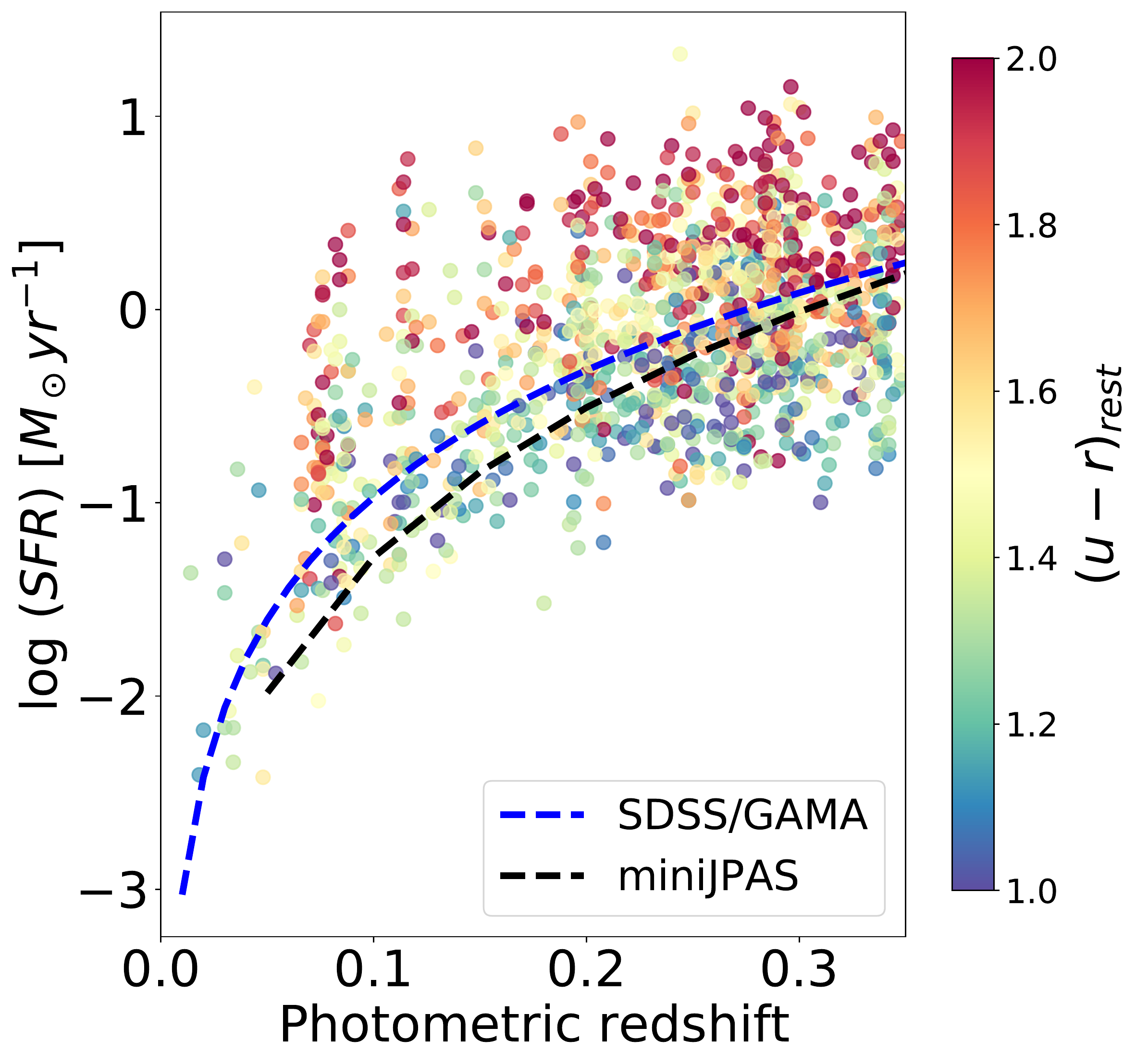}
        \caption{\tiny{Relation between the SFR derived from \Ha and redshift for the galaxy sample described in Sect.~ \ref{subsec:SFselection}. The blue dotted line is the approximate SFR completeness limit for GAMA and SDSS galaxies \citep{2013MNRAS.433.2764G}, and the dotted black line is the $95 \%$ completeness limit from blue galaxies in miniJPAS. Galaxies are color-coded with their (u-r) rest-frame color. }}
\label{fig:SFR_coverage}
\end{figure}

 The SFR coverage of J-PAS will be at least as competitive as that of the SDSS or GAMA surveys. In Fig. \ref{fig:SFR_coverage} we show the SFR as a function of the resdshift for our SF galaxy sample. The dotted blue line is the approximate SFR completeness limit assuming a flux limit of $F_{H\alpha} = 10^{-18} W m^{-2}$ for GAMA and SDSS galaxies \citep{2013MNRAS.433.2764G}. The dotted black line represents the 95 $\%$ completeness limit of miniJPAS for blue galaxies (DÃ­az-GarcÃ­a et. al. in prep). We used the best fit obtained in Sect.~ \ref{subsec:Bayesianroutine} to transform the completeness limit in mass into SFR. 
\par
\begin{figure}
    \centering
        \includegraphics[width=\hsize]{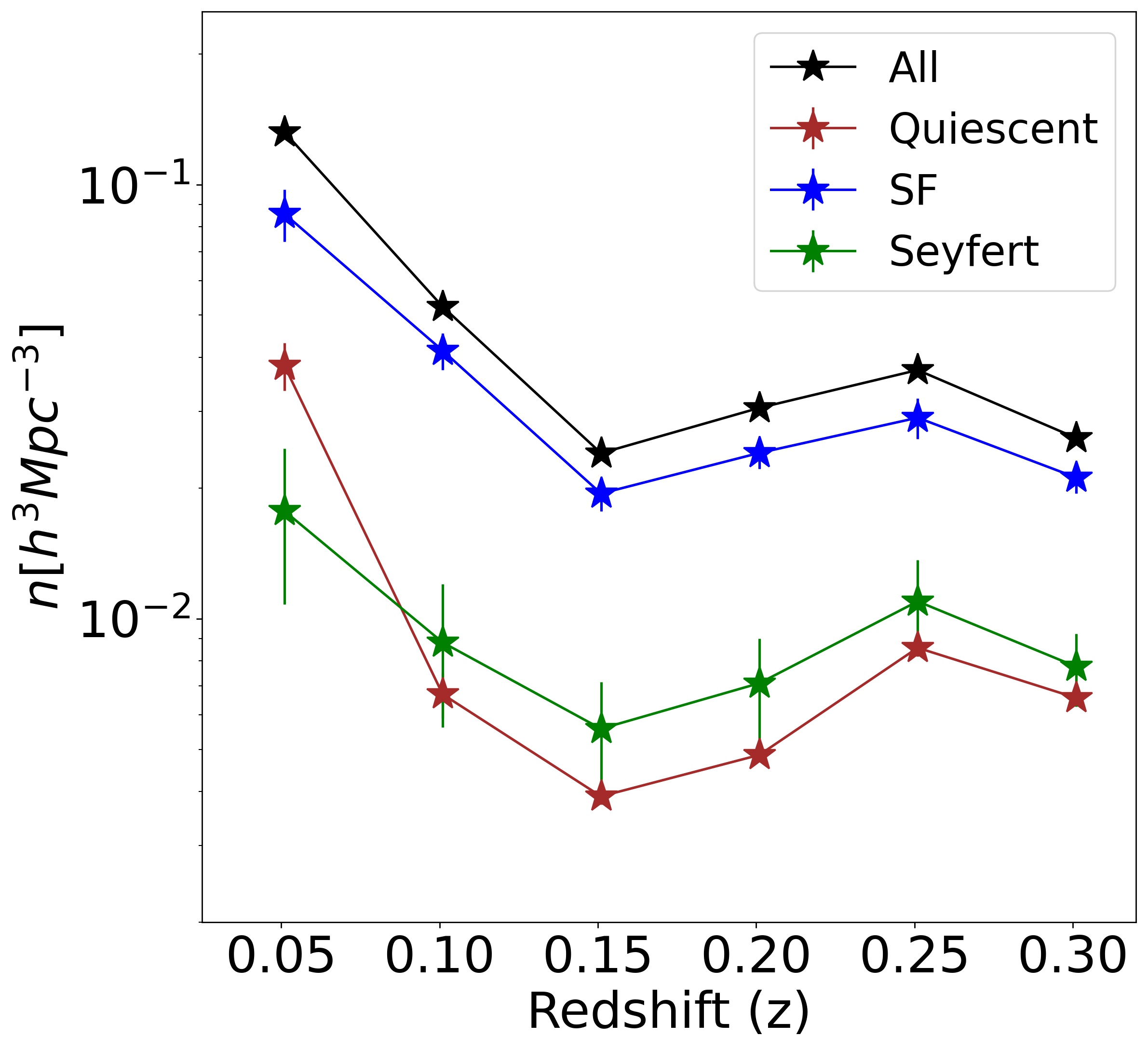}
        \caption{\tiny{Comoving number density of galaxies in miniJPAS as a function of redshift. The total galaxy population (black star) is broken into star-forming (blue stars), AGN-like (green stars), and quiescent galaxies (red stars). We used the WHAN diagram with the Ka03 dividing line to separate AGN and SF galaxies. Quiescent galaxies include LINERs and passive galaxies. The uncertainty due to the cosmic variance is not included in the error budget.}}
\label{fig:ndensity}
\end{figure}

 Finally, in Fig. \ref{fig:ndensity} we show the comoving number density of galaxies in miniJPAS as a function of redshift for the total galaxy population (black stars) for the star-forming galaxies (blue stars), for AGN-like galaxies (green stars), and for quiescent galaxies (red stars). Error bars represent the variation in the number density when a different division line in the WHAN diagram is considered, for example, k03, Ke01, or S08.

\section{Summary and conclusion}\label{sec:conclusions}
We analyzed a subsample of galaxies (a total of 2154) from the AEGIS field observed by miniJPAS with redshift below 0.35 in detail. The method developed in MS21 used ANN trained with CALIFA and MaNGA in order to measure and detect the main emission lines in the J-spectrum: \Ha, \Hb, \oiii, and \nii. 
\par We used a criterion based on the mass and color of the galaxy. We estimated that $83$ \% and $17$ \% in the sample are blue and red galaxies, respectively. With the ANN classifier, which is based on the EW of the emission lines, we found that $82$ \% of the sample are strong ELs amd $18$ \% are weak ELs.
\par We employed the BPT and WHAN diagrams to classify galaxies according to the main source of ionization and to select star-forming galaxies. We obtained that of the galaxies with reliable EW values (2000 galaxies in total), $72.8 \pm 0.4$ \%, $17.7 \pm 0.4$ \%, and $9.4 \pm 0.2$ \% are SF, Seyfert, and passive or LINER galaxies, respectively, using
the WHAN diagram and the Ka03 separation line. One hundred and fifty-four galaxies from the parent sample remain unclassified because of high uncertainties in the measurement of the emission lines. $\text{Ninety-four}$  percent of the SF galaxies and 97 \% of the LINER or passive galaxies are classified with the color criterion as blue and red, respectively.
\par The analyses of the properties of the stellar population performed in  \cite{2021arXiv210213121G} allowed us to compare and complement the information of the emission lines. For instance, we showed in color-mass diagrams that blue (red) galaxies are composed of a younger (older) stellar population, respectively, and present stronger (weaker) emission lines. This synergy between the properties of the gas and the stellar populations also appears in the BPT diagram, where galaxies become more massive as they evolve through the SF-wing.  
\par We derived the SFR from the flux of \Ha and relied on the Balmer decrement to correct for the extinction produced by interstellar dust. Subsequently, we fit the slope, zeropoint, and the intrinsic scatter of the SFMS, obtaining  $0.90^{+ 0.02}_{-0.02}$, $-8.85^{+ 0.19}_{-0.20}$ and,  $0.20^{+ 0.01}_{-0.01}$, respectively. We tested the turnover-mass hypothesis by fitting a quadratic and a broken power law. However, we did not observe a flattening of the slope at high mass. We argue that this is likely produced by our selection criteria of SF galaxies together with the limitation of the method to detect very weak emission lines in comparison with spectroscopic surveys. The results we obtained are compatible with those of other studies.
\par Finally, we computed the cosmic evolution of the $\rho_{\text{SFR}}$ within three redshift bins: $0 < z \leq 0.15$, $0.15 < z \leq 0.25$, and $0.25 < z \leq 0.35$. We found agreement with previous measurements based on the \Ha emission line. Nevertheless, we found an offset compared to the studies that derived $\rho_{\text{SFR}}$ from the SED fitting of the stellar continuum. We discussed the origin of this discrepancy in detail, which is most probably a combination of several factors, such as the correction for dust attenuation, the SFR tracer, or the escape of ionizing photons.
\par The work presented in this paper builds the foundation upon which the analysis of ELGs in J-PAS will be conducted as soon as hundreds of squares degrees are mapped in the northern sky in the next years.

\begin{acknowledgements} 
G.M.S., R.G.D., R.G.B., L.A.D.G., J.R.M., and E.P. acknowledge financial support from the State Agency for Research of the Spanish MCIU through the "Center of Excellence Severo Ochoa" award to the Instituto de Astrof\'isica de Andaluc\'i a (SEV-2017-0709), and to the AYA2016-77846-P and PID2019-109067-GB100. SDP is grateful to the Fonds de Recherche du Qu\'ebec - Nature et Technologies. J.C.M. and S.B. acknowledge financial support from Spanish Ministry of Science, Innovation, and Universities through the project PGC2018-097585-B-C22. JVM and SDP acknowledge financial support from the Spanish Ministerio de Econom\'ia y Competitividad under grants AYA2016-79724-C4-4-P and PID2019-107408GB-C44, from Junta de Andaluc\'ia Excellence Project P18-FR-2664, and the funding of the "Center of Excellence Severo Ochoa" award to the Instituto de Astrof\'\i sica de Andaluc\'\i a (SEV-2017-0709). AE and JAFO acknowledges the financial support from the Spanish Ministry of Science and Innovation and the European Union -NextGenerationEU through the Recovery and Resilience Facility project ICTS-MRR-2021-03-CEFCA. LSJ acknowledges support from CNPq (304819/2017-4) and FAPESP (2019/10923-5). R.A.D. acknowledges partial support support from CNPq grant 308105/2018-4. V.M. thanks CNPq (Brazil) for partial financial support. This project has received funding from the European Union's Horizon 2020 research and innovation programme under the Marie Skłodowska-Curie grant agreement No 888258. Based on observations made with the JST/T250 telescope and PathFinder camera for the miniJPAS project at the Observatorio Astrof\'isico de Javalambre (OAJ), in Teruel, owned, managed, and operated by the Centro de Estudios de F\'isica del  Cosmos de Arag\'on (CEFCA). We acknowledge the OAJ Data Processing and Archiving Unit (UPAD) for reducing and calibrating the OAJ data used in this work. Funding for OAJ, UPAD, and CEFCA has been provided by the Governments of Spain and Arag\'on through the Fondo de Inversiones de Teruel; the Arag\'on Government through the Research Groups E96, E103, and E16\_17R; the Spanish Ministry of Science, Innovation and Universities (MCIU/AEI/FEDER, UE) with grant PGC2018-097585-B-C21; the Spanish Ministry of Economy and Competitiveness (MINECO/FEDER, UE) under AYA2015-66211-C2-1-P, AYA2015-66211-C2-2, AYA2012-30789, and ICTS-2009-14; European FEDER funding (FCDD10-4E-867, FCDD13-4E-2685); the Brazilian agencies FINEP, FAPESP, FAPERJ and by the National Observatory of Brazil. Additional funding was also provided by the Tartu Observatory and by the J-PAS Chinese Astronomical Consortium. The authors acknowledge the following people for providing valuable comments and suggestions on the first draft of this paper: Iris Breda, Ana Chies Santos, Maria Luiza Linhares Dantas, Alejando Lumbreras Calle, Elmo Tempel, Jos\'e Eduardo Telles, and Rahna P T. We also thank the anonymous referee for many useful comments and suggestions.
\end{acknowledgements}

\bibliographystyle{aa}
\bibliography{aa}

\appendix

\section{AGN selection criteria}
In Table \ref{Tab:comparison_SFMS} we show the best-fitting parameter as a function of the separation curves, the redshift bin, and the fitting equation we used to fit the SFMS. The results are discussed in the main text (Sect.~ \ref{sub:AGN_criteria})
\begin{table*}[ht]
\caption{\tiny{Parameters of the SFMS derived in different redshift bins with the models described in Sects. \ref{subsec:Bayesianroutine} and \ref{subsec:turn_over} using different selection criteria (see Sect. \ref{sub:AGN_criteria}) }}
\begin{tabular}{|l|l|l|l|l|l|l|l|l|l|}
\hline
\hline
Sample &  Size&  \nii / \Ha& Fitting equation &   $\alpha$ & $\beta$ & $\sigma_{int}$ & $\gamma$ & $M_{0}$ \\ 
&   &   &  & &  & &  &  \\   \hline
 \multirow{9}{*}{$0 < z \leq 0.35$} &   \multirow{3}{*}{1361} & \multirow{3}{*}{$ \le 0.79$ (Ke01)}  & Power law &  $0.88^{+ 0.02}_{-0.02}$ & $-8.69^{+ 0.18}_{-0.18}$ & $0.20^{+ 0.01}_{-0.01}$  & - & - \\ 
 &   &   & Broken power law &  $0.84^{+ 0.03}_{-0.03}$ & $-0.89^{+ 0.09}_{-0.07}$  & $0.20^{+ 0.01}_{-0.01}$  & - & $10.75^{+ 0.18}_{-0.14}$\\ 
  &   &   & Quadratic power law & $2.49^{+ 0.34}_{-0.35}$ & $-15.50^{+ 1.71}_{-1.76}$  & $0.20^{+ 0.01}_{-0.01}$  & $0.09^{+ 0.02}_{-0.02}$ & -\\  
&   &   &  & &  & &  &  \\   
 &   \multirow{3}{*}{1178} & \multirow{3}{*}{$ \le 0.48$ (Ka03)}  & Power law &  $0.90^{+ 0.02}_{-0.02}$ & $-8.85^{+ 0.19}_{-0.20}$ & $0.20^{+ 0.01}_{-0.01}$ & - & -  \\ 
 &   &   & Broken power law &  $0.82^{+ 0.03}_{-0.03}$ & $-0.99^{+ 0.12}_{-0.09}$ & $0.20^{+ 0.01}_{-0.01}$ & - & $10.93^{+ 0.22}_{-0.17}$  \\ 
  &   &   & Quadratic power law & $2.21^{+ 0.33}_{-0.33}$ & $-14.18^{+ 1.61}_{-1.61}$ & $0.20^{+ 0.01}_{-0.01}$ & $0.08^{+ 0.02}_{-0.02}$ &  \\  
    &   &   &  & &  & &  &  \\    
 &   \multirow{3}{*}{1026} & \multirow{3}{*}{$ \le 0.40$ (Ke01)}  & Power law &  $0.92^{+ 0.02}_{-0.02}$ & $-8.99^{+ 0.20}_{-0.20}$ & $0.20^{+ 0.01}_{-0.01}$ & - & -  \\ 
 &   &   & Broken power law &  $0.82^{+ 0.03}_{-0.04}$ & $-1.08^{+ 0.18}_{-0.13}$  & $0.20^{+ 0.01}_{-0.01}$ & - & $11.01^{+ 0.32}_{-0.21}$ \\ 
  &   &   & Quadratic power law & $2.11^{+ 0.37}_{-0.36}$ & $-13.80^{+ 1.75}_{-1.75}$ & $0.19^{+ 0.01}_{-0.01}$  & $0.07^{+ 0.02}_{-0.02}$ & - \\  
  &   &   &  & &  & &  &  \\   \hline
    &   &   &  & &  & &  &  \\   
 \multirow{3}{*}{$0 < z \leq 0.15$} &  220  & $ \le 0.79$ (Ke01)  &  \multirow{3}{*}{Power law} &  $0.84^{+ 0.04}_{-0.03}$ & $-8.40^{+ 0.33}_{-0.34}$ & $0.20^{+ 0.02}_{-0.02}$ & - & -  \\ 
&  197 & $ \le 0.48$ (Ka03)  &  & $0.85^{+ 0.04}_{-0.04}$ & $-8.54^{+ 0.34}_{-0.38}$ & $0.21^{+ 0.02}_{-0.02}$ & - & -  \\  
&171  & $ \le 0.40$ (S08)  &  &  $0.90^{+ 0.04}_{-0.04}$ & $-8.97^{+ 0.41}_{-0.42}$ & $0.21^{+ 0.02}_{-0.02}$ & - & -  \\  
  &   &   &  & &  & &  &  \\   \hline
    &   &   &  & &  & &  &  \\   
 \multirow{3}{*}{$0.15 < z \leq 0.25$} &  461  & $ \le 0.79$ (S08)  &  \multirow{3}{*}{Power law} &  $0.77^{+ 0.04}_{-0.04}$ & $-7.52^{+ 0.36}_{-0.37}$ & $0.18^{+ 0.02}_{-0.02}$ & - & -  \\ 
&  384 & $ \le 0.48$ (Ka03)  &  &  $0.77^{+ 0.04}_{-0.03}$ & $-7.54^{+ 0.36}_{-0.37}$ & $0.17^{+ 0.02}_{-0.02}$ & - & -  \\ 
&336  & $ \le 0.40$ (Ke01)  &  &  $0.81^{+ 0.04}_{-0.04}$ & $-7.88^{+ 0.39}_{-0.42}$ & $0.17^{+ 0.02}_{-0.02}$ & - & -  \\  
  &   &   &  & &  & &  &  \\   \hline
 
    &   &   &  &  &  & &  &  \\   
 \multirow{3}{*}{$0.25 < z \leq 0.35$} &  641  & $ \le 0.79$ (S08)  &  \multirow{3}{*}{Power law} &  $0.81^{+ 0.04}_{-0.04}$ & $-7.94^{+ 0.35}_{-0.38}$ & $0.06^{+ 0.04}_{-0.06}$ & - & -  \\ 
&  561 & $ \le 0.48$ (Ka03)  &  &  $0.85^{+ 0.03}_{-0.03}$ & $-8.26^{+ 0.35}_{-0.36}$ & $0.00^{+ 0.06}_{-0.00}$ & - & -  \\ 
&488  & $ \le 0.40$ (Ke01)  &  & $0.82^{+ 0.04}_{-0.04}$ & $-7.98^{+ 0.41}_{-0.42}$ & $0.00^{+ 0.01}_{-0.00}$ & - & -  \\ 
  &   &   &  &  &  &  &  &  \\   \hline
\label{Tab:comparison_SFMS}
\end{tabular}

\end{table*}
\end{document}